\documentclass[preprint]{aastex}
\usepackage[]{verbatim}
\usepackage[]{float}
\usepackage[caption=false]{subfig}
\usepackage[]{amsmath}
\usepackage{graphics,graphicx}
\usepackage[]{natbib}
\usepackage{latexsym,amsfonts,amsmath,amssymb}
\usepackage{gensymb}
\newcommand{\upperRomanNumeral}[1]{\uppercase\expandafter{\romannumeral#1}}

\begin{document}
\title{ALMA Observations of Circumstellar Disks in the Upper Scorpius OB Association}

\author{Scott A. Barenfeld}
\affil{California Institute of Technology, Department of Astronomy, MC 249-17, Pasadena, CA 91125, USA}

\author{John M. Carpenter}
\affil{Joint ALMA Observatory, Av. Alonso de C{\'o}rdova 3107, Vitacura, Santiago, Chile}

\author{Luca Ricci}
\affil{Harvard-Smithsonian Center for Astrophysics, 60 Garden Street, Cambridge, MA 02138, USA}

\author{Andrea Isella}
\affil{Department of Physics and Astronomy, Rice University, 6100 Main Street, Houston, TX 77005, USA}

\begin{abstract}

We present ALMA observations of 106 G-, K-, and M-type stars in the Upper Scorpius OB Association hosting 
circumstellar disks. With these data, we measure the 0.88 mm continuum and $^{12}$CO $J$ = 3$-$2 
line fluxes of disks around low-mass ($0.14-1.66$ $M_{\odot}$) stars at an age of 5-11 Myr.  
Of the 75 primordial disks in the sample, 53 are detected in the dust continuum and 26 in CO.  Of the 31 disks 
classified as debris/evolved transitional disks, 5 are detected in the continuum and none in CO.  The lack of CO 
emission in approximately half of the disks with detected continuum emission can be explained if CO is optically thick but has a 
compact emitting area ($\lesssim 40$ au), or if the CO is heavily 
depleted by a factor of at least $\sim1000$ relative to interstellar medium abundances and is optically thin.  The continuum 
measurements are used to estimate the dust mass of the disks.  We find a correlation between disk dust mass and 
stellar host mass consistent with a power-law relation of $M_{\mathrm{dust}}\propto M_*^{1.67\pm0.37}$.  Disk dust masses in Upper Sco are 
compared to those measured in the younger 
Taurus star-forming region to constrain the evolution of disk dust mass.  We find that the difference in the mean of $\log(M_{\mathrm{dust}}/M_*)$ 
between Taurus and Upper Sco is $0.64\pm0.09$, such that $M_{\mathrm{dust}}/M_*$ is lower in Upper Sco by a factor of $\sim4.5$.

\end{abstract}

\section{Introduction}
The lifetime of protoplanetary disks is closely linked to planet formation. In the core accretion theory of planet formation, the formation of gas giant 
planets is a race to accumulate a solid core large enough to rapidly accrete gas before the gas and dust 
in the disk disappear.  A key step in this process is the growth of solid material from micron-sized dust grains to kilometer-sized planetesimals, which can then collisionally grow into 
the cores of gas giants \citep{Mordasini2010}. The ability to form these planetesimals depends strongly on 
conditions within the disk, and in particular on the disk mass in solids. 
The time available for planetesimals to form is therefore set by the decline in disk dust mass as the disk 
evolves. 

The past decade has seen tremendous growth in our understanding of circumstellar disks. 
Infrared observations with the \emph{Spitzer Space Telescope} \citep{Werner2004} and the \emph{Wide-field Infrared Survey 
Explorer} \citep[\emph{WISE,}][]{Wright2010}
have cataloged hundreds of disks in nearby star-forming regions, revealing spectral energy distributions 
indicative of optically thick, irradiated dust disks surrounding an exposed stellar photosphere \citep[e.g.][]{Gutermuth2004,Hartmann2005,Megeath2005,Carpenter2006,Lada2006,
Sicilia-Aguilar2006,Balog2007,Barrado2007,Cieza2007,Dahm2007,Hernandez2007a,Hernandez2007b,Flaherty2008,Gutermuth2008,Hernandez2008,Luhman2012}.  
Collectively, these surveys have shown that disks surround $\sim$80\% of K- and M-type stars at an age of $\sim$1 Myr, but by an age of 
$\sim$5 Myr, only $\sim$20\% of stars retain a disk as traced by infrared dust emission.

Submillimeter observations complement this picture by revealing disk dust masses.  While infrared data 
probe only the warm dust within 1 au of the star, most of the solid mass in disks will be in the outer regions.  
To study this colder dust, submillimeter observations are required.  
At these wavelengths, dust emission in disks is generally optically thin, providing a measure of the total surface area 
of millimeter-sized grains in the disk \citep[e.g.][]{Ricci2010}. Combined with assumptions about the disk temperature and 
dust opacity, this can be used to derive the total mass of solids in the disk \citep[e.g.][]{Beckwith1990,Andre1994,Motte1998,Andrews2005,Andrews2007}.  
By further obtaining spatially resolved images of the disk with interferometers, the surface density of the disk can be inferred \citep[e.g.][]{Kitamura2002,Andrews2007,Andrews2009,Isella2009,Isella2010,Guilloteau2011}. 
\citet{Andrews2013} combined new observations and literature values to create a catalog of disk fluxes at 1.3 mm for 179 stars earlier than M8.5 in the 1-2 Myr old Taurus star-forming region. 
The authors found a statistically significant, approximately linear correlation between disk mass and stellar mass, with the disk mass 
typically between 0.2\% and 0.6\% of the stellar host mass. 

To study the \emph{evolution} of disks, it is necessary to compare disk properties in Taurus to disk properties in regions of different ages.  
However, observational constraints on older disks remain relatively sparse at submillimeter wavelengths. 
Surveys of IC 348 \citep[age $\sim$2-3 Myr,][]{Carpenter2002,Lee2011}, Lupus \citep[age $\sim$3 Myr,][]{Nuernberger1997}, 
$\sigma$ Orionis \citep[age $\sim$3 Myr,][]{Williams2013}, $\lambda$ Orionis \citep[age $\sim$5 Myr,][]{Ansdell2015}, 
and the Upper Scorpius OB association \citep[age $\sim$5-11 Myr,][]{Mathews2012} 
suggest that these older regions lack disks as bright as the most luminous objects in younger regions. However, the small number of detected objects in these 
surveys make it difficult to draw firm conclusions about the nature of disks at older ages or how disk properties change in time \citep[see discussion in][]{Andrews2013}. 
Of the older systems studied thus far, the Upper Scorpius OB association (hereafter Upper Sco) in particular represents an ideal sample for studying disk evolution. The 5-11 Myr age of 
Upper Sco \citep{Preibisch2002,Pecaut2012} places its disks at the critical stage when infrared observations indicate that disk 
dissipation is nearly complete. \citet{Carpenter2014} presented results of an ALMA 0.88 mm continuum survey of 20 disk-bearing stars 
in Upper Sco, achieving an order of magnitude improvement in sensitivity over previous surveys. By comparing their results with the \citet{Andrews2013} Taurus catalog, 
they found that, on average, disk dust masses in Upper Sco are lower than in Taurus. However, due to the small size of the Upper Sco sample, the difference 
was not statistically significant. 

We present additional ALMA observations of disks in Upper Sco, expanding the \citet{Carpenter2014} sample to 106 stars. 
This study represents the largest survey of its kind for 5-11 Myr old stars. With these data, we measure continuum and CO line emission to establish 
the demographics of disk luminosities at an age of 5-11 Myr, when disks are in the final stages of dissipation. 
We then compare the distribution of disk dust masses in Upper Sco to that in Taurus in order 
to quantify the evolution of dust mass in disks between an age of 1-2 Myr and 5-11 Myr. In a future paper, we will analyze the CO 
measurements in detail to study the gas in disks at the end of their evolution.

\section{Stellar Sample}
Our ALMA sample consists of 106 stars in Upper Sco between spectral types 
of M5 and G2 (inclusive) that are thought to be surrounded by a disk based on the presence of 
excess infrared emission observed by \emph{Spitzer} or \emph{WISE} \citep{Carpenter2006,Luhman2012}. Twenty of these 
stars were observed in ALMA Cycle 0 using the disk sample obtained by 
\citet{Carpenter2006} who used \emph{Spitzer} observations.  The remaining stars were observed in Cycle 2 based 
on the compilation of 235 stars with disks in Upper Sco identified by \citet{Luhman2012}.
\footnote{One 
star in this compilation, 2MASS J16113134-1838259 (AS 205), has been previously considered a member of the 
$\rho$ Ophiuchus region by numerous authors \citep[e.g.][]{Prato2003,Eisner2005,Andrews2009}. More recently, 
\citet{Reboussin2015} considered AS 205 to be a member of Upper Sco, and this star was included in the 
\citet{Luhman2012} Upper Sco disk catalog. Given the fact that AS 205 is well separated from the main $\rho$ 
Ophiuchus clouds \citep[see Figure 1 of][]{Reboussin2015}, we consider this star to be a member of Upper Sco 
and include it in our sample.}
The combined ALMA Cycle 0 and Cycle 2 observations observe all 100 disk-host candidates in 
\citet{Luhman2012} with spectral types between M4.75 and G2, as well as six M5 stars. The ALMA sample 
is not complete at M5.

\begin{figure}[!h]
\centerline{\includegraphics[scale=1.0]{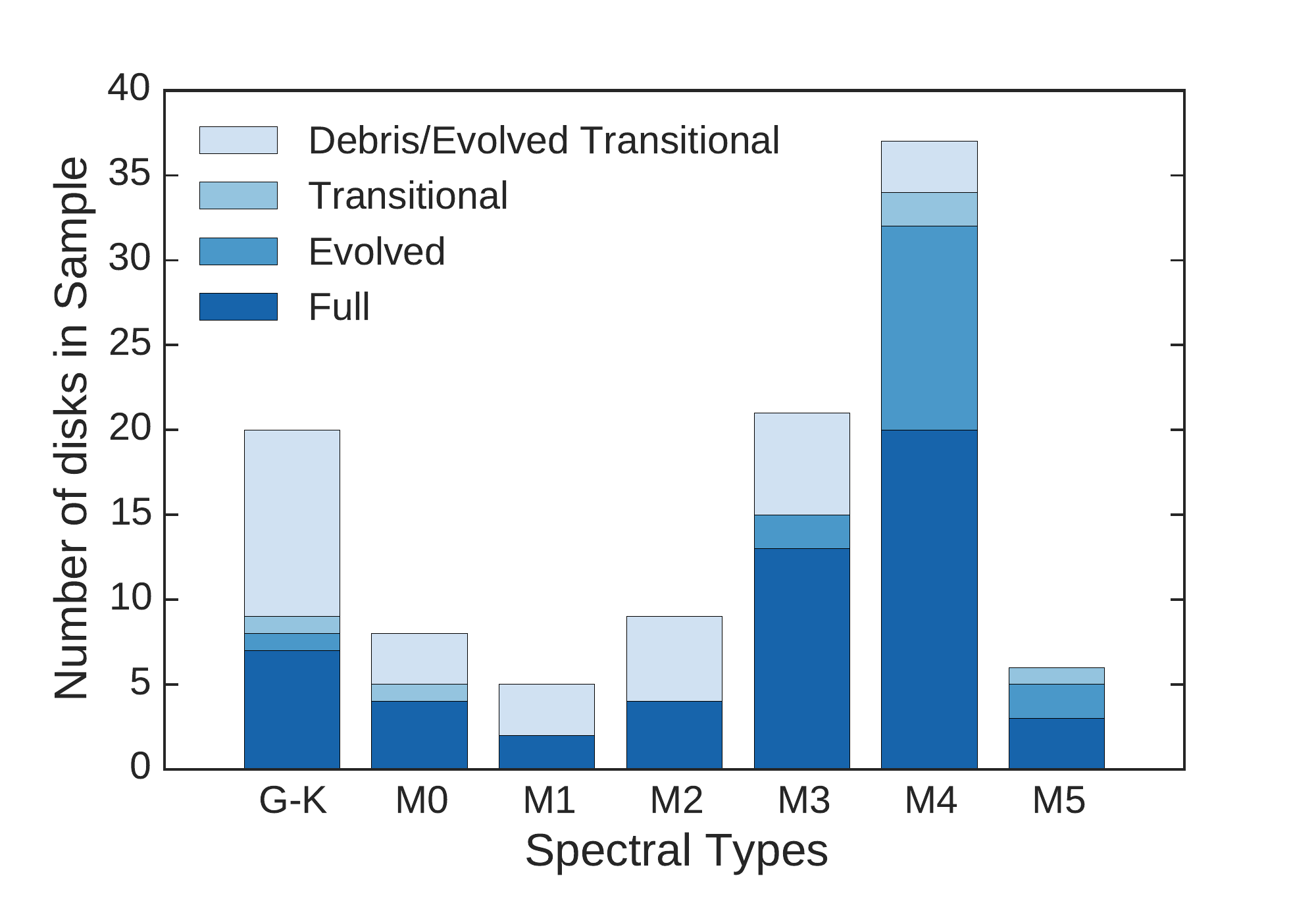}}
\caption{Distribution of disk types, as defined by \citet{Luhman2012}, in the Upper Sco sample grouped by spectral type.}
\label{fig:sample}
\end{figure}

Of our sources, 75 are classified by \citet{Luhman2012} 
as ``full'' (optically thick in the infrared with an SED that shows no evidence of disk clearing, 53 sources), 
``transitional'' (with an SED showing evidence for gaps and 
holes, 5 sources), or ``evolved'' (becoming optically thin in the infrared, but no evidence of clearing, 17 sources) disks. 
We consider these systems to be ``primordial'' disks.  The remaining 31 sources have infrared SEDs indicative 
of an optically thin disk with a large inner hole.  These are classified by \citet{Luhman2012} as 
``debris/evolved transitional'' disks and represent either young debris disks composed of second-generation 
dust originating from the collisional destruction of planetesimals, or the final phase of primordial disk evolution.
Figure \ref{fig:sample} shows the distribution of disk types in our sample.

Stellar luminosities ($L_*$), effective temperatures ($T_*$), and masses ($M_*$) were determined as described 
in \citet{Carpenter2014}.  Briefly, luminosity was estimated using J-band photometry from the Two Micron All Sky 
Survey \citep[2MASS,][]{Cutri2003,Skrutskie2006} and bolometric corrections for 5-30 Myr stars from \citet{Pecaut2013}. Visual 
extinction ($A_V$) was calculated using DENIS $I-J$ colors \citep{DENIS2005}, intrinsic colors from \citet{Pecaut2013}, 
and the \citet{Cardelli1989} extinction law. Effective temperatures were estimated from spectral type as in \citet{Andrews2013} using the 
temperature scales of \citet{SchmidtKaler1982}, \citet{Straizys1992}, and \citet{Luhman1999}.  Spectral types were taken from 
\citet{Luhman2012}, with an assumed uncertainty of $\pm$1 subclass. Stellar masses 
were then determined from $T_*$ and $L_*$ using the \citet{Siess2000} pre-main-sequence evolutionary tracks with a 
metallicity of $Z=0.02$ and no convective overshoot.  Uncertainties in stellar mass reflect uncertainties in luminosity 
(incorporating photometric, bolometric correction, and extinction uncertainties, as well as a $\pm$20 pc uncertainty in the distance 
to Upper Sco of 145 pc) and temperature (reflecting uncertainty in spectral type).
The derived stellar properties are given in Table \ref{tab:starProp}.

\section{ALMA Observations}

ALMA observations were obtained in Cycle 0 and Cycle 2 using the 12 m array.  Twenty sources were observed in 
Cycle 0 between 2012 August and 2012 December.  Eighty-seven sources were observed in 2014 June and 2014 July.
2MASS J16064385-1908056 was observed in 
Cycle 0 and had a marginal (2.5$\sigma$) continuum disk detection. Since the Cycle 0 observations did not achieve the 
requested sensitivity, the source was re-observed in Cycle 2. The Cycle 2 data have a factor of 2.8 better signal-
to-noise for this source than the Cycle 0 data; therefore, the Cycle 2 data are used throughout the paper for this source.

All observations used band 7 with the correlator configured to record dual polarization.   
Spectral windows for Cycle 2 were centered at 334.2, 336.1, 346.2, and 348.1 GHz for a mean frequency of 341.1 GHz (0.88 mm). 
The bandwidth of each window is 1.875 GHz.  The 345.8 GHz window has channel widths of 0.488 MHz (0.429 km s$^{-1}$) to 
observe the $^{12}$CO $J$ = 3$-$2 line. The spectral resolution is twice the 
channel width.
Table \ref{tab:obs} summarizes the observations, showing the number of antennas, baseline range, precipitable water vapor (pwv), and calibrators for 
each day. Cycle 0 observations used between 17 and 28 antennas with maximum baselines of $\sim$400 m, for an angular 
resolution of $\sim$0$\overset{''}{.}$55.  Cycle 2 observations used between 34 and 36 antennas with baselines extending out to 
650 m, corresponding to an angular resolution of 0$\overset{''}{.}$34.  The full-width-at-half-maximum (FWHM) primary beam size of the observations is 18$\overset{''}{.}$5.  
The typical on-source integration times were 5.5 minutes for Cycle 0 observations and 2.5 minutes for Cycle 2 observations.

The data were calibrated using the Common Astronomy Software Applications (CASA) package \citep{McMullin2007}.  The reduction scripts were kindly provided by the ALMA project. 
Data reduction steps include atmospheric calibration using the 183 GHz water vapor radiometers, bandpass calibration, flux calibration, and gain 
calibration. The calibrators for each observation date are listed in Table \ref{tab:obs}. We assume a 1$\sigma$ calibration uncertainty of 10$\%$. 

We rescaled the uncertainties of the visibility measurements to reflect the empirical scatter in the data so that the appropriate values of 
the uncertainties are used in model fitting (see Section \ref{sec:cont_flux}).  For each source, the visibilities were placed on a grid in \emph{uv} 
space for each spectral window and polarization.  At every grid cell, a scale factor was calculated to match the $\sigma$ 
values of the visibilities within that cell to their empirical scatter. The median scaling factor of the cells with at least 
10 visibilities was then applied to all $\sigma$ values for that polarization and spectral window.

\section{ALMA Results}

In this section, we use the ALMA observations described above to measure the 0.88 mm continuum and $^{12}$CO 
$J$ = 3$-$2 line fluxes of the 106 Upper Sco targets in our sample.

\subsection{Continuum Fluxes}
\label{sec:cont_flux}

To measure the submillimeter continuum flux density, the four spectral windows were combined after excluding 
a -15 to +30 km s$^{-1}$ region about the $^{12}$CO $J$ = 3$-$2 rest frequency in the frame of the local standard of rest (LSR). 
This safely excludes CO emission at the expected 0-10 km s$^{-1}$ LSR radial velocities \citep{deZeeuw1999,Chen2011,Dahm2012} of our Upper Sco targets. 
Flux densities were determined by first fitting a point source to the visibility 
data using the \emph{uvmodelfit} routine in CASA. The point-source model contains three free parameters: 
the integrated flux density and the right ascension and declination offsets from the phase center. 
If the flux density of a source is less than three times 
its statistical uncertainty, the source is considered a non-detection and we re-fit a point source to the visibilities with the 
offset position fixed at the expected stellar position. Expected positions were estimated using 
stellar positions from 2MASS \citep{Cutri2003,Skrutskie2006} and proper motions from the PPMXL catalog 
\citep{Roeser2010}.  For stars lacking PPMXL measurements, the median proper motion of the remainder of the sample (-11.3 km s$^{-1}$, -24.9 km s$^{-1}$) 
was used. 2MASS J16041740-1942287 
has a PPMXL proper motion discrepant from the median proper motion of Upper Sco. However, this star may be blended with two neighboring stars, calling 
into question the PPMXL data, which may compromise the measured proper motion.  We therefore also adopt the sample median proper motion for this star.

If the source was detected, an 
elliptical Gaussian model was also fit with \emph{uvmodelfit}.  This model includes an additional three parameters: 
the FWHM, aspect ratio, and position angle of the major axis. 
To determine which model best describes the data, we used the 
Bayesian Information Criterion (BIC) test. This test evaluates the relative 
probabilities of models describing a data set, while penalizing models for having additional free parameters. 
For each source, if the probability of a point-source model relative to an elliptical Gaussian model is $<0.0027$ ($3\sigma$ confidence), 
we adopt the latter model for the source. Otherwise, we adopt the point-source model.  Nine sources were fit with elliptical Gaussians, 
with deconvolved FWHM disk sizes ranging from 0$\overset{''}{.}$140 to 0$\overset{''}{.}$492, corresponding to $\sim$20$-$70 au at the 145 pc distance of Upper Sco. 
Two additional sources, 2MASS J15583692-2257153 and 
2MASS J16042165-2130284, were well-resolved and showed centrally depleted cavities that were not well described by either a point source or elliptical Gaussian 
at the resolution of our data. We measured the flux of 2MASS J15583692-2257153 using 
aperture photometry with a 0$\overset{''}{.}$6 radius circular aperture. For 2MASS J6042165-2130284, we adopt a flux of $218.76\pm0.81$ mJy measured by \citet{Zhang2014} using 
a 1$\overset{''}{.}$5 radius circular aperture.  At the distance of Upper Sco, these apertures correspond to radii of 87 and 218 au, respectively.

Unlike in the image domain, it is not possible to specify a boundary within which to fit the brightness profile of a source when 
fitting visibilities directly.  Thus, if there is a second bright source in the field, this could potentially bias the 
fit of a single source.
To account for possible contamination to the measured flux from sources elsewhere in the field, we searched each field in the 
image domain for any pixels (not including those associated with the target star) brighter than five times the RMS noise of the image. 
Ten such sources are were detected toward 9 of the 106 fields (see Table \ref{tab:sec_fluxes}).  For these sources, multiple-component models of a point 
source or elliptical Gaussian (determined as described above) were fit to each source using the \emph{uvmultifit} Python library 
\citep{Marti-Vidal2014}. Point-source models were used to fit all secondary sources.  
Fluxes and positions determined in this way for the secondary detections are listed in Table \ref{tab:sec_fluxes}. 
A search of the NASA/IPAC Extragalactic Database reveals that no known background galaxies are present at the positions of the secondary sources.

The secondary sources in the fields 
of 2MASS J16113134-1838259 and 2MASS J16135434-2320342 are also detected in CO at the expected radial velocity of Upper Sco. 2MASS J16113134-1838259 
is a known hierarchical triple system, in which the southern source is itself a spectroscopic binary.  
The southern binary is separated by 1$\overset{''}{.}$31 from the northern single star \citep{Eisner2005}. 
2MASS J16135434-2320342 has not been previously classified as a multiple system.  The fitted continuum positions of the two components reveal a separation 
of $0\overset{''}{.}61\pm0\overset{''}{.}19$ ($88\pm28$ au).  \citet{Luhman2012} classify both systems as single stars since their multiplicity is unresolved by 2MASS and the United Kingdom Infrared 
Telescope (UKIRT) Infrared Deep Sky Survey \citep[UKIDSS,][]{Lawrence2007}. We therefore only consider the brighter continuum component of these sources throughout the 
remainder of this paper, so as not to bias our sample by including additional stars found only because of their 880 $\mu$m continuum emission.

\begin{figure}[!h]
\centerline{\includegraphics[width=\textwidth]{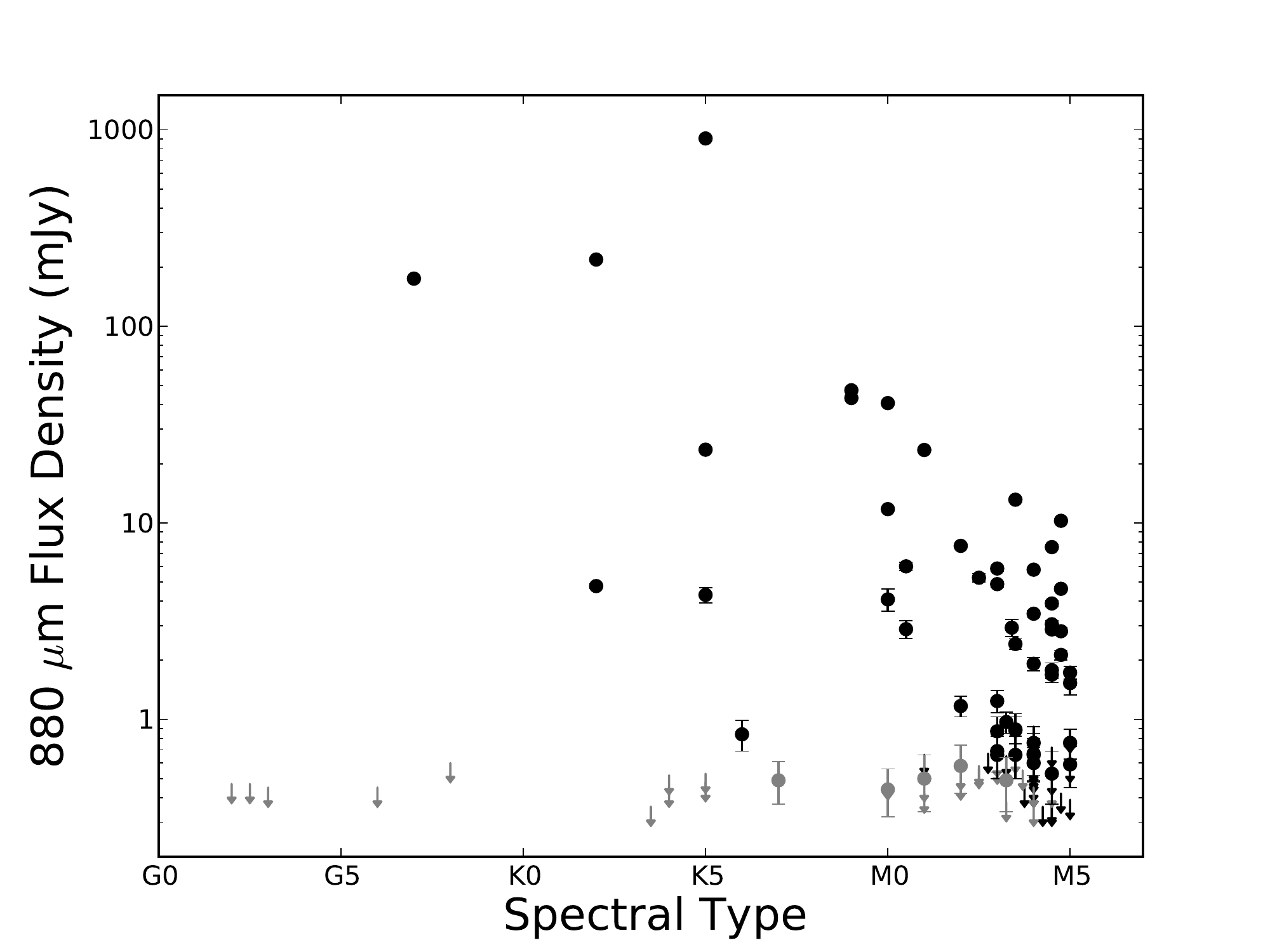}}
\caption{Continuum flux density at 0.88 mm as a function of spectral type for all targets in our sample. Black symbols show the primordial disks, while gray symbols 
represent the debris/evolved transitional disks. Arrows represent 3$\sigma$ upper limits.}
\label{fig:flux_vs_spt}
\end{figure}

\begin{figure}[!h]
\centering
\centerline{\includegraphics[width=\textwidth, angle=-90]{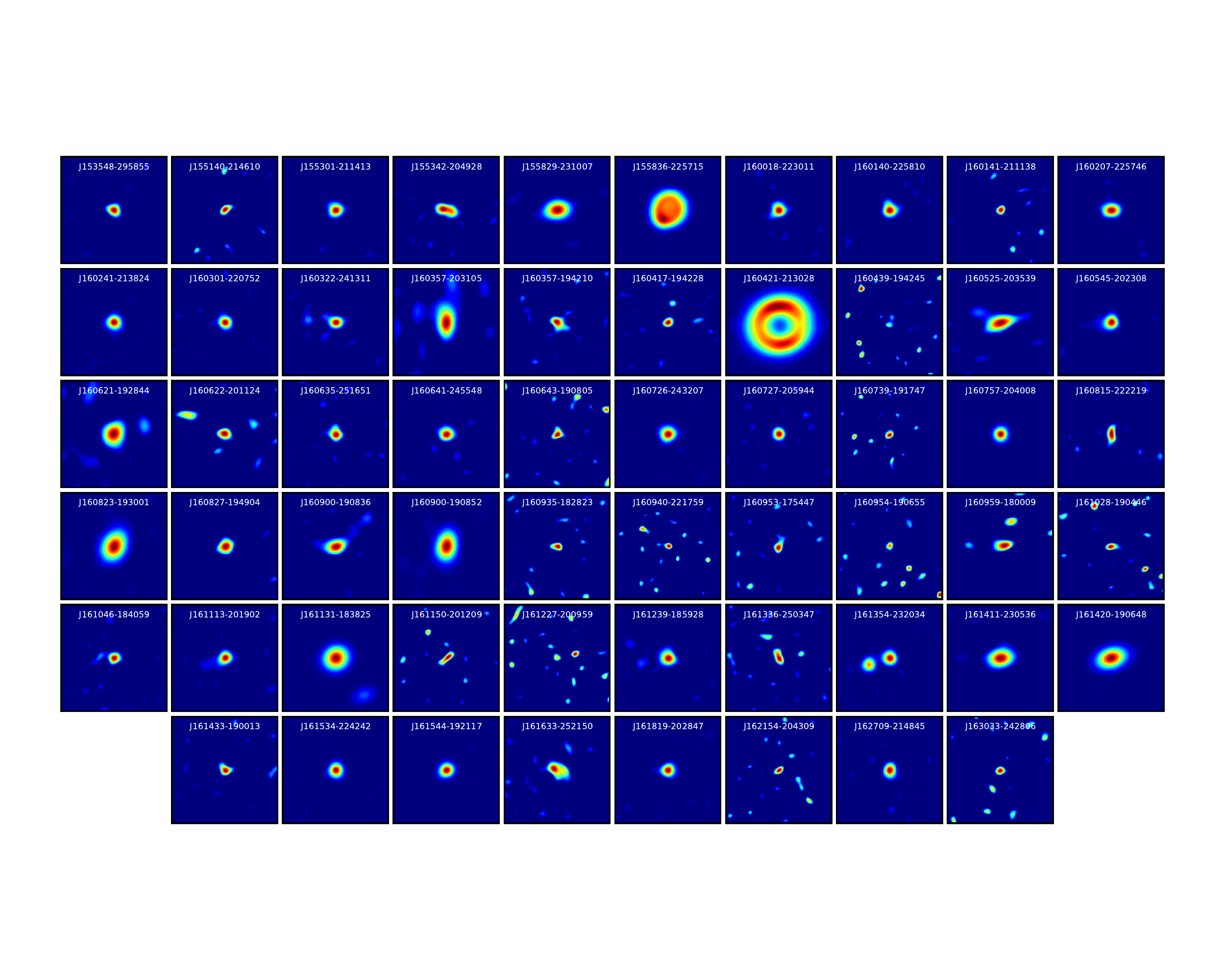}}
\caption{Images of the 0.88 mm continuum for the 58 primordial and debris/evolved transitional disks detected ($>3\sigma$) in the Upper Sco sample.  
Each image is centered on the fitted position of the source and is $3''\times 3''$ in size.}
\label{fig:coloredImages}
\end{figure}

The measured continuum flux for each source is listed in Table \ref{tab:fluxes} and plotted against spectral type in Figure \ref{fig:flux_vs_spt}. We detect 53 of 75 
primordial and 5 of 31 debris/evolved transitional sources at $>3\sigma$. Images of all (primordial and debris/evoloved transitional) continuum detections are shown 
in Figure \ref{fig:coloredImages}. 
The real part of the visibilities as a function of baseline length for all primordial and debris/evolved transitional sources are shown in the left columns of Figures 
\ref{fig:images} and \ref{fig:imagesd0}.
Most of the sources show flat visibility profiles indicating that these sources are compact relative to the beam size of $\sim$0$\overset{''}{.}$35 (50 au).  This agrees with our visibility fitting, for which only 11 sources 
were conclusively spatially resolved.  The compact nature of the majority of the dust disks in our sample matches previous findings in younger-star forming 
regions that faint disks tend to be radially compact.  \citet{Andrews2010} observed a correlation between disk mass (and flux density) and disk radius for sources in the 
Ophiuchus star-forming region, while \citet{Pietu2014} found dust disk sizes of tens of astronomical units or less among faint disks in Taurus.  Note that the faintest sources in Upper Sco 
detected with ALMA are an order of magnitude less luminous than the faintest disks detected by these authors.

The second column of Figure \ref{fig:images} shows continuum images of the 75 primordial disks in the sample. Images of the 
31 debris/evolved transitional disks are shown in the second column of Figure \ref{fig:imagesd0}; the five detected debris/evolved transitional disks are 2MASS J16043916-1942459, 
2MASS J16073939-1917472, 2MASS J16094098-2217594, 
2MASS J16095441-1906551, and 2MASS J16215466-2043091.  All detected sources are consistent with the expected 
stellar position, with the exception of 2MASS J15534211-2049282, 2MASS J16113134-1838259, and 2MASS J16153456-2242421. These three sources 
are offset from the expected stellar position by slightly more than three times the uncertainty in the offset (see Table \ref{tab:fluxes}). However, $^{12}$CO 
$J$ = 3$-$2 emission is detected in all three sources at a velocity consistent with Upper Sco. We therefore assume these continuum sources are associated with the target 
Upper Sco stars.

\subsection{CO Line Fluxes}
CO line fluxes were determined by first subtracting the continuum dust emission using the \emph{uvcontsub} routine in 
CASA, which removes a linear fit to the continuum in the spectral window containing the CO line. Fluxes were then measured using 
aperture photometry of the cleaned, continuum subtracted images.  Measuring line fluxes in this way can be problematic 
due to the need to balance simultaneously choosing a velocity range and aperture size that 
include all emission, but are not so large that they add unnecessary noise to the measurement. On the other hand, it is also possible to 
select a velocity range too narrow and include only a portion of the spectrum, biasing the flux measurement. To avoid these potential pitfalls, 
we first identify the appropriate velocity range of the CO emission for each source, and then measure the optimal aperture size that includes all of the CO 
emission to within the noise.

We started with a circular aperture 0$\overset{''}{.}$5 in radius (large enough to enclose regions emitting at a range 
of velocities) centered on the expected stellar position 
or the center of continuum emission if it is detected.  The spectrum within this aperture was computed with 0.5 km s$^{-1}$ velocity sampling. 
Since the expected host star
radial velocity relative to the LSR of our Upper Sco targets is approximately between 0 and 10 km s$^{-1}$ \citep{deZeeuw1999,Chen2011,Dahm2012}, 
we searched each spectrum between -5 and 15 km s$^{-1}$ for emission exceeding three times the RMS of an emission-free region of the spectrum.  
If a source had at least two channels in this velocity range exceeding this threshold, we considered the source a candidate detection and 
selected the velocity range surrounding these channels, bounded by the emission falling to zero. Next, the flux was measured over 
the appropriate velocity range with increasing aperture size to determine the radius at which the flux becomes constant to within the uncertainty.  
The field of 2MASS J16113134-1838259 contains two sources with continuum and CO detections; for this star, we used an aperture of 0$\overset{''}{.}$8 in radius 
to ensure that only emission from the primary star is included.

This procedure was done using the clean components and residuals directly rather than 
the clean image to avoid the need to use larger apertures that enclose emission smeared to a larger area by convolution with the clean beam. 
To estimate the uncertainty in the measured flux, we measured the flux within 20 circular apertures 
of the same radius and over the same velocity range randomly distributed around the clean component and residual images (with the region containing the 
source itself excluded). We adopt the standard deviation of these measurements as the uncertainty. 

For all sources analyzed in this way, if the measured line flux exceeds five times its 
uncertainty, we consider the source a detection. We adopt a higher detection threshold for the CO than the continuum since the proceedure to estimate the 
line flux selects the velocity range and aperture size that maximizes the signal, and thus may produce false detections.
To validate our procedure, we repeated our measurements with a 0$\overset{''}{.}$3 aperture for velocities 
between 50 and 62 km s$^{-1}$, a region of the spectrum that should contain no emission.  No 5$\sigma$ detections were identified in this velocity range, but one 3$\sigma$ detection was made.  
We therefore expect our 5$\sigma$ threshold to yield a reliable list of detections.

For sources that were not detected at $\geq5\sigma$ using the above method, we measured the flux using a 0$\overset{''}{.}$3 radius aperture between the velocity range of 
$-$1.5 km s$^{-1}$ and 10.5 km s$^{-1}$.  These velocities correspond to the median edges of the velocity ranges of the detected sources.
For any sources with measured flux greater than five times its uncertainty, either from 
the initial 5$\sigma$ cut or from sources measured with a 0$\overset{''}{.}$3 aperture and median velocity range, we repeated the flux measurement proceedure described above, with 
the aperture centered on the centroid of the CO emission.

We detect 26 of the 75 primordial disks with $>5\sigma$ significance and an 
additional 5 primordial disks between $3\sigma$ and $5\sigma$.  None of the debris/evolved transitional disks are detected. Of the $5\sigma$ CO detections, 24 were also detected in the continuum, along with 4 of the CO detections between $3\sigma$ and $5\sigma$.
Our final CO line flux measurements are listed in Table \ref{tab:fluxes}. 
The aperture size and velocity range used is also indicated.  Moment 0 (integrated intensity) maps for each source are shown in the 
third columns of Figures \ref{fig:images} and \ref{fig:imagesd0}.  
Moment 1 (mean velocity) maps are shown for 5$\sigma$ detections in Figure \ref{fig:mom1}. The right columns of Figures \ref{fig:images} and 
\ref{fig:imagesd0} show the spectrum of each source around the CO line, with the velocity range used indicated for 5$\sigma$ detections.

The CO spectra show a variety of line shapes.  Some sources, such as 2MASS J16142029-1906481, 
show the characteristic broad, double-peaked emission of an inclined, Keplerian disk.  Others, such as 2MASS J16041265-2130284 and 2MASS J16113134-1838259, exhibit narrow, single-peaked lines indicative of face-on disks.  
2MASS J16001844-2230114 has a single-peaked line at the expected velocity of Upper Sco, with a tail of weaker emission at 
higher velocity; this high-velocity tail appears to be coming from just to the northwest of the center of the disk emission.  In the moment 0 map of 2MASS J16001844-2230114, the high-velocity tail region can be seen as a wider extension 
of the disk on the northwest side relative to the southeastern side.

\section{Disk Properties in Upper Sco}
In this section, we derive disk dust masses from continuum flux densities. We then investigate 
the dependence of dust mass on stellar mass for the primordial disks in our sample. Finally, we use a 
stacking analysis to determine the mean dust mass of the debris/evolved transitional disks. 

\subsection{Primordial Disk Dust Masses}
\label{sec:dust_masses}

In the present study, we are primarily interested in the bulk dust masses of the disks in our sample.  For optically thin, isothermal dust emission, the dust mass is given by
\begin{equation}
\label{eq:flux}
M_d = \frac{S_{\nu}d^2}{\kappa_{\nu}B_{\nu}(T_d)},
\end{equation}
where $S_{\nu}$ is the continuum flux density, $d$ is the distance, $\kappa_{\nu}$ is the dust 
opacity, and $B_{\nu}(T_d)$ is the Planck function for the dust temperature $T_d$.  We adopt $d=145$ pc, 
which is the mean distance to the OB stars of Upper Sco \citep{deZeeuw1999}. For consistency, we follow 
the opacity and temperature assumptions of \citet{Andrews2013}, assuming a dust opacity of $\kappa_{\nu}=2.3$ cm$^2$g$^{-1}$ at 
230 GHz which scales with frequency as $\nu^{0.4}$. We estimated the dust temperature using the stellar luminosity as 
$T_d=25\mathrm{ K}\times(L_*/L_{\odot})^{0.25}$, which represents 
the characteristic temperature of the dust in the disk contributing to the continuum emission \citep[see the discussion in][]{Andrews2013}. 
\citet{VanDerPlas2016} emphasized that systematic variations in disk size can modify the $T_d$-$L_*$ 
relation. However, without direct measurements of disk sizes and how they may vary between Taurus and Upper Sco, we adopt the \citet{Andrews2013} relation.
Given the assumptions regarding dust opacity and temperature, relative dust masses within the sample may be more accurate than the absolute 
dust masses if dust properties are similar within Upper Sco.

The derived dust masses are listed in Table \ref{tab:dust_masses}.  For sources not detected in the 
continuum, dust mass upper limits were estimated using the upper limit of the measured continuum flux density, 
calculated as three times the uncertainty plus any positive measured flux density.  Uncertainties in dust masses 
include uncertainties in the measured flux density and in the assumed distance uncertainty, which we 
take to be $\pm$20 pc \citep[][p. 235]{Preibisch2008}.  Statistical uncertainties in the dust temperature implied from luminosity uncertainties 
are negligible (of the order of 1 K).  Potential systematic uncertainties in dust temperatures and opacities are not included in the dust mass uncertainties.
Among the 53 primordial disks detected in the continuum, detected dust masses range from 0.17 to 126 $M_{\oplus}$, with a median of 0.52 $M_{\oplus}$.
The five detected debris/evolved transitional disks have dust masses ranging from 0.10 $M_{\oplus}$ to 0.27 $M_{\oplus}$.

\subsection{Stellar Mass Dependence}
\label{sec:mass_dependence}
Derived dust masses are plotted against stellar mass for the primordial disks in Figure \ref{fig:disk_masses}. Visual inspection of this figure 
shows a spread in dust masses over two orders of magnitude at a given stellar mass.  This scatter far exceeds the uncertainties in the 
individual dust mass measurements and indicates large variations in either the dust opacity or dust mass of the disks in Upper Sco.  
Despite this scatter, Figure \ref{fig:disk_masses} reveals 
a trend that more massive stars tend to have more massive disks.  The distribution of upper limits also supports this; only 36 of 57 sources are detected below a stellar mass 
of 0.35 $M_{\odot}$, compared to 17 of 18 above.  We used the Cox proportional hazard test for censored data, implemented with the 
\emph{R Project for Statistical Computing} \citep{R}, to evaluate the significance of this correlation. We find the probability of 
no correlation to be $2.12\times10^{-4}$.  We thus conclude there is strong evidence that disk dust mass increases with 
stellar mass in Upper Sco. 

Following \citet{Andrews2013}, we fit a power law to dust mass as a function of stellar mass using the Bayesian approach of 
\citet{Kelly2007}, which incorporates uncertainties in both parameters, intrinsic scatter about the relation, and observational 
upper limits. The resulting best-fit relation is $\log (M_{\mathrm{dust}}/M_{\oplus}) = (1.67\pm0.37)\log(M_*/M_{\odot}) + (0.76\pm0.21)$ 
with an intrinsic scatter of $0.69\pm0.08$ dex in $\log (M_{\mathrm{dust}}/M_{\oplus})$.

\subsection{Debris/Evolved Transitional Disks}

Of the 31 stars classified by \citet{Luhman2012} as debris/evolved transitional disks, 5 were detected in the continuum.  
For the remaining stars, we 
performed a stacking analysis to determine their average disk properties.  The fields of four of these remaining stars (2MASS J15584772-1757595, 
2MASS J16020287-2236139, 2MASS J16025123-2401574, and 2MASS J16071971-2020555) contain contain a submillimeter continuum source that is offset from the 
stellar position but within the $6''$ resolution of the 24 $\mu$m \emph{Spitzer} observations used by \citet{Luhman2012}. 
Thus, it is possible 
the 24 $\mu$m excess seen for these stars is due to a background source, and not a disk associated with the star. These four stars were excluded 
from our stacking analysis. 

Images of each source were generated from the visibilities using a circular Gaussian synthesized beam with an FWHM of  
0$\overset{''}{.}$4. Since none of these sources were individually detected, we centered the image of each 
on the expected stellar position to generate the stacked image. Each pixel of the stacked image was calculated as the mean of the corresponding pixels of the source images, 
weighted by the RMS noise of each image. Figure \ref{fig:stacked_images} shows the resulting mean image. The measured flux density in a 0$\overset{''}{.}$4 diameter 
aperture at the center of the stacked image is $0.03\pm0.05$ mJy.
We determined the dust mass of the stacked disk in the same way as described in Section \ref{sec:dust_masses}, assuming a median dust temperature of 18 K, and find a 
$3\sigma$ upper limit to the dust mass of 0.06 $M_{\oplus}$.

\section{Comparison Between Upper Sco and Taurus}

It has been well established that the statistical properties of the disks in Upper Sco and Taurus are different.  
While $\sim65\%$ of low-mass stars in Taurus host an optically thick inner disk 
\citep{Hartmann2005}, this fraction has decreased to $\sim19\%$ in Upper Sco 
\citep{Carpenter2006}. The frequency of disks showing signs of accretion drops even more rapidly, 
and accretion rates of disks in Upper Sco that still show signs of accretion have dropped by an order of magnitude 
relative to accreting disks in Taurus \citep{Dahm2009,Dahm2010,Fedele2010}.  Such observations have been interpreted as evidence for 
disk evolution between Taurus and Upper Sco.
However, for the disks still present in Upper Sco, the question 
remains whether they differ significantly in dust mass from younger Taurus disks.  

The Taurus star-forming region is ideally suited for such a comparison.  Decades of study have led to a 
nearly complete census of the stars with and without disks in the region \citep[see][]{Luhman2010,Rebull2010}, 
along with an abundance of stellar data that allow for a comparison with Upper Sco 
over the same stellar mass range. In addition, the proximity of Taurus provides improved sensitivity of 
submillimeter observations. Indeed, most disks around stars in Taurus with spectral type M3 or earlier have 
been detected in the submillimeter continuum \citep{Andrews2013}. 

\subsection{Relative Ages}

The age of Upper Sco has become a subject of controversy in the past several years. \citet{Pecaut2012} derived an age of $11\pm2$ Myr through a 
combination of isochronal ages of B, A, F, and G stars, along with the M supergiant Antares, and a kinematic expansion age.  
The masses and radii of several eclipsing binaries recently discovered in Upper Sco by the \emph{K2} extended \emph{Kepler} mission 
\citep{Howell2014} favor an age of $\sim$10 Myr when compared to pre-main-sequence models \citep{David2015,Kraus2015,Lodieu2015}. This is in conflict with 
the canonical age of $\sim5$ Myr based on the HR diagram positions of lower mass stars \citep{deGeus1989,Preibisch2002,Slesnick2008}.
More recently, \citet{Herczeg2015} used the latest stellar models of \citet{Tognelli2011}, \citet{Baraffe2015}, and \citet{Feiden2015} to find an age of $\sim4$ Myr from the HR diagram positions 
of low-mass stars and brown dwarfs. 

In contrast, the mean age of stars in Taurus is $\sim$1-2 Myr based on HR diagram positions of member stars \citep{Kenyon1995,Hartmann2001,Bertout2007,Andrews2013}, indicating 
that Taurus is younger than Upper Sco. However, ages determined using different methods with different samples of stars are not always comparable. \citet{Herczeg2015} 
showed that isochronal ages depend systematically on not only the evolutionary models used, but also on the stellar mass range observed. These issues are apparent 
in the differing age estimates for Upper Sco.  Ages inferred for Taurus and Upper Sco using the same stellar models and spectral type range 
indicate that Upper Sco is older than Taurus on a relative basis. 
Also, the 
late-type members of Upper Sco have spectral lines indicating stronger surface gravity than stars in Taurus and thus an older age 
\citep[e.g.][]{Slesnick2006}. Therefore, despite the uncertainties associated with determining the absolute ages of young stars, on a relative basis, it is clear that Upper 
Sco is older than Taurus.

\subsection{Relative Dust Masses}
\label{sec:disk_comp}
The sample of Taurus sources we use for our comparison of disk dust masses was compiled by \citet{Luhman2010} and \citet{Rebull2010}. 
A catalog of submillimeter fluxes of these sources was published by \citet{Andrews2013}, who used new observations and 
literature measurements to estimate the flux density of these sources at 1.3 and 0.89 mm. For our comparison, we use the 0.89 mm flux densities, scaled to our mean 
wavelength of 0.88 mm assuming $S_{\nu}\propto\nu^{2.4}$, which is the frequency dependence adopted by \citet{Andrews2013} to generate the Taurus 
catalog.  Among our Upper Sco sample, we only consider the 75 
full, evolved, and transitional disks for this comparison. The debris/evolved transitional disks may 
represent second-generation systems that are in a different evolutionary phase than the disk sources in Taurus, and thus 
would not be suitable for a comparison to study \emph{primordial} disk evolution.

Note that our upper limits were not calculated in the same way as those of \citet{Andrews2013}.  Taurus upper limits are reported as 
three times the RMS of the measurement, while our Upper Sco upper limits are three times the RMS plus any positive flux density. However, since the dust 
masses may be expected to be lower in Upper Sco relative to Taurus, the inconsistent treatment of upper limits strengthen our conclusions by bringing 
the samples closer together. 

Figure \ref{fig:disk_masses} shows disk dust mass as a function of stellar mass for the Upper Sco and Taurus samples. Taurus stellar masses 
were estimated using the stellar temperatures and luminosities reported by \citet{Andrews2013} and the same interpolation method used for the Upper 
Sco sample.  Taurus disk masses were calculated as described in Section \ref{sec:dust_masses} using the flux densities from \citet{Andrews2013} scaled to a wavelength of 
0.88 mm. Figure \ref{fig:disk_masses} shows seemingly lower dust masses in Upper Sco than in Taurus, particularly at low stellar masses.  Across the entire range of 
stellar masses, the upper envelope of Upper Sco disk masses is lower than that of Taurus. 
These differences could in principle be quantified by the cumulative dust mass distributions in Taurus and Upper Sco.  However, as emphasized by 
\citet{Andrews2013}, since dust mass is correlated with stellar mass, such a comparison requires that there is no bias in the stellar mass 
distributions between the two samples.  Based on the log-rank and Peto \& Peto Generalized Wilcoxon two-sample tests in \emph{R}, which estimate the 
probability that two samples have the same parent distribution, we find that the probability that the Taurus and Upper Sco sample 
have the same stellar mass distribution to be between $3.1\times10^{-6}$ and $3.2\times10^{-5}$.  Thus the dust masses in the two samples cannot be compared 
without accounting for this bias.

To account for the dependence of disk dust mass on stellar mass, we compare the ratio of dust mass to stellar mass between the Taurus and Upper Sco samples. 
This ratio is shown as a function of stellar mass in Figure \ref{fig:mdms}.  To test for a correlation between this ratio and stellar mass, we used the Cox 
proportional hazard test; we find $p$ values of 0.19 and 0.49 for Taurus and Upper Sco, respectively, consistent with no correlation.  Thus, 
the ratios of disk dust mass to stellar mass in Taurus and Upper Sco can be safely compared. 
Using the log-rank and Peto \& Peto Generalized Wilcoxon tests,  
we find a probability between $1.4\times10^{-7}$ and $4.8\times10^{-7}$ that $M_{\mathrm{dust}}/M_*$ in Taurus and Upper Sco are drawn from the same distribution, strong evidence that dust masses 
are different in Upper Sco and Taurus.  Figure \ref{fig:mdms_dist_all} shows the distributions of $M_{\mathrm{dust}}/M_*$ in Taurus and Upper Sco found using the Kaplan$-$Meier estimator for censored data.  
We find a mean ratio of dust mass 
to stellar mass of $\langle\log(M_{\mathrm{dust}}/M_*)\rangle = -4.44 \pm 0.05$
in Taurus and $\langle\log(M_{\mathrm{dust}}/M_*)\rangle = -5.08 \pm 0.08$ in Upper Sco.  Thus, $\Delta\langle\log(M_{\mathrm{dust}}/M_*)\rangle = 0.64\pm0.09$ (Taurus-Upper Sco), such that the $M_{\mathrm{dust}}/M_*$ is lower in Upper Sco 
by a factor of $\sim 4.5$.

Having shown that the ratio of disk dust mass to stellar mass is lower in Upper Sco than in Taurus, we now examine how this difference depends on stellar 
mass by comparing the power-law slope of dust mass versus stellar mass in Taurus and Upper Sco.  As mentioned above, \citet{Andrews2013} found a significant 
correlation between dust mass and stellar mass in Taurus. The authors performed a power-law fit using stellar masses 
from three different stellar models. The weighted mean of the resulting fit parameters gives a power-law slope of 1.2$\pm$0.4 and intrinsic scatter 
of 0.7$\pm$0.1 dex for stellar masses between $\sim0.1$ and $\sim10$ $M_{\odot}$. 
Our results for Upper Sco are consistent with this slope and scatter. Restricting the Andrews sample over the range of Upper Sco stellar masses, we use our 
derived Taurus dust and stellar masses to find a power-law slope of 1.45$\pm$0.30 and scatter of 0.69$\pm$0.06 dex over the range of 0.14-1.66 
$M_{\odot}$, also consistent with our Upper Sco results and the \citet{Andrews2013} result for the full Taurus sample. While disk dust masses in 
Upper Sco are significantly lower than those in Taurus, the power-law slopes of dust mass versus stellar mass are in agreement.
This is consistent with evolution in dust mass between Taurus and Upper Sco  being independent of stellar mass within 
our stellar mass range, though we note that the uncertainties are large.

\section{Discussion}

\subsection{Dust Mass Evolution}
While it has already been established that the fraction of stars with disks is lower in Upper Sco than in 
Taurus \citep{Carpenter2006,Luhman2012}, we have shown that for the Upper Sco primordial disks that remain, the ratio of disk dust mass to stellar 
mass is significantly lower than for disks in Taurus \citep[see also][]{Mathews2012,Mathews2013,Carpenter2014}.  This conclusion assumes the dust emission 
is optically thin and the dust opacity is the same between the two regions, such that differences in the measured continuum flux can be interpreted as variations in the 
disk dust mass.  However, from Equation \ref{eq:flux}, the 0.88 mm flux density is proportional to
the product of dust mass and dust opacity.  Thus, difference in flux density could be due to changes in dust mass, grain 
size/composition or some combination of the two. For a distribution of dust 
grain sizes described by $\frac{dn}{da}\propto a^{-p}$, the opacity varies with the maximum grain size as 
$\kappa \propto (a_{max})^{p-4}$ \citep{Draine2006}.  Assuming $p$ = 3.5, 
an increase in maximum grain size by a factor of $\sim$20, for example from 1 mm to 2 cm, could fully explain the 
apparent decrease in dust mass by a factor of 4.5 between Taurus and Upper Sco.  Such a change in 
the maximum grain size would change the slope of the dust opacity between wavelengths of 1 mm and 7 mm 
from $\beta$ = 1.8$-$1.9 to $\beta$ = 1.0$-$1.5, depending on the grain composition model assumed \citep{Natta2004}.  

No compelling evidence for variations in $\beta$ with stellar age has been found to date.  
\citet{Ricci2010} found no correlation between $\beta$ and age for individual stars in Taurus.  However, 
much of the apparent age spread within Taurus can be attributed to measurement uncertainties and the effects of binarity 
\citep{Hartmann2001}.  Comparison between clusters with different ages should yield more robust results, but the 
sample sizes remain limited and no conclusive evidence for variations in $\beta$ have been found \citep[][p. 339]{Ubach2012,Testi2014}
However, none of these results compare $\beta$ in systems with ages as different as Taurus and Upper Sco.
Thus, we cannot exclude the possibility that the disk mass distribution is the same, but the underlying particle size distribution 
differs.  To break this degeneracy, observations at multiple (sub)millimeter wavelengths of both Upper Sco and Taurus are required.

\subsection{The Relationship between Gas and Dust}

Our combination of CO $J$ = 3$-$2 and dust continuum observations allows us to probe both the gaseous and solid 
material in the disks of Upper Sco. 
Figure \ref{fig:co_cont} shows CO line flux plotted against continuum 
flux density for the 75 primordial disks in our sample. This figure shows that CO flux is correlated with continuum flux 
over $\sim3$ orders of magnitude. The optically thin continuum flux is proportional to the mass of solid material in the 
disk, while the CO emission, if it is optically thick, is a proxy for the projected area of the gas in the disk. Thus, the total 
mass of solids in a disk seems to trace the spatial extent of the gas in the disk. Both continuum and CO 
flux depend on the temperature of the disk, but this should not vary by a factor of more than a few and not enough to explain 
the trend between continuum and CO flux over $\sim3$ orders of magnitude. Instead, it appears that in Upper Sco, stars still 
surrounded by relatively large quantities of dust also maintain extended gas disks.  This is consistent with the fact that the 
six brightest continuum sources are also spatially resolved.
In a future paper, we will use the spatial information provided by the high angular resolution of 
our continuum and CO observations to obtain more quantitative measurements 
of dust and gas disk sizes in Upper Sco.

While 53 of the 75 primordial disks are detected in the 0.88 mm continuum, only 26 are detected in CO. Similarly, 
\citet{VanDerPlas2016} surveyed seven brown dwarfs in the 0.88mm continuum and CO J=3$-$2 with a sensitivity and angular resolution 
comparable to our survey; while six brown dwarfs were detected in the continuum, only one was detected in CO.  
Among the non-detections in the present study, the median $5\sigma$ sensitivity in the integrated spectra is 72 mJy per channel, which corresponds to a 
brightness temperature of $\sim$9 K.
The gas temperature in the disk where the CO is present is expected to be $>20$ K, as CO will freeze out onto dust grains at lower temperatures 
\citep{Collings2003,Bisschop2006}.  Given that the brightness temperature limit of the observations is much less than 20 K, the lack of detectable 
CO in half of the continuum sources can be attributed to two possibilities:
the CO is optically thick but does not fill the aperture, or the CO is optically thin.

If the CO emitting region is smaller than the aperture size, the $\gtrsim$20 K physical temperature can be diluted to a 9 K observed brightness temperature. 
This will depend on the projected area of the emitting region, given by 
\begin{equation}
A_{CO} = \pi {R_{CO}}^2 \cos i, 
\end{equation}
where $R_{\mathrm{CO}}$ is the outer radius of the CO emission and $i$ is the disk inclination.  The 0$\overset{''}{.}$3 radius aperture corresponds to a physical radius of 43.5 au at the distance of 
Upper Sco.  Thus, assuming an inclination of 60$\degree$, the $5\sigma$ brightness 
temperature upper limit of 9 K sets an upper limit on $R_{\mathrm{CO}}$ of $\sim$40 au to dilute the brightness temperature from 20 K.  
While extensive measurements of CO disk radii of comparably low-mass disks are not available, such small disk sizes are not unprecedented.  
\citet{Woitke2011} measured a CO disk radius of 10 au for the disk around ET Cha based on analysis 
of the continuum and the lack of CO $J$ = 3$-$2 emission.  \citet{Pietu2014} measured CO radii as small as 60 au for a sample of five disks in Taurus, 
although these disks are at least a factor of five greater in dust mass than our median dust mass of CO non-detections.

An alternative explanation for the lack of CO detections is that gaseous CO in the disk has been depleted or dispersed 
to the point of becoming optically thin. 
The upper limit on the CO optical depth ($\tau_{\mathrm{CO}}$) can be related to the brightness temperature upper limit ($T_b$) 
and the physical CO temperature $T_{\mathrm{CO}}$ by the expression 
\begin{equation}
B_{\nu}(T_b) = [B_{\nu}(T_{\mathrm{CO}})-B_{\nu}(T_{\mathrm{CMB}})] \left(1 - \mathrm{e}^{-\tau_{\mathrm{CO}}}\right), 
\end{equation}
where $B_{\nu}(T)$ is the Planck function and $T_{\mathrm{CMB}}$ is the background temperature of the cosmic microwave background \citet{Mangum2015}.
Again assuming a minimum physical temperature of 20 K 
for the CO, we place a $5\sigma$ upper limit on the CO $J$ = 3$-$2 optical depth of $\tau_{\mathrm{CO}}=0.28$ if the CO emission fills the aperture used to measure the flux. 
Such an optical depth would 
require substantial CO depletion in these disks. \citet{Mangum2015} give an expression for the total column density of a molecule given the integrated intensity 
of its spectrum, assuming optically thin emission:
\begin{equation}\label{eq:Nthin}
N_{tot} = \left(\frac{8\pi\nu^3}{c^3A_{\mathrm{ul}}}\right) \left(\frac{Q_{\mathrm{rot}}}{2J+1}\right)\frac{\exp\left(\frac{E_{\mathrm{u}}}{k T_{\mathrm{ex}}}\right)}{\exp\left(\frac{h\nu}{kT_{\mathrm{ex}}}\right)-1} \frac{1}{(J_{\nu}(T_{\mathrm{ex}}) - J_{\nu}(T_{\mathrm{CMB}}))}\int T_B\, \mathrm{d}v.
\end{equation}
In this expression, $\nu$ is the frequency of the transition (345.79599 GHz for $^{12}$CO $J=3-2$), $c$ is the speed of light, $k$ is Boltzmann's constant, $h$ 
is Planck's constant, $T_{\mathrm{ex}}$ is the excitation temperature of the gas, and $T_{\mathrm{CMB}}$ is the temperature of the cosmic microwave 
background radiation.  $A_{\mathrm{ul}}$ is the Einstein $A$ coefficient and $E_{\mathrm{u}}$ is the energy of the upper level 
of the transition \citep[$A_{\mathrm{ul}} = 2.497\times10^{-6}$ s$^{-1}$ and $\frac{E_{\mathrm{u}}}{k} = 33.19$ K for $^{12}$CO $J$ = 3$-$2,][]{Muller2001,Muller2005}. $Q_{\mathrm{rot}}$ is the partition function, which can be approximated as 
\begin{equation}
Q_{\mathrm{rot}} = \frac{kT}{hB_0}\exp\left(\frac{hB_0}{3kT}\right),
\end{equation}
where $B_0 = 5.8\times10^{10}$ s$^{-1}$ \citep{Huber1979}.  $J_{\nu}$ is defined as
\begin{equation}
J_{\nu}\equiv \frac{\frac{h\nu}{k}}{\exp(\frac{h\nu}{kT})-1}.
\end{equation}
Finally the integral in Equation \ref{eq:Nthin} is simply the integrated line flux in terms of brightness temperature. 

To estimate an upper limit on the $^{12}$CO column density if it is optically thin, we assume an excitation temperature of 20 K.  For the 
CO non-detections, our median $5\sigma$ upper limit on the integrated flux density is 202 mJy km s$^{-1}$.  This corresponds to a CO column density upper limit of 
$3.5\times10^{15}$ cm$^{-2}$.
This value can be compared to that expected for a typical disk 
in our sample given our measured dust masses.  Assuming a gas to dust mass ratio of 100, a disk radius of 43.5 au to fill the measurement aperture, and the median 
dust mass of our CO non-detections of 0.4 
$M_{\oplus}$, 
the column density of molecular hydrogen would be $5.3\times10^{22}$ cm$^{-2}$. For a $^{12}$CO abundance relative to H$_{2}$ of $7\times10^{-5}$ \citep[][and references therein]{Beckwith1993,Dutrey1996}, 
the CO column density would be $3.7\times10^{18}$ cm${^2}$, a factor of $\sim$1000 higher than the limit we observe. For the disks in our sample to have spatially 
extended CO that fills the aperture and not be detected, the abundance of gaseous CO relative dust must be drastically reduced by depletion of CO specifically (for example, through 
freeze out onto dust grains) or of the gas as a whole.  

Previous observations \citep[e.g.]{Dutrey2003,Chapillon2008,Williams2014} have found evidence for CO depletion 
in Taurus disks by factors of up to $\sim$100 relative to the interstellar medium.  Based on a lack of C\upperRomanNumeral{1} emission toward the disk 
around CQ Tau, \citet{Chapillon2010} concluded that the weak CO emission previously observed for this disk is due to depletion of the gas as a whole, not just of CO. Focusing 
on disks later in their evolution, \citet{Hardy2015} observe 24 sources with ALMA lacking signs of ongoing accretion, but still showing infrared excesses indicative of dust.  While four 
of these sources are detected in the 1.3mm continuum, none are detected in $^{12}$CO $J$ = 2$-$1.  Assuming interstellar medium gas to dust ratios and CO abundances, the CO in the four 
continuum-detected disks should have been easily detected, again implying substantial depletion of CO.  Given that the Upper Sco disks in the present study represent 
the final phase of primordial disk evolution, similar or greater levels of CO depletion may be plausible.

\section{Summary}
We have presented the results of ALMA observations of 106 stars in the Upper Scorpius OB association classified as circumstellar disk 
hosts based on infrared excess. We constructed a catalog of the 0.88 mm continuum and $^{12}$CO $J$ = 3$-$2 fluxes of these stars.  Continuum 
emission was detected toward 53 of 75 primordial disks and 5 of 31 debris/evolved transitional disks, while CO was detected in 26 of the 
primordial disks and none of the debris/evolved transtional disks.  The continuum observations were used to measure the dust mass in the disks assuming the emission 
is optically thin and isothermal.
We compared these masses to dust masses of disks in Taurus measured using the flux catalog compiled by \citet{Andrews2013} in order to investigate the 
evolution of disk dust mass and how this evolution depends on stellar mass.  Within Upper Sco itself, we analyzed the 
dependence of disk mass on stellar host mass and the relationship between gas and dust in primordial disks.
The key conclusions of this paper are as follows:

\begin{enumerate}

\item There is strong evidence for systematically lower dust masses 
in Upper Sco relative to Taurus. For the stellar mass range of $0.14-1.66\ M_{\oplus}$, we find that the ratio of disk dust masses to stellar masses
in Upper Sco are a factor of $\sim 4.5$ lower than in Taurus, with a probability between $1.4\times10^{-7}$ and $4.8\times10^{-7}$ 
that the dust masses in Taurus and Upper Sco are drawn from the same distribution. 

\item There is a statistically significant correlation between disk dust mass and stellar host mass for primordial 
disks in Upper Sco. Fitting a power law, we find $M_{\mathrm{dust}} \propto M_*^{1.67\pm0.37}$. Within uncertainties, the power-law slope of this relation is in agreement with the slope of the power-law relation found for Taurus dust and 
stellar masses by \citet{Andrews2013}, indicating that dust mass evolution is consistent with being independent of stellar mass. 

\item Only about half of the primordial disks detected in the continuum were detected in CO.  The lack of CO detections could be explained if the CO is 
optically thick and has an emitting area with a radius of $\lesssim 40$ au, or if the CO has an optical depth of $\lesssim 0.28$ and is more extended.
Continuum flux and $^{12}$CO flux are correlated over $\sim3$ orders of magnitude for primordial disks in Upper Sco, suggesting that the same stars have 
maintained relatively large gas and dust disks. 

\end{enumerate}

\acknowledgements

We thank the referee for their useful comments, which improved this manuscript.
We are grateful to Sean Andrews for his advice on the comparison 
of Upper Sco and Taurus disk masses, to Trevor David for valuable input 
on the age of Upper Sco, to Ivan Marti-Vidal for clarification regarding 
the use of \emph{uvmultifit}, and to Nick Scoville for providing an original version 
of the aperture photometry code that was adapted for use in this work.
We also thank the ALMA staff for their assistance in the data
reduction. The National Radio Astronomy Observatory is a facility of the
National Science Foundation operated under cooperative agreement by Associated
Universities, Inc. This paper makes use of the following ALMA data:
ADS/JAO.ALMA\#2011.0.00966.S and ADS/JAO.ALMA\#2013.1.00395.S. ALMA is a partnership of ESO (representing its
member states), NSF (USA) and NINS (Japan), together with NRC (Canada) and NSC
and ASIAA (Taiwan), in cooperation with the Republic of Chile. The Joint ALMA
Observatory is operated by ESO, AUI/NRAO, and NAOJ. A.I. and J.M.C. acknowledge
support from NSF awards AST-1109334 and AST-1140063. This publication makes use
of data products from the Two Micron All Sky Survey, which is a joint project
of the University of Massachusetts and the Infrared Processing and Analysis
Center/California Institute of Technology, funded by the National Aeronautics
and Space Administration and the National Science Foundation. This publication 
makes use of data products from the \emph{Wide-field Infrared Survey Explorer}, which 
is a joint project of the University of California, Los Angeles, and the Jet 
Propulsion Laboratory/California Institute of Technology, funded by the National 
Aeronautics and Space Administration.  This research has made use of the NASA/IPAC 
Extragalactic Database (NED) which is operated by the Jet Propulsion Laboratory, 
California Institute of Technology, under contract with the National Aeronautics and 
Space Administration. This work is based [in part] on observations 
made with the Spitzer Space Telescope, which is operated by the Jet Propulsion Laboratory, 
California Institute of Technology under a contract with NASA.

\begin{figure}[!h]
\centering

\subfloat{
\includegraphics[width=0.9\textwidth]{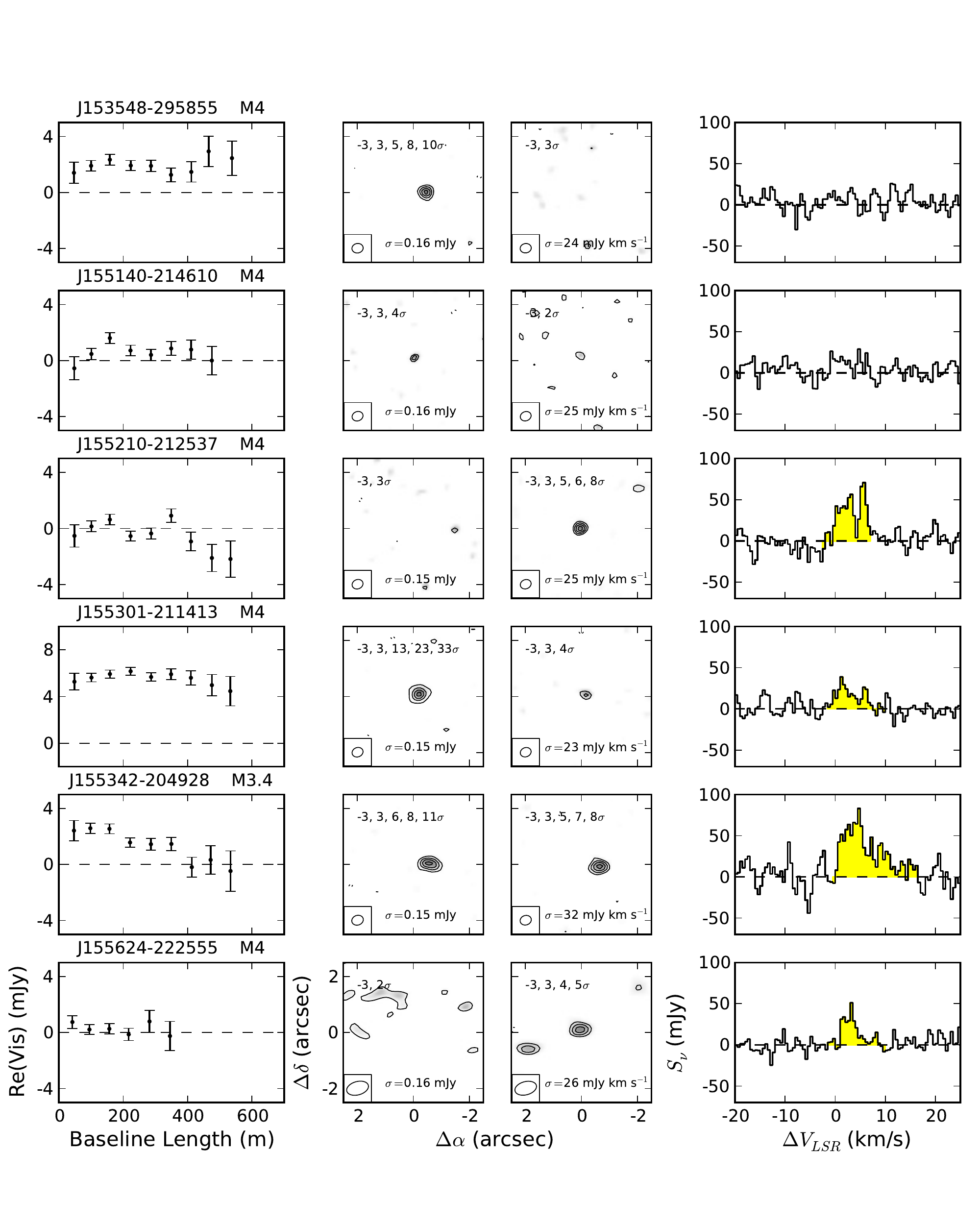}
}
\caption{Left: real part of the visibilities as a function of projected baseline length for the 75 primordial disks in the Upper Sco sample. The phase center has been 
shifted to the centroid of the continuum for each source, or to the expected stellar position in the case of non-detections. The host star and its spectral type are given above 
each plot.
Center: images of the 0.88 mm continuum and integrated CO $J$ = 3$-$2 line, centered on the expected stellar position. Contour levels are given at the top of each image, with 
the point-source sensitivity at the bottom. 
Right: spectra of the CO $J$ = 3$-$2 line.  The yellow shaded region indicates, for 5$\sigma$ detections, the velocity range given in Table \ref{tab:fluxes}
over which the line is integrated to measure the flux and generate the integrated intensity map. The aperture radii used to make the spectra are also given in Table \ref{tab:fluxes}.}
\label{fig:images}
\end{figure}

\begin{figure}
\ContinuedFloat
\centering
\subfloat{
\includegraphics[width=0.9\textwidth]{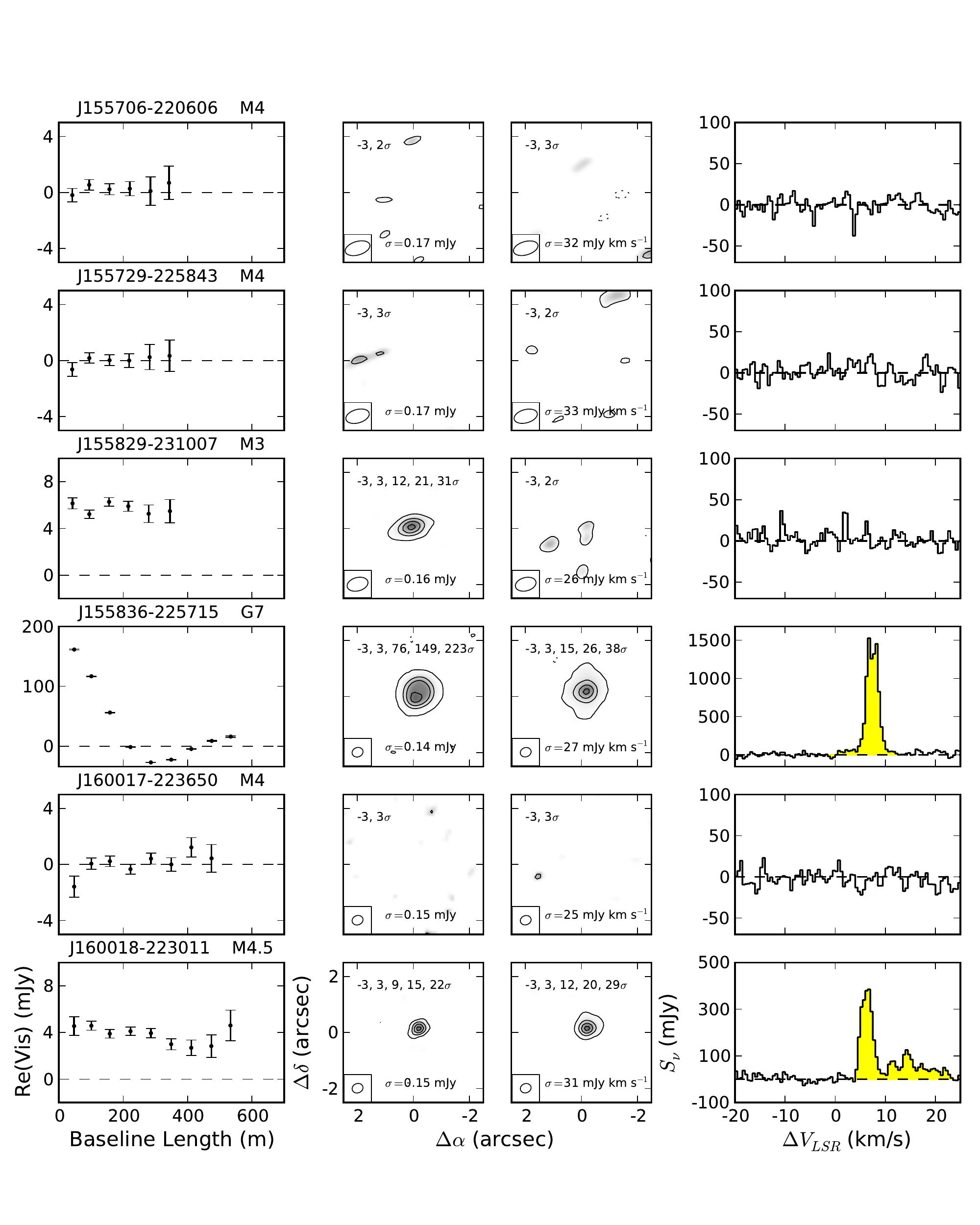}
}
\caption{Continued.}
\label{fig:images}

\end{figure}

\begin{figure}
\ContinuedFloat
\centering
\subfloat{
\includegraphics[width=0.9\textwidth]{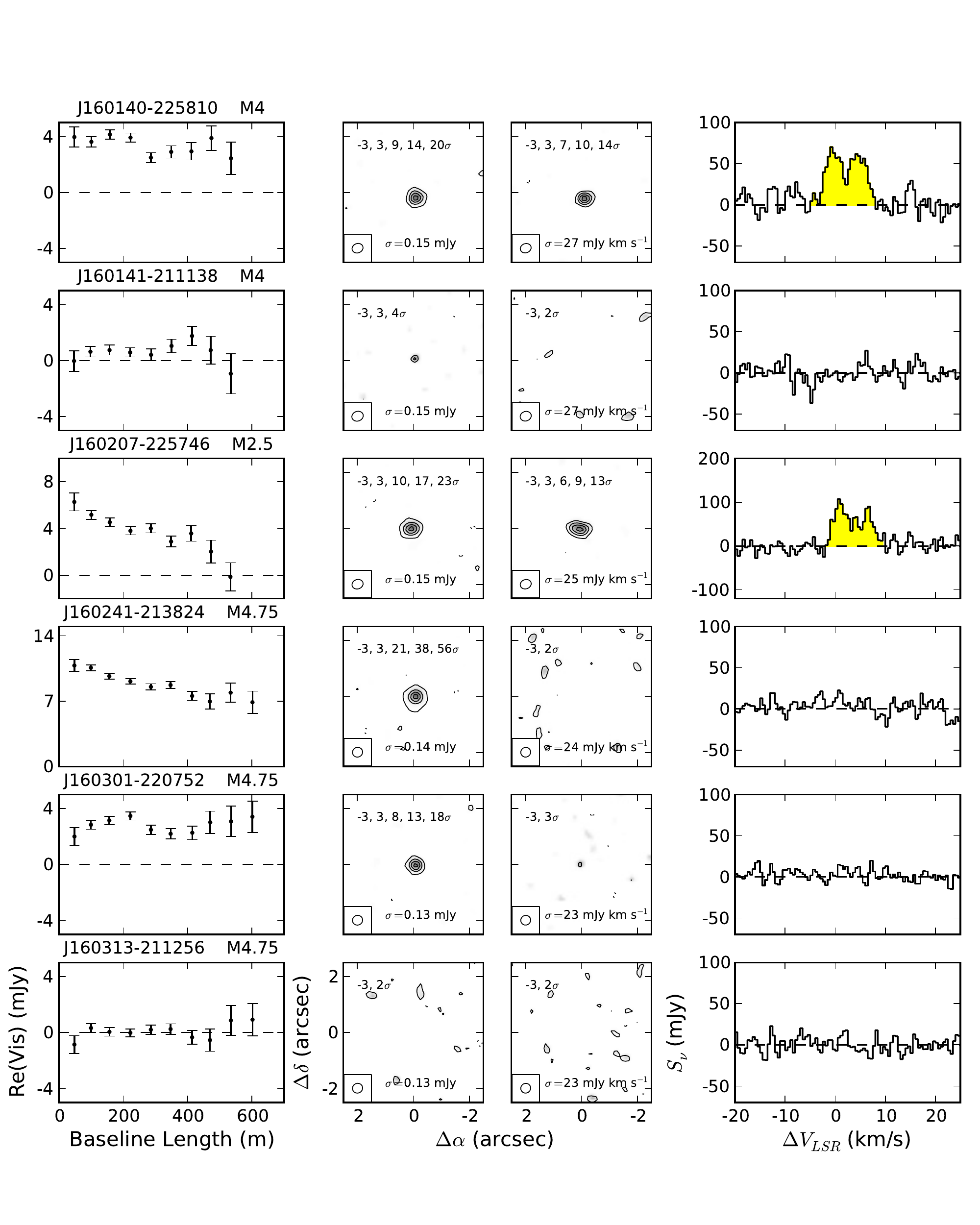}
}
\caption{Continued.}
\label{fig:images}
\end{figure}

\begin{figure}
\ContinuedFloat
\centering
\subfloat{
\includegraphics[width=0.9\textwidth]{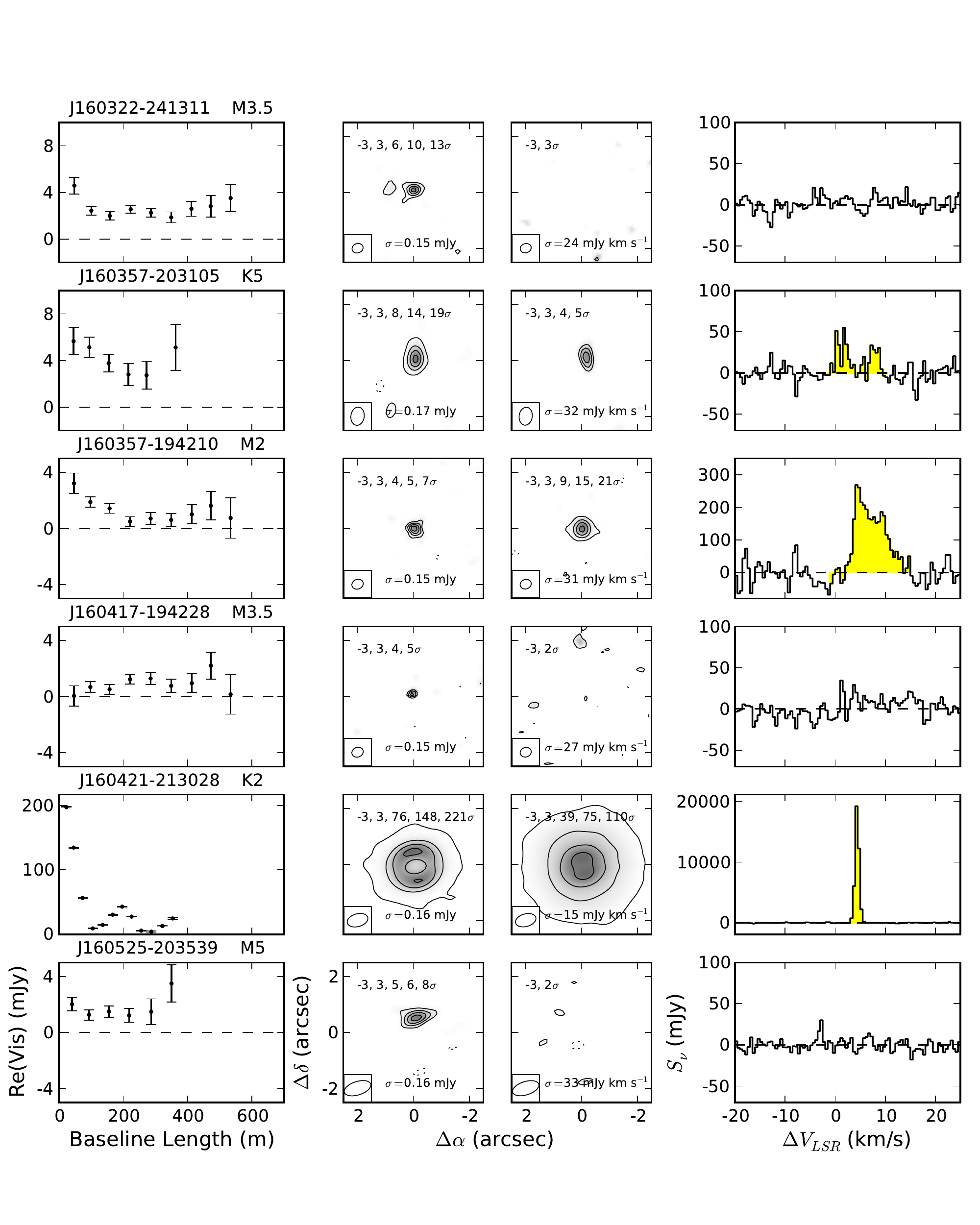}
}
\caption{Continued.}
\label{fig:images}
\end{figure}

\begin{figure}
\ContinuedFloat
\centering
\subfloat{
\includegraphics[width=0.9\textwidth]{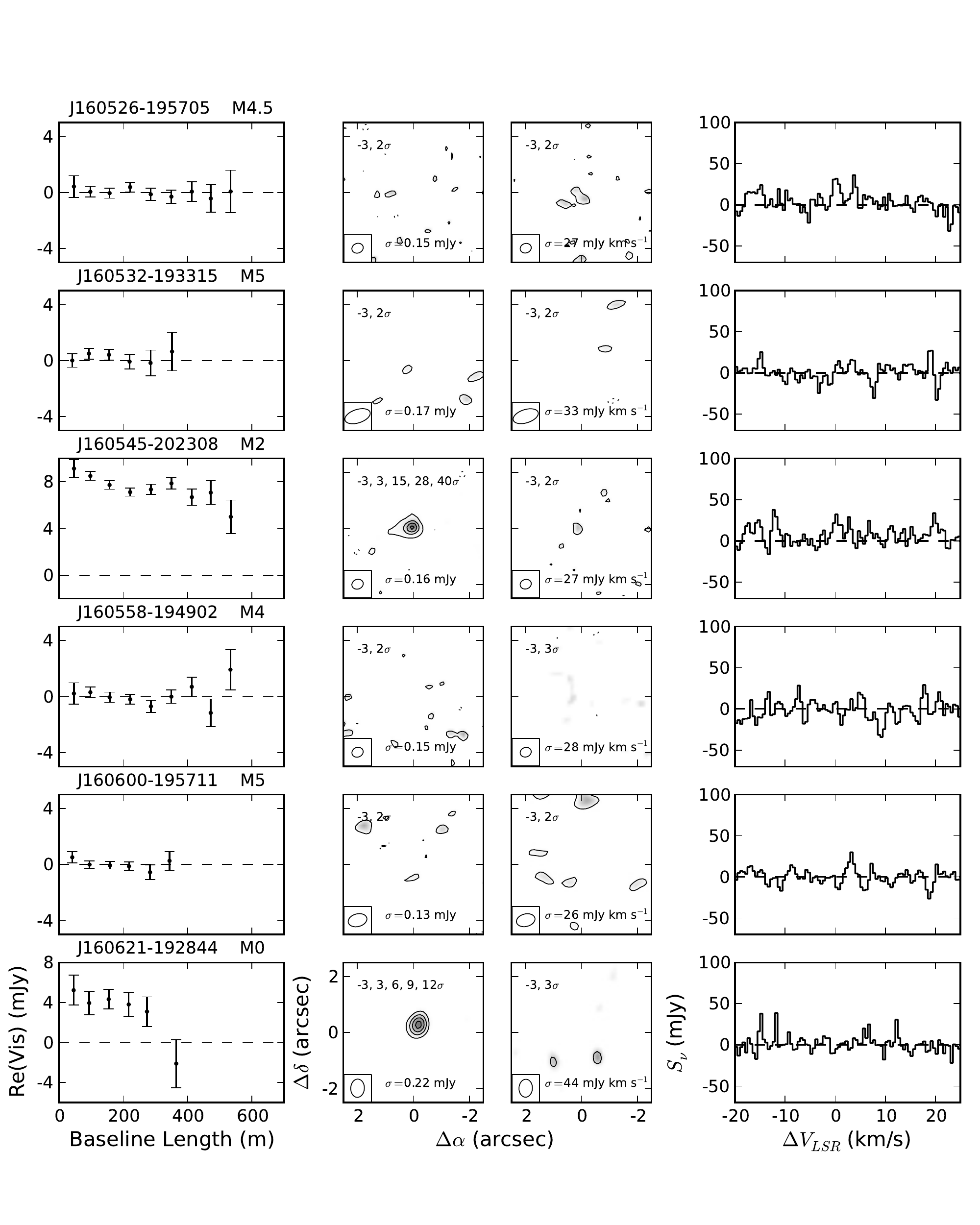}
}
\caption{Continued.}
\label{fig:images}
\end{figure}

\begin{figure}
\ContinuedFloat
\centering
\subfloat{
\includegraphics[width=0.9\textwidth]{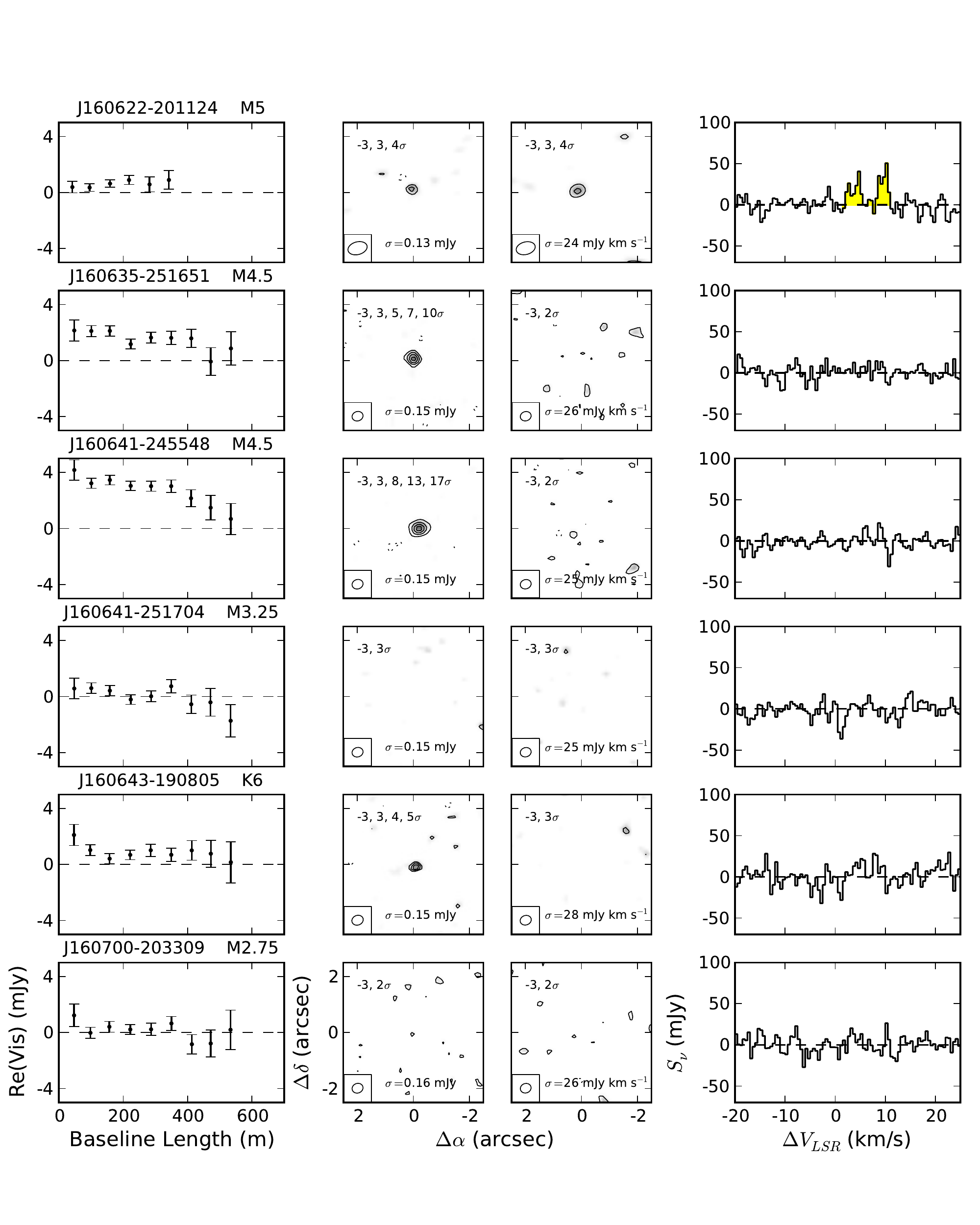}
}
\caption{Continued.}
\label{fig:images}
\end{figure}

\begin{figure}
\ContinuedFloat
\centering
\subfloat{
\includegraphics[width=0.9\textwidth]{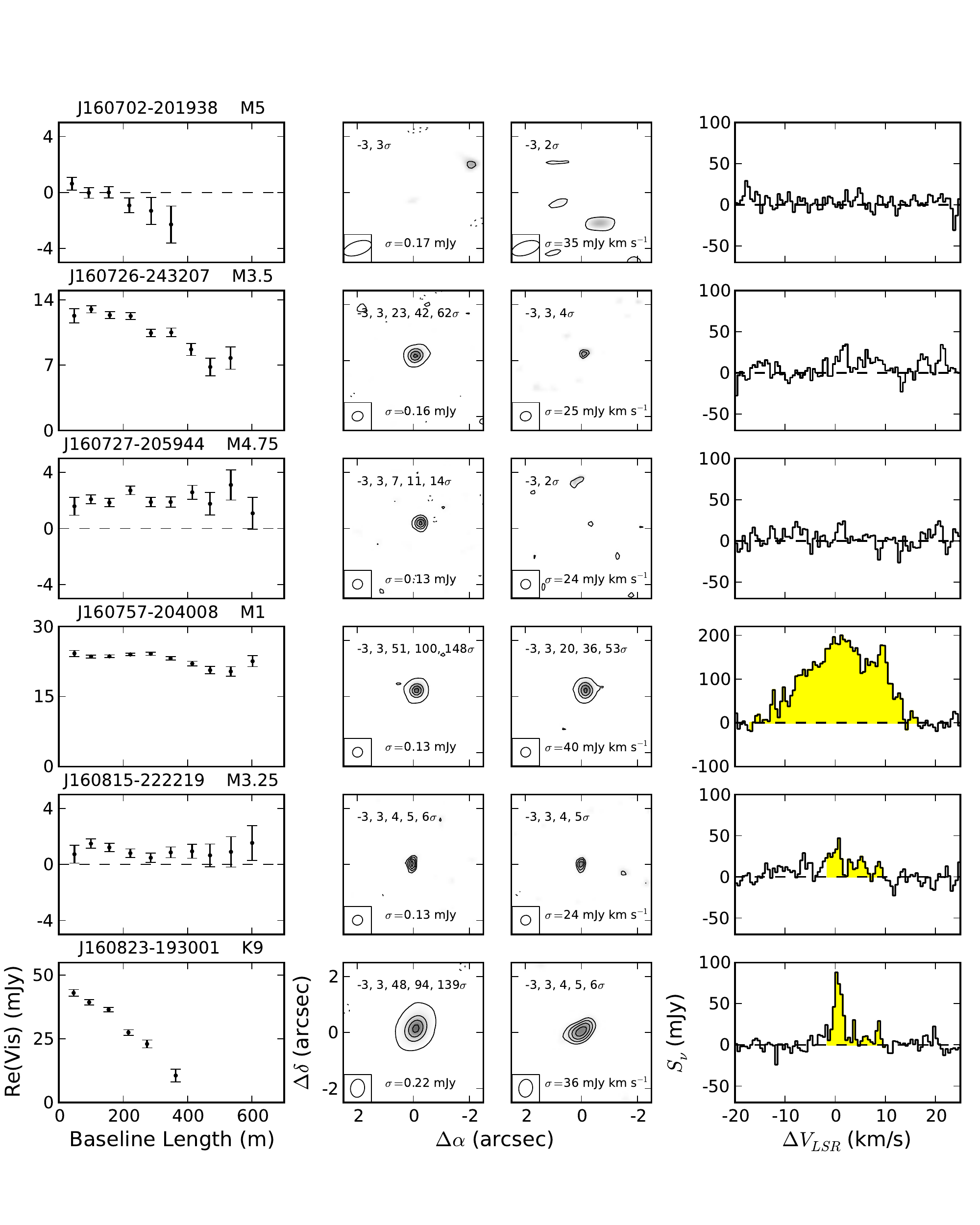}
}
\caption{Continued.}
\label{fig:images}
\end{figure}

\begin{figure}
\ContinuedFloat
\centering
\subfloat{
\includegraphics[width=0.9\textwidth]{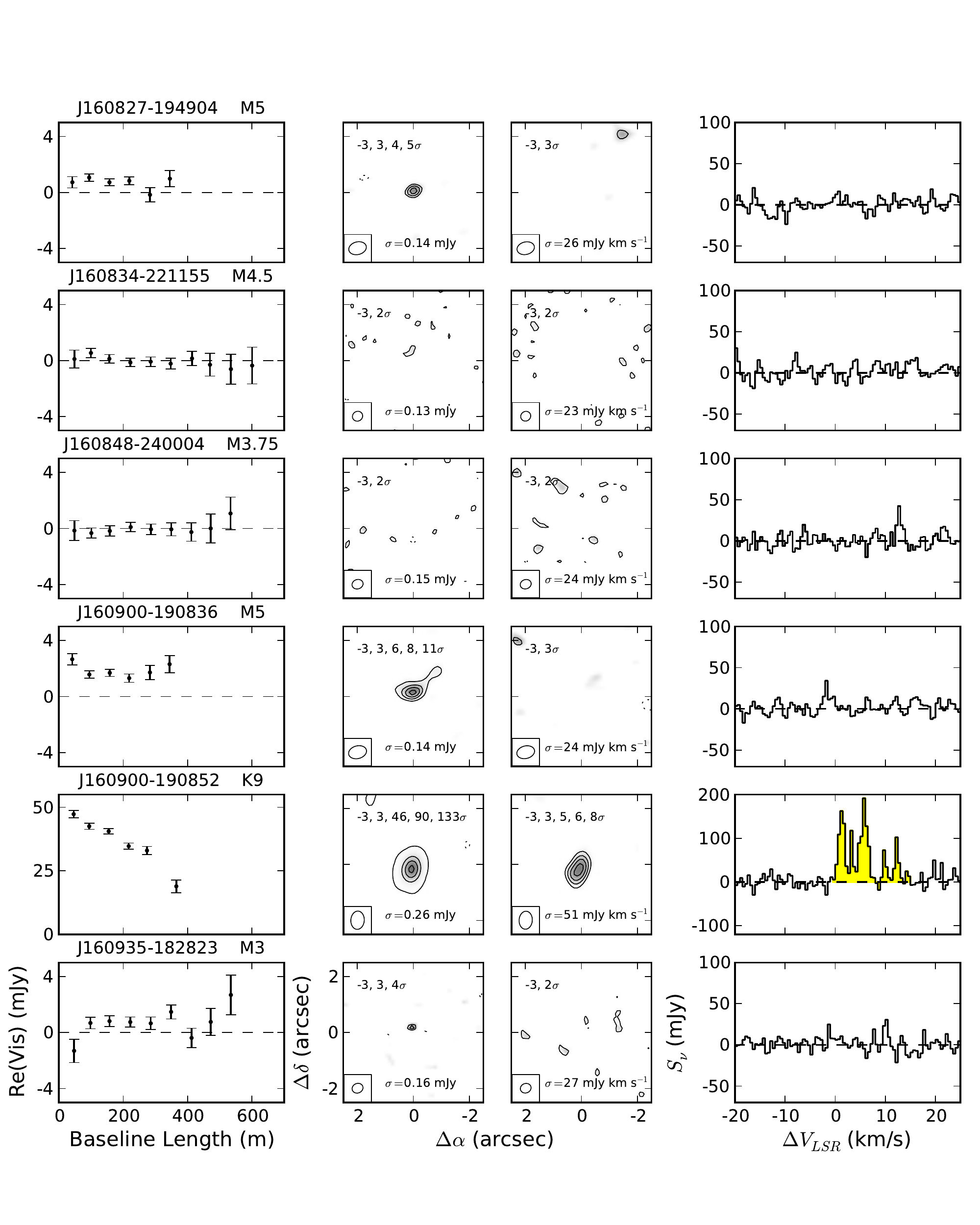}
}
\caption{Continued.}
\label{fig:images}
\end{figure}

\begin{figure}
\ContinuedFloat
\centering
\subfloat{
\includegraphics[width=0.9\textwidth]{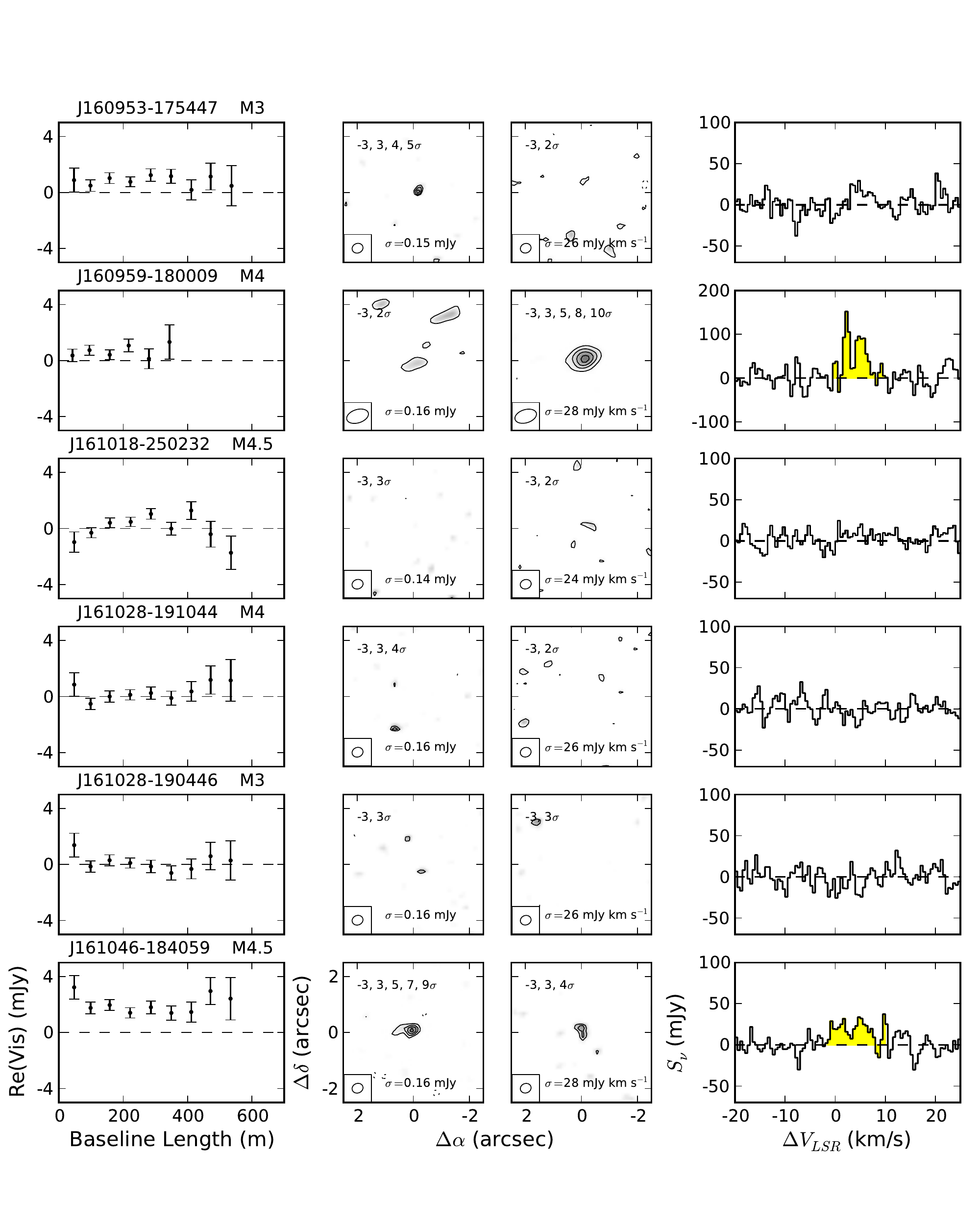}
}
\caption{Continued.}
\label{fig:images}
\end{figure}

\begin{figure}
\ContinuedFloat
\centering
\subfloat{
\includegraphics[width=0.9\textwidth]{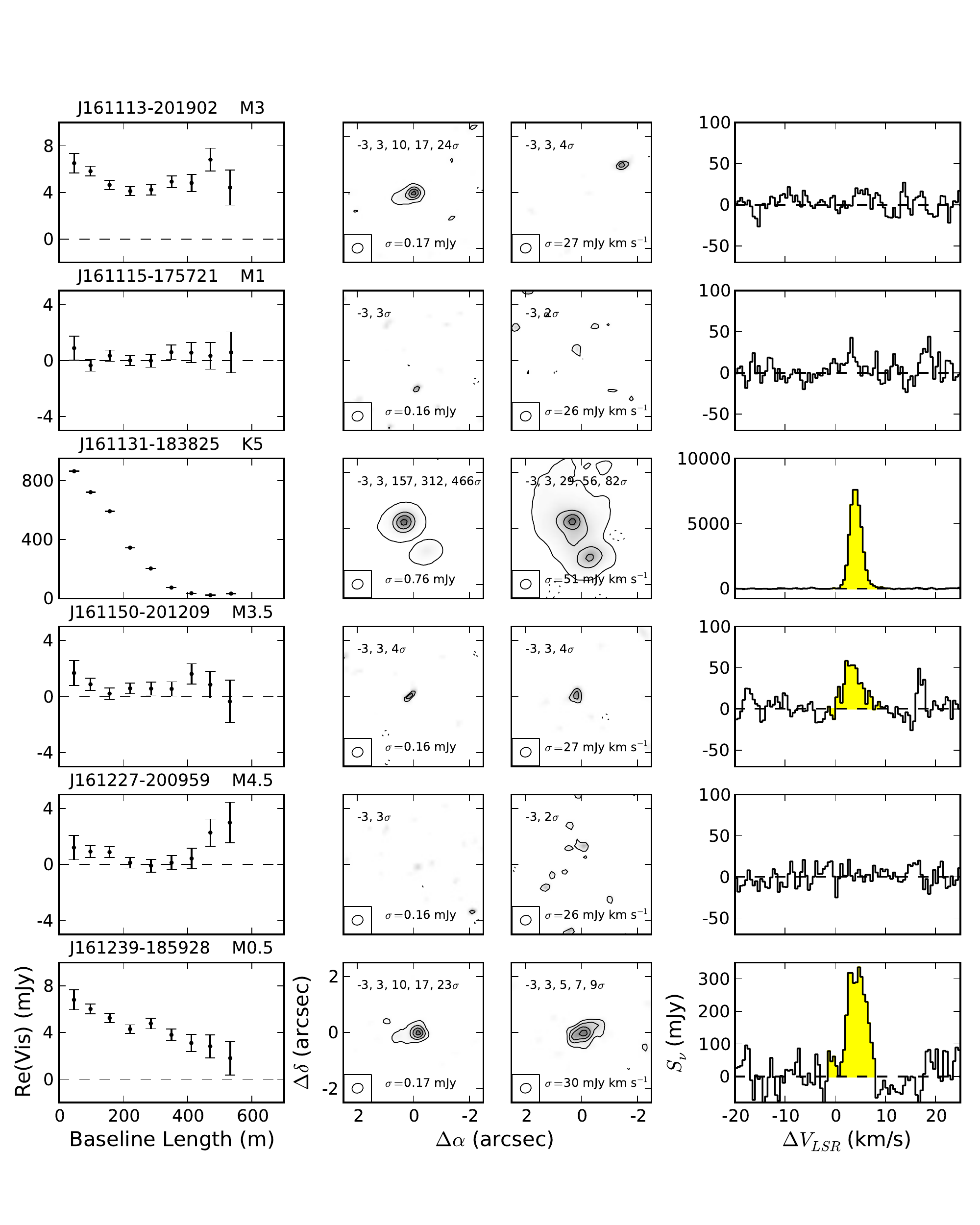}
}
\caption{Continued.}
\label{fig:images}
\end{figure}

\begin{figure}
\ContinuedFloat
\centering
\subfloat{
\includegraphics[width=0.9\textwidth]{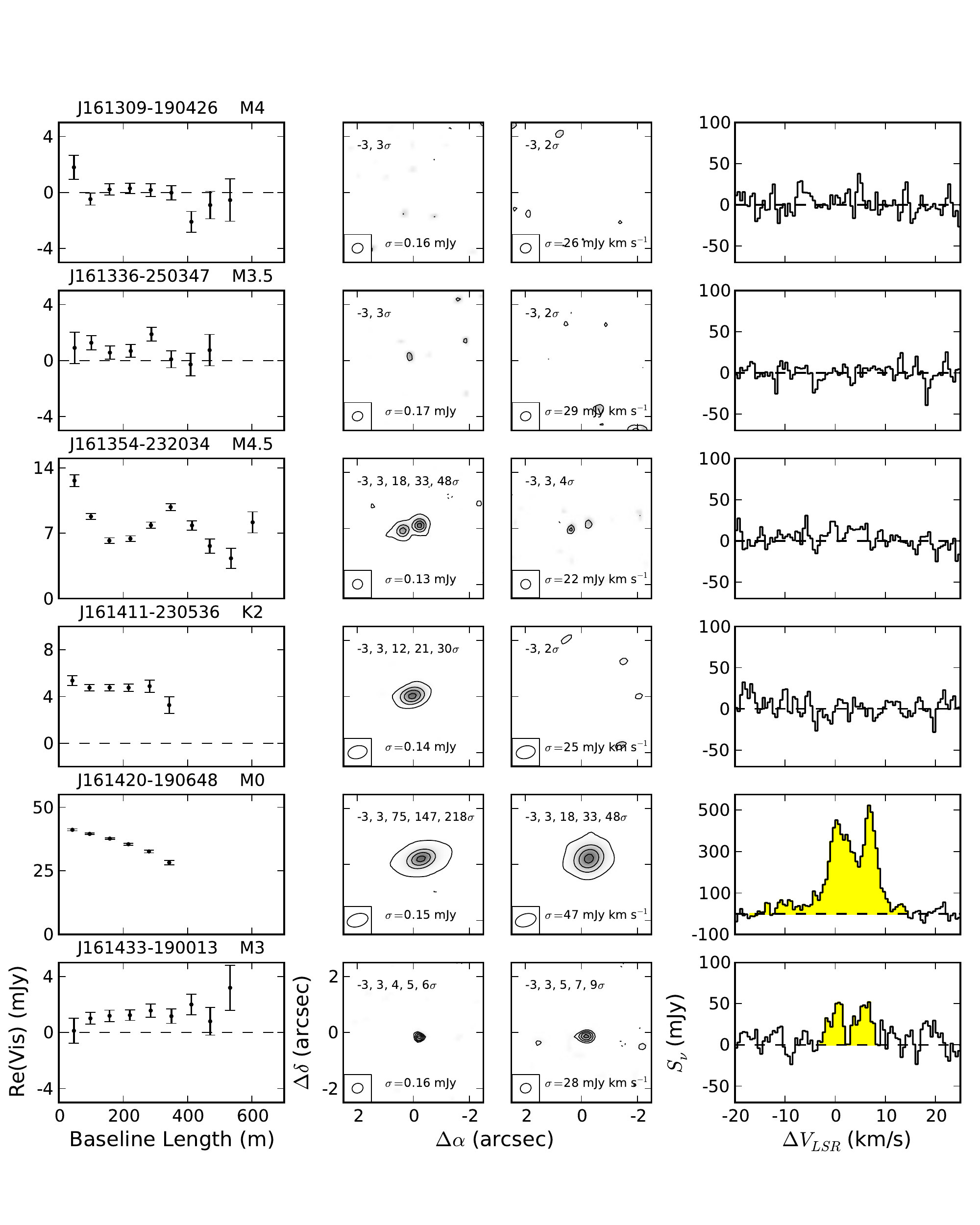}
}
\caption{Continued.}
\label{fig:images}
\end{figure}

\begin{figure}
\ContinuedFloat
\centering
\subfloat{
\includegraphics[width=0.9\textwidth]{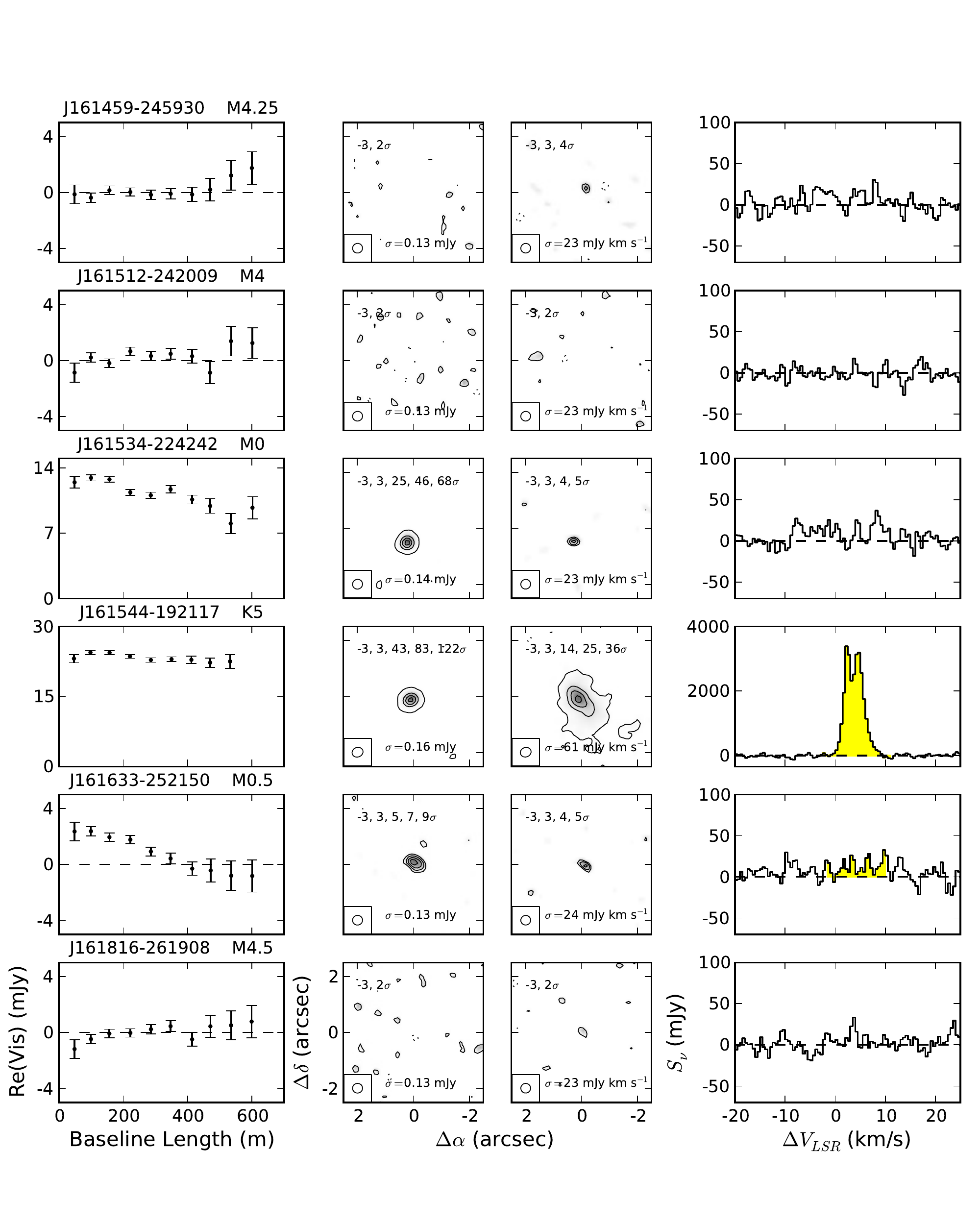}
}
\caption{Continued.}
\label{fig:images}
\end{figure}

\begin{figure}
\ContinuedFloat
\centering
\subfloat{
\includegraphics[width=0.9\textwidth]{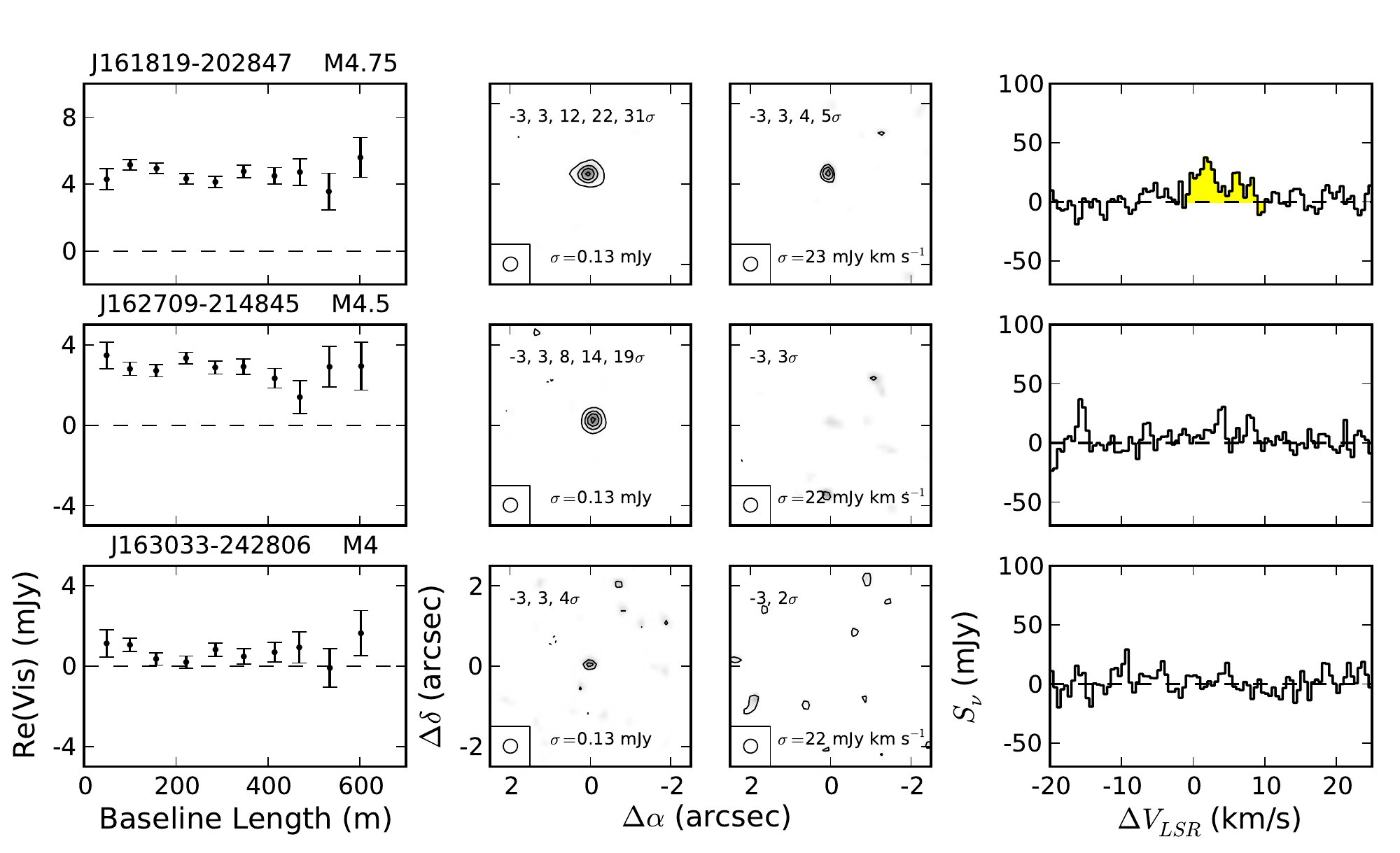}
}
\caption{Continued.}
\label{fig:images}
\end{figure}

\clearpage

\begin{figure}[!h]
\centering

\subfloat{
\includegraphics[width=0.9\textwidth]{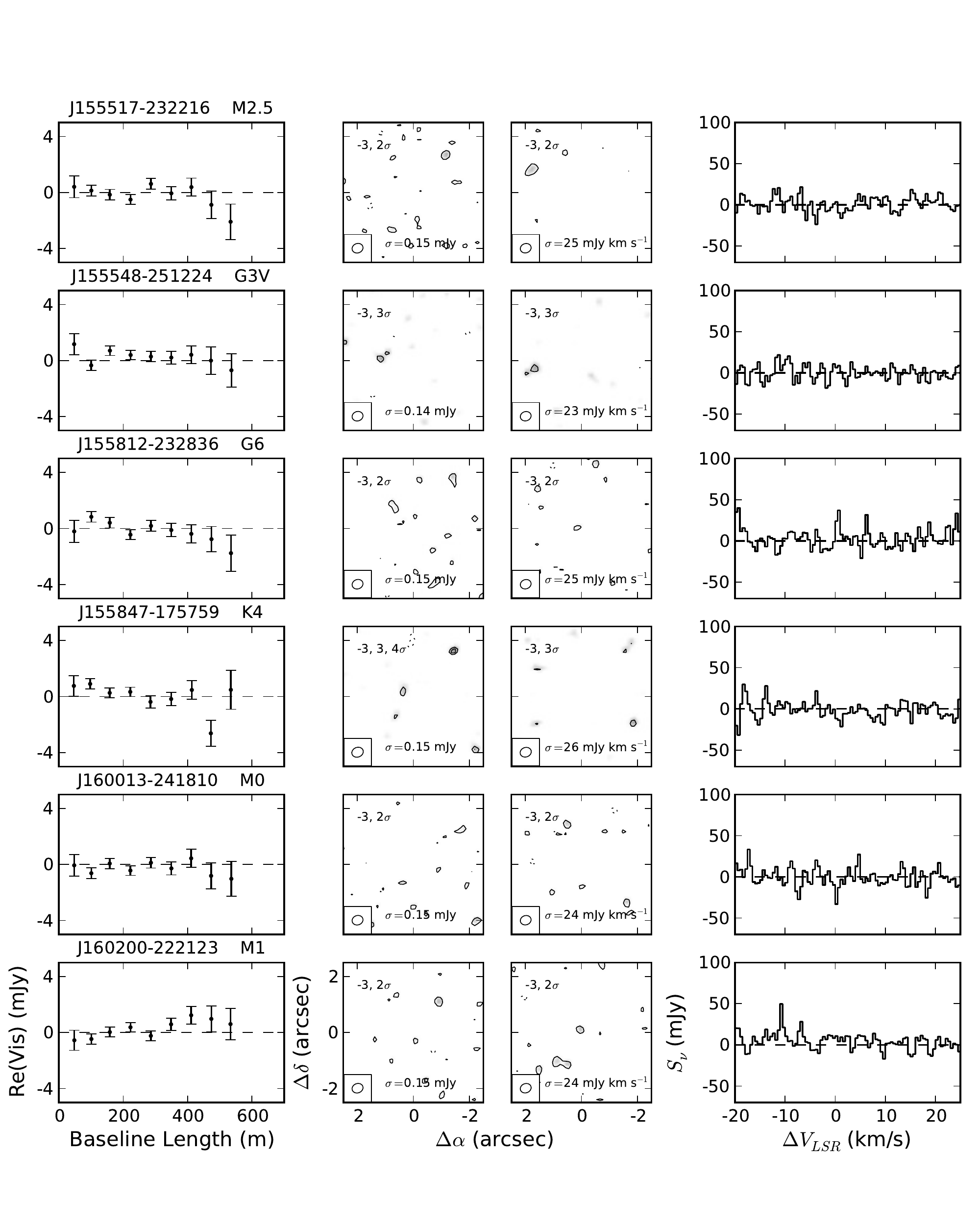}
}
\caption{Same as Figure \ref{fig:images}, but for the 31 debris/evolved transitional disks in the Upper Sco sample.
\label{fig:imagesd0}}
\end{figure}

\begin{figure}
\ContinuedFloat
\centering
\subfloat{
\includegraphics[width=0.9\textwidth]{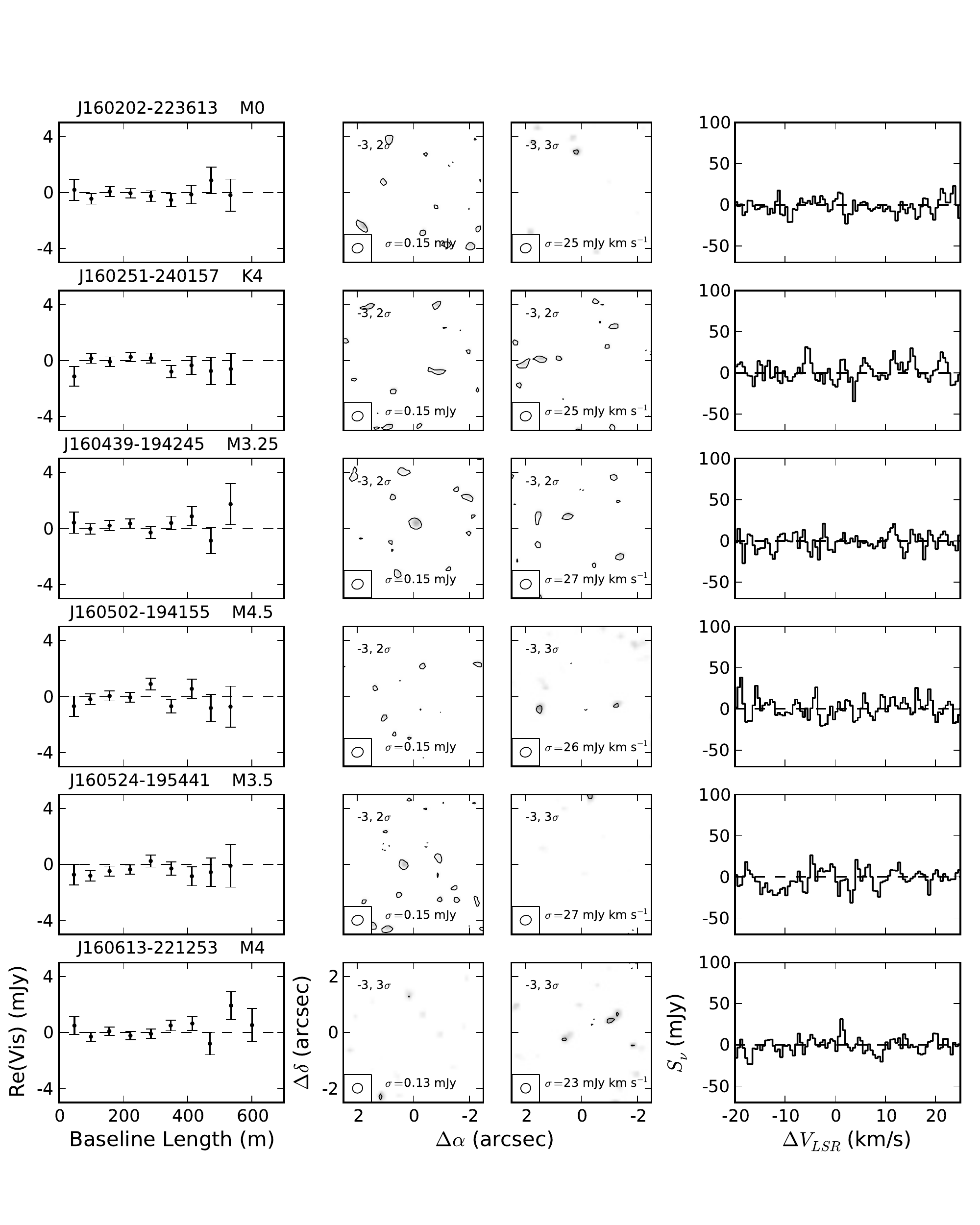}
}
\caption{Continued.}
\label{fig:imagesd0}
\end{figure}

\begin{figure}
\ContinuedFloat
\centering
\subfloat{
\includegraphics[width=0.9\textwidth]{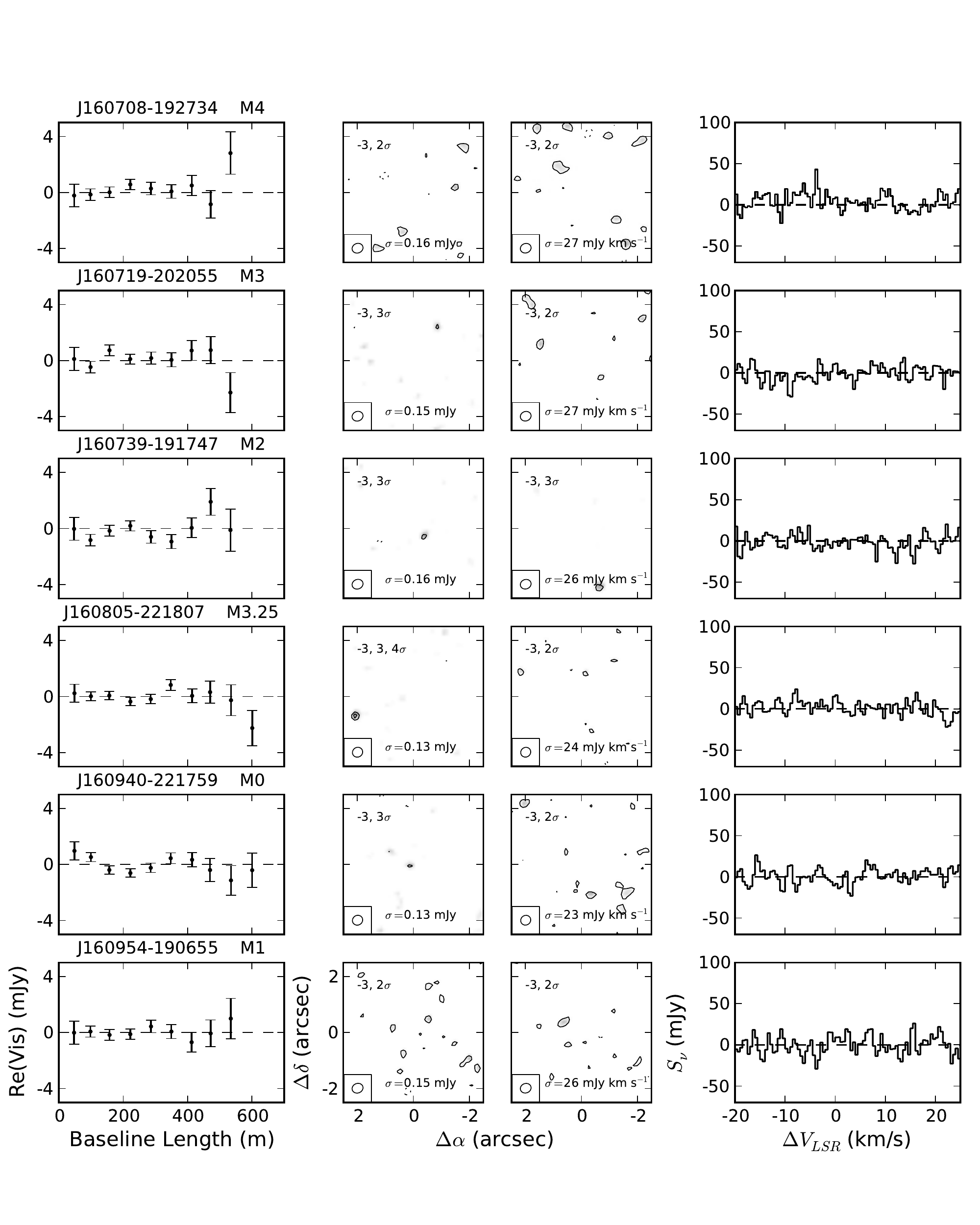}
}
\caption{Continued.}
\label{fig:imagesd0}
\end{figure}

\begin{figure}
\ContinuedFloat
\centering
\subfloat{
\includegraphics[width=0.9\textwidth]{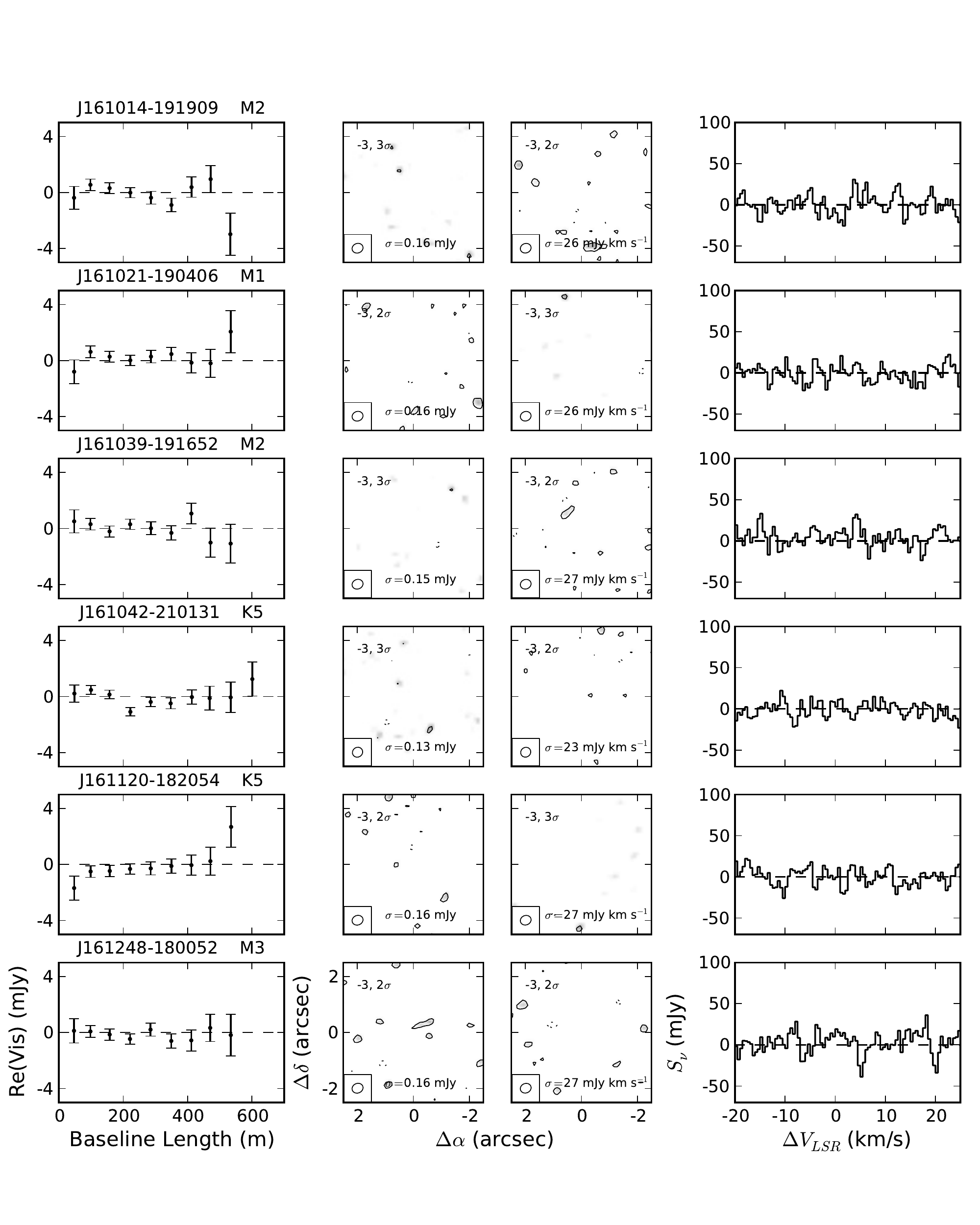}
}
\caption{Continued.}
\label{fig:imagesd0}
\end{figure}

\begin{figure}
\ContinuedFloat
\centering
\subfloat{
\includegraphics[width=0.9\textwidth]{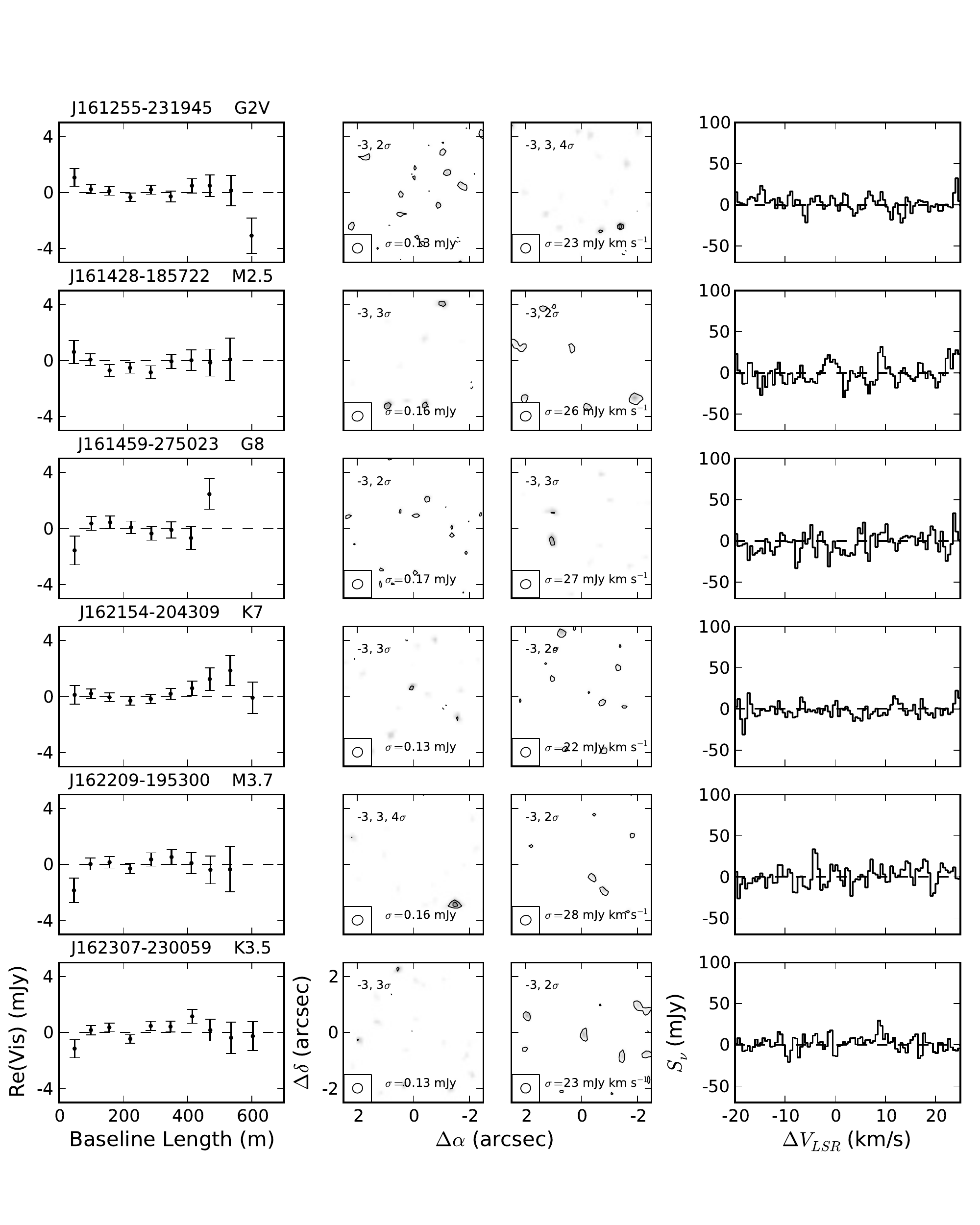}
}
\caption{Continued.}
\label{fig:imagesd0}
\end{figure}

\begin{figure}
\ContinuedFloat
\centering
\subfloat{
\includegraphics[width=0.9\textwidth]{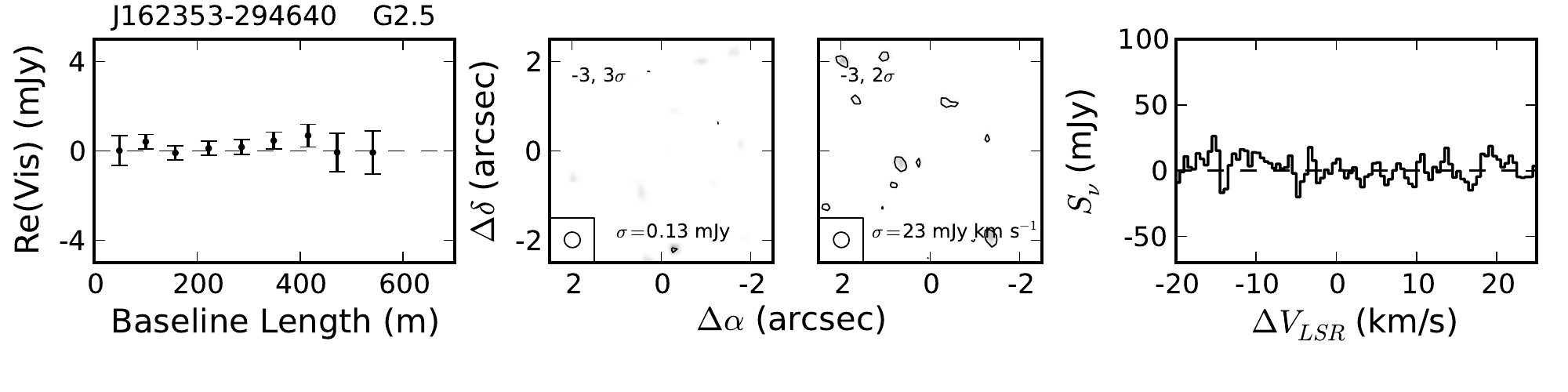}
}
\caption{Continued.}
\label{fig:imagesd0}
\end{figure}

\clearpage
\begin{figure}[!h]
\centerline{\includegraphics[width=\textwidth]{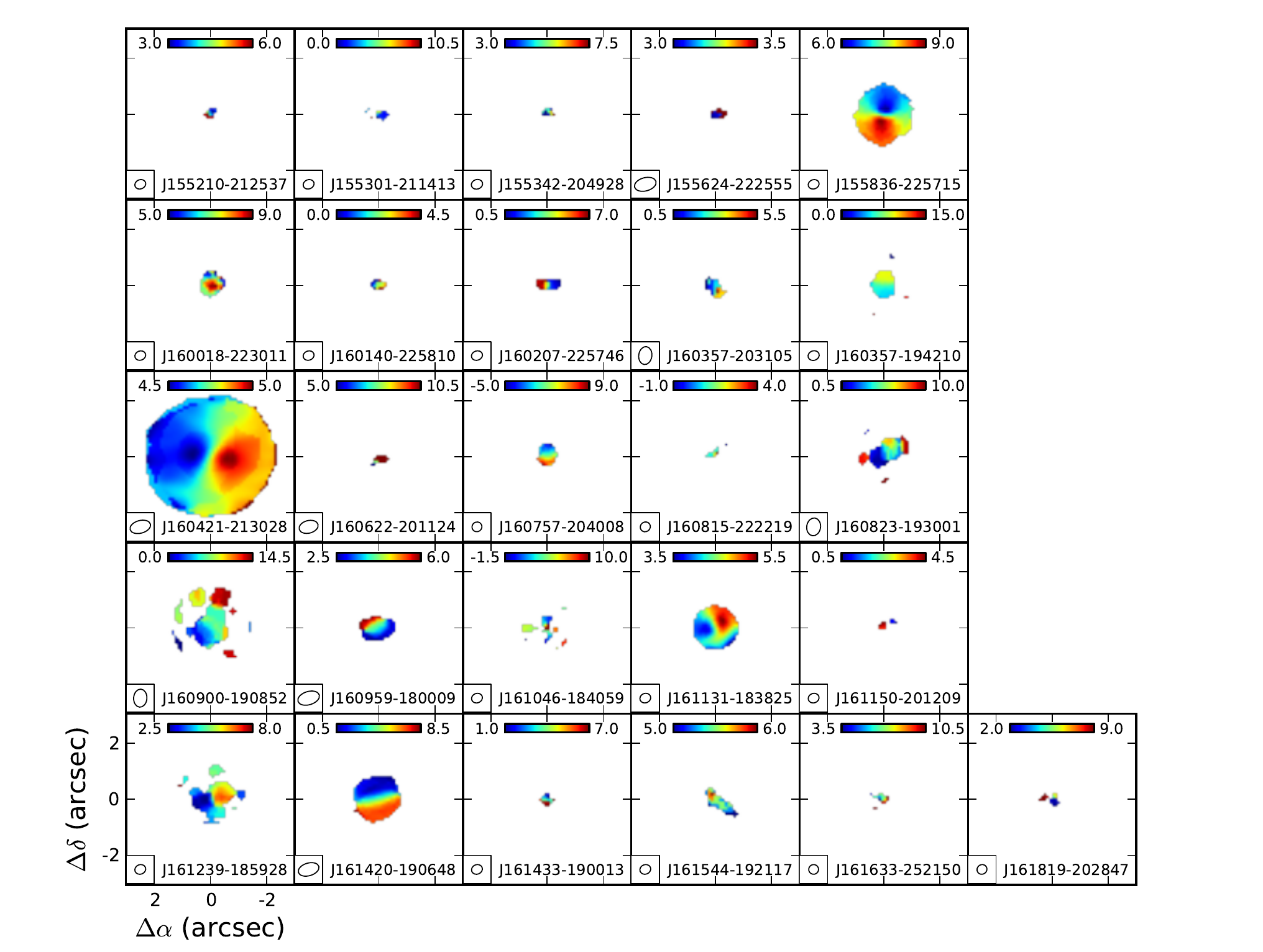}}
\caption{Moment 1 maps showing the mean LSRK velocity of the $^{12}$CO $J$ = 3$-$2 line for all sources detected ($>5\sigma$) in CO.  Each image is centered on the expected stellar 
position.  A color bar indicating the velocity range of each map in km s$^{-1}$ is shown at the top of each map.}
\label{fig:mom1}
\end{figure}

\begin{figure}[!h]
\centerline{\includegraphics[width=\textwidth]{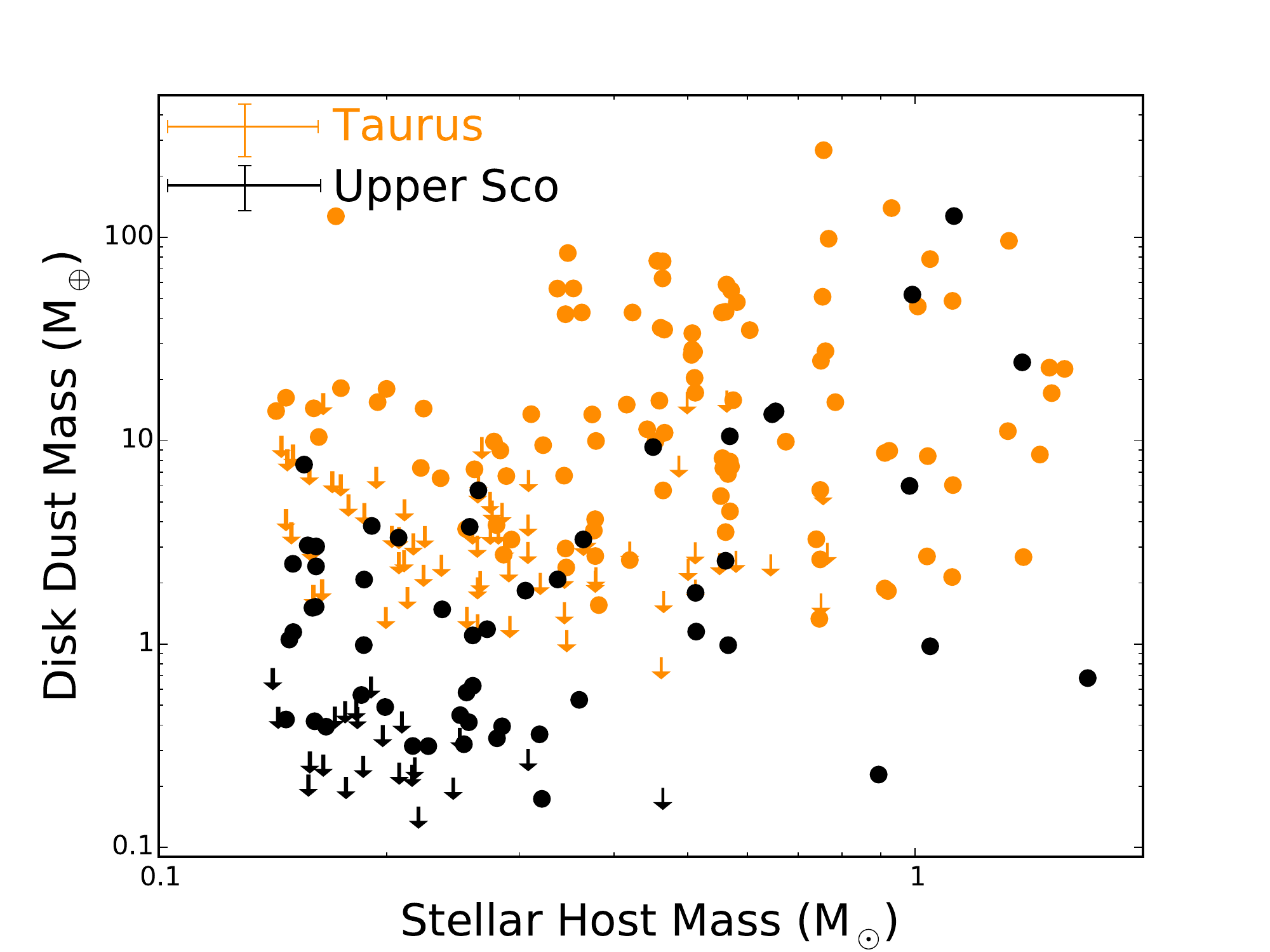}}
\caption{Disk dust mass as a function of stellar mass for the Taurus (orange) and Upper Sco (black) primordial disk samples.  Upper limits (3$\sigma$) are plotted as 
arrows. Typical error bars are shown in the upper left.}
\label{fig:disk_masses}
\end{figure}

\begin{figure}[!h]
\centerline{\includegraphics[width=\textwidth]{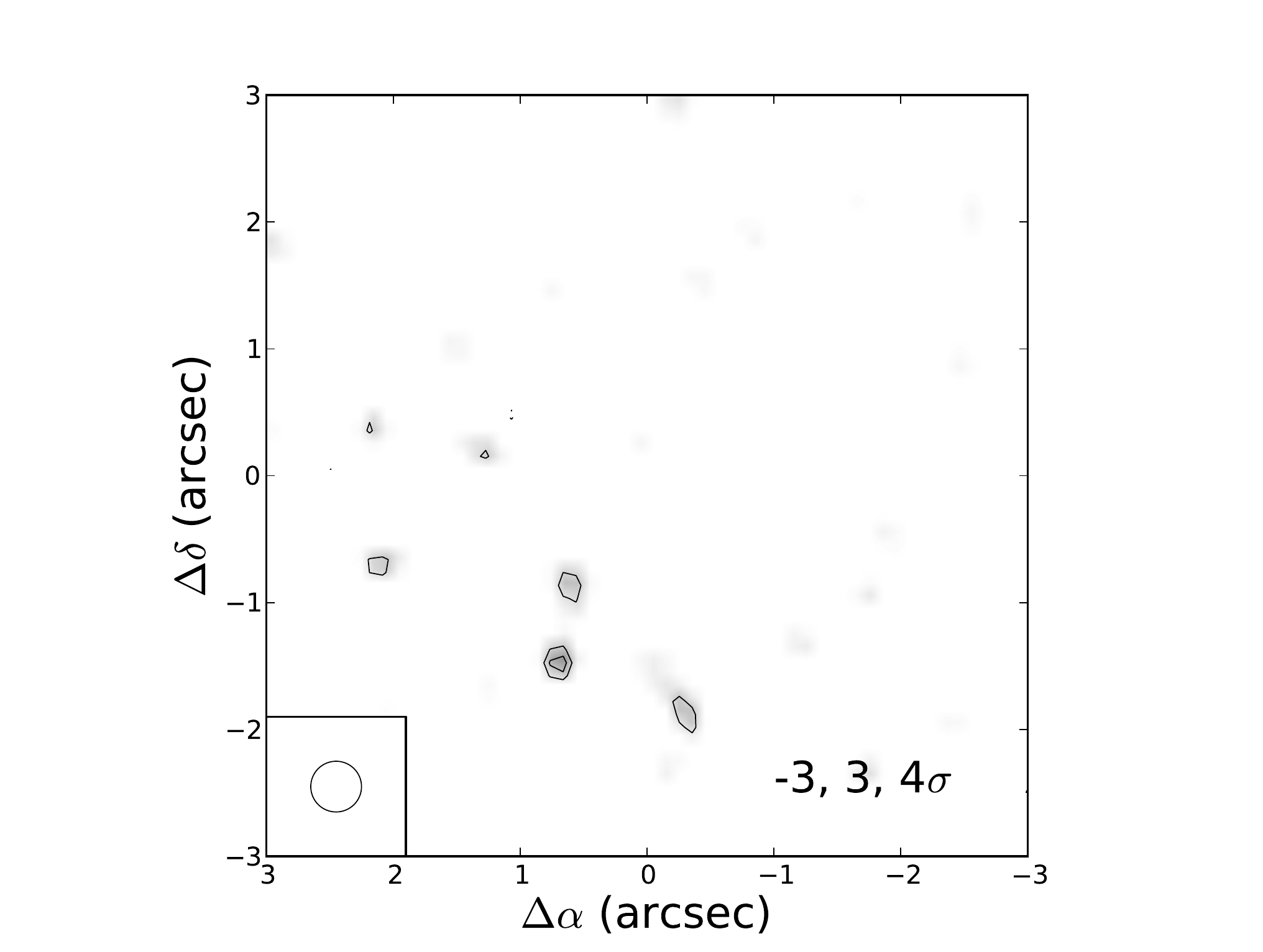}}
\caption{Stacked continuum image of the debris/evolved transitional disks which are not detected. Four sources were excluded due to the possibility of being identified as disks due to contamination from background 
sources (see the text). The flux density inside a 0$\overset{''}{.}$4 radius aperture at the center of the image is $0.03\pm0.05$ mJy. }
\label{fig:stacked_images}
\end{figure}

\begin{figure}[!h]
\centerline{\includegraphics[width=\textwidth]{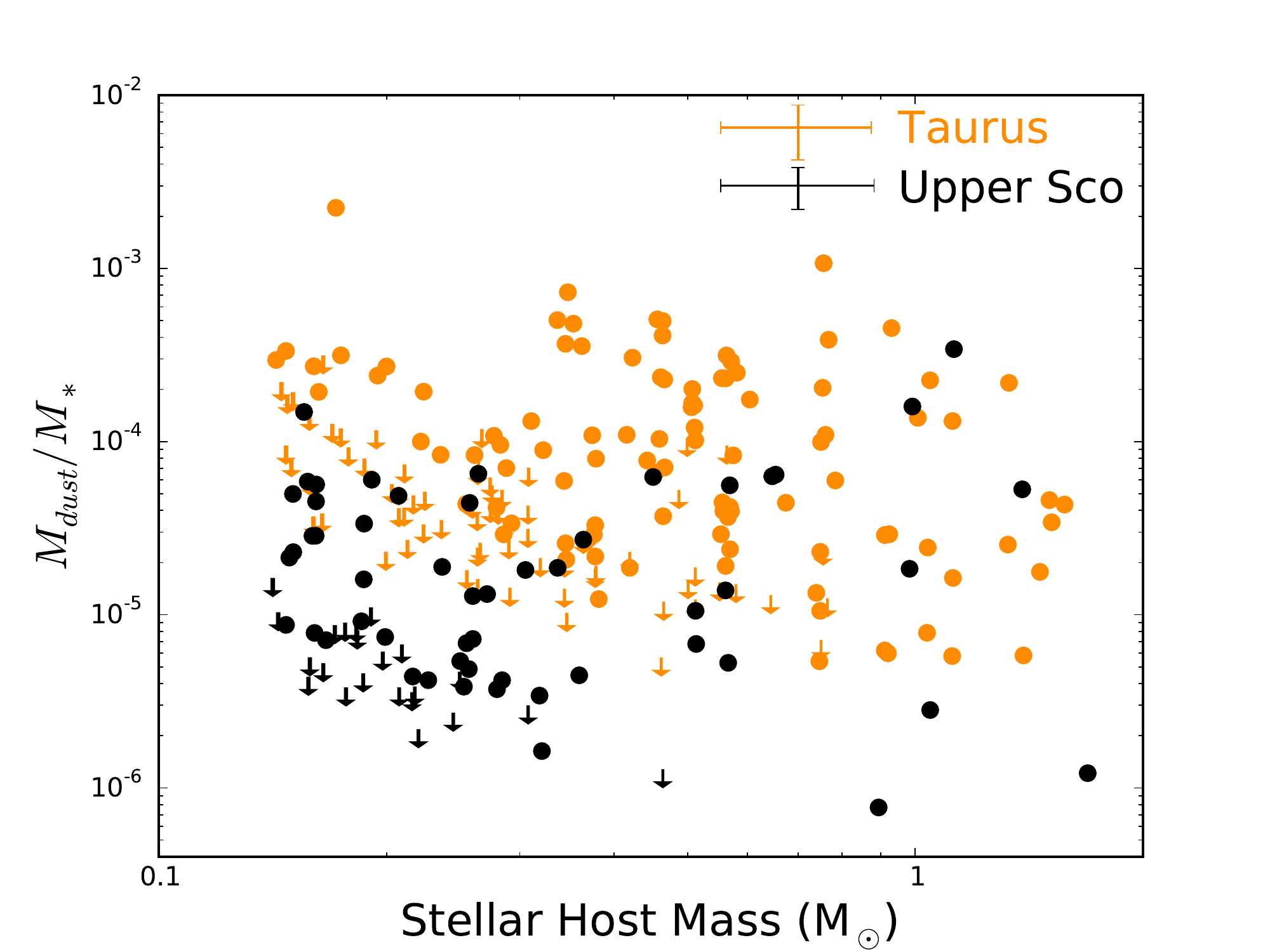}}
\caption{Ratio of disk dust mass to stellar mass as a function of stellar mass for the Taurus (orange) and Upper Sco (black) primordial disk samples. 
Upper limits (3$\sigma$) are 
plotted as arrows. Typical error bars are shown in the upper right. The probability that the dust mass over stellar mass values in each sample are 
drawn from the same distribution is $p = 1.4\times10^{-7}-4.8\times10^{-7}$.} 
\label{fig:mdms}
\end{figure}

\begin{figure}[!h]
\centerline{\includegraphics[width=\textwidth]{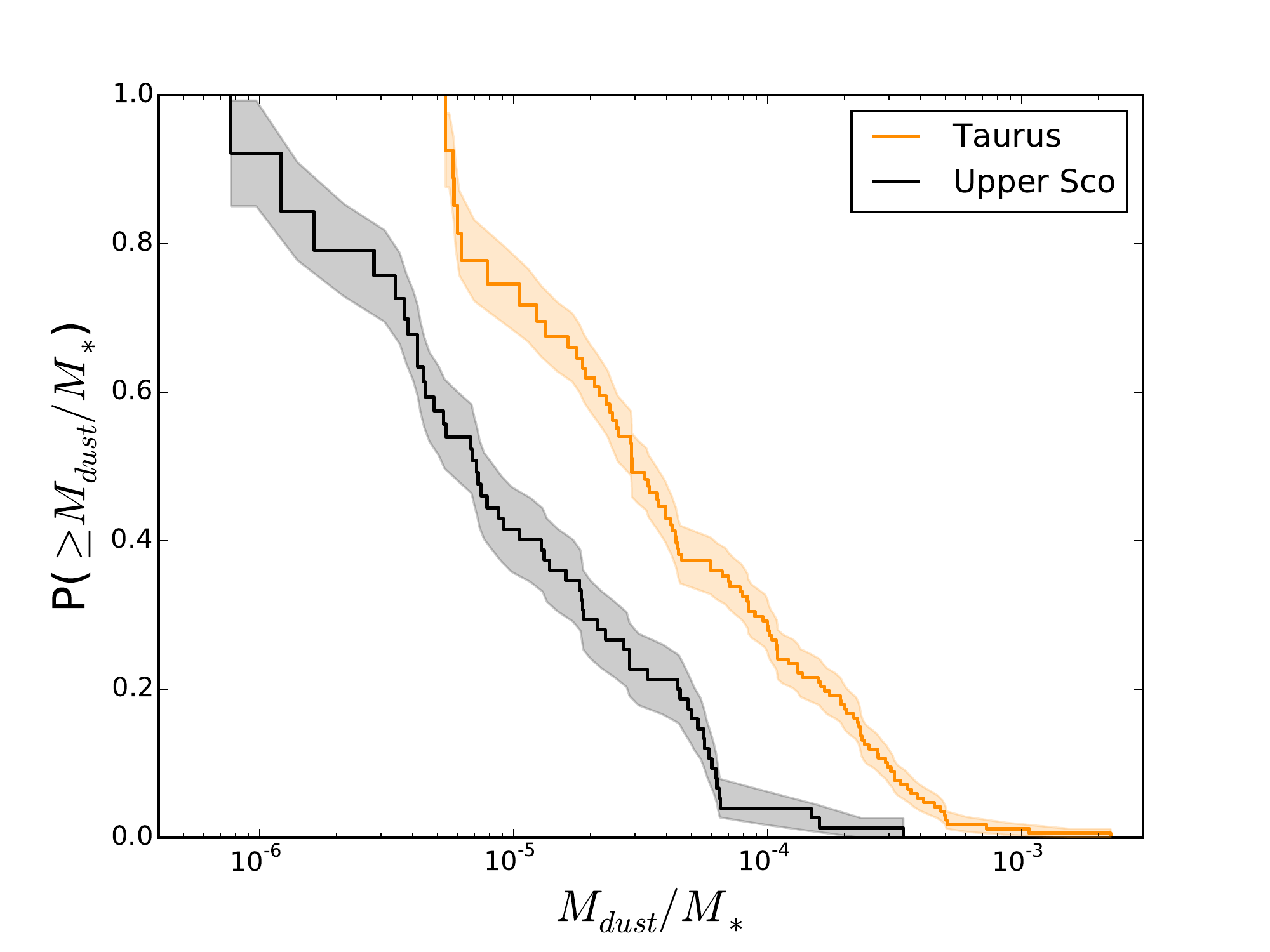}}
\caption{Cumulative distribution of the ratio of disk dust mass to stellar mass in Taurus and Upper Sco for the primordial disks. The shaded regions show the 
68.3\% confidence intervals of the distributions.  Using the Kaplan$-$Meier estimate of the mean of $\log(M_{\mathrm{dust}}/M_*)$ in Taurus and Upper Sco, we find that 
$\Delta\langle\log(M_{\mathrm{dust}}/M_*)\rangle = 0.64\pm0.09$, with $(M_{\mathrm{dust}}/M_*)$ a factor of $\sim 4.5$ lower in Upper Sco than in Taurus.
}
\label{fig:mdms_dist_all}
\end{figure}

\begin{figure}[!h]
\centerline{\includegraphics[width=\textwidth]{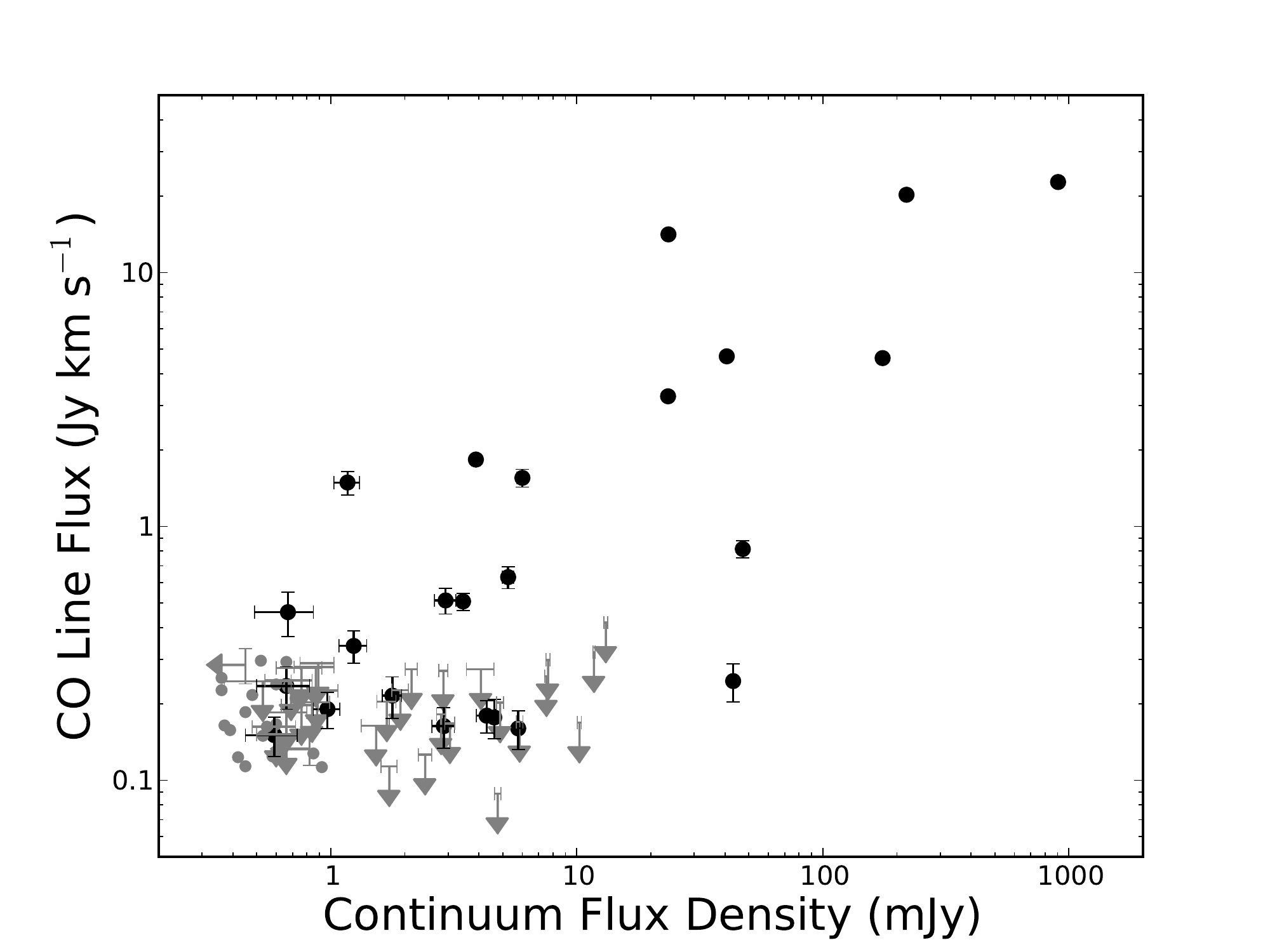}}
\caption{$^{12}$CO $J = 3-2$ flux versus 0.88 mm continuum flux density for the primordial disks in our Upper Sco sample.  Upper limits in the CO and 
continuum flux are shown with arrows.  The gray circles are upper limits for both the CO and continuum. Black points show CO and continuum detections.}
\label{fig:co_cont}
\end{figure}

\newpage
\begin{deluxetable}{lcccccc}
\tabletypesize{\footnotesize}
\tablecolumns{7} 
\tablewidth{0pt} 
\tablecaption{Stellar Properties}
\tablehead{
  \colhead{Source}   &
  \colhead{SpT}  &
  \colhead{Disk Type} &
  \colhead{Av}  &
  \colhead{$\log (T_*/K)$}  &
  \colhead{$\log (L_*/L_{\odot})$}  &
  \colhead{$\log (M_*/M_{\odot})$}
}
\startdata
2MASS J15354856-2958551 & M4    &  Full 	     &   0.7$\pm$0.5 & 3.51$\pm$0.02 & -0.60$\pm$0.15 & -0.58$(-0.09,+0.09)$ \\
2MASS J15514032-2146103 & M4    &  Evolved	     &   0.38$\pm$0.36 & 3.51$\pm$0.02 & -1.31$\pm$0.14 & -0.70$(-0.12,+0.13)$ \\
2MASS J15521088-2125372 & M4    &  Full 	     &   3.32$\pm$0.48 & 3.51$\pm$0.02 & -1.81$\pm$0.14 & -0.75$(-0.13,+0.13)$ \\
2MASS J15530132-2114135 & M4    &  Full 	     &   1.27$\pm$0.40 & 3.51$\pm$0.02 & -1.2$\pm$0.14 & -0.68$(-0.12,+0.13)$ \\
2MASS J15534211-2049282 & M3.5  &  Full 	     &   1.71$\pm$0.38 & 3.52$\pm$0.02 & -0.84$\pm$0.14 & -0.57$(-0.09,+0.10)$ \\
2MASS J15551704-2322165 & M2.5  &  Debris/Ev. Trans.   &   0.7$\pm$0.5 & 3.54$\pm$0.02 & -0.54$\pm$0.15 & -0.46$(-0.08,+0.08)$ \\
2MASS J15554883-2512240 & G3    &  Debris/Ev. Trans.   &   0.7$\pm$0.5 & 3.77$\pm$0.00 & 0.37$\pm$0.15 & 0.07$(-0.05,+0.04)$ \\
2MASS J15562477-2225552 & M4    &  Full 	     &   0.71$\pm$0.37 & 3.51$\pm$0.02 & -1.18$\pm$0.14 & -0.68$(-0.12,+0.13)$ \\
2MASS J15570641-2206060 & M4    &  Full 	     &   0.7$\pm$0.5 & 3.51$\pm$0.02 & -1.44$\pm$0.15 & -0.72$(-0.13,+0.14)$ \\
2MASS J15572986-2258438 & M4    &  Evolved	     &   0.7$\pm$0.5 & 3.51$\pm$0.02 & -1.33$\pm$0.15 & -0.70$(-0.12,+0.13)$ \\
2MASS J15581270-2328364 & G6    &  Debris/Ev. Trans.   &   0.7$\pm$0.5 & 3.76$\pm$0.00 & 0.40$\pm$0.15 & 0.10$(-0.06,+0.05)$ \\
2MASS J15582981-2310077 & M3    &  Full 	     &   1.10$\pm$0.41 & 3.53$\pm$0.02 & -1.31$\pm$0.14 & -0.59$(-0.11,+0.12)$ \\
2MASS J15583692-2257153 & G7    &  Full 	     &   0.7$\pm$0.5 & 3.75$\pm$0.00 & 0.47$\pm$0.15 & 0.14$(-0.05,+0.05)$ \\
2MASS J15584772-1757595 & K4    &  Debris/Ev. Trans.   &   0.7$\pm$0.5 & 3.65$\pm$0.01 & -0.01$\pm$0.15 & 0.08$(-0.04,+0.05)$ \\
2MASS J16001330-2418106 & M0    &  Debris/Ev. Trans.   &   0.7$\pm$0.5 & 3.59$\pm$0.01 & -0.56$\pm$0.15 & -0.24$(-0.05,+0.05)$ \\
2MASS J16001730-2236504 & M4    &  Full 	     &   0.7$\pm$0.5 & 3.51$\pm$0.02 & -0.82$\pm$0.15 & -0.61$(-0.10,+0.11)$ \\
2MASS J16001844-2230114 & M4.5  &  Full 	     &   0.7$\pm$0.5 & 3.50$\pm$0.02 & -1.13$\pm$0.15 & -0.73$(-0.12,+0.14)$ \\
2MASS J16014086-2258103 & M4    &  Full 	     &   0.83$\pm$0.35 & 3.51$\pm$0.02 & -0.90$\pm$0.14 & -0.63$(-0.11,+0.11)$ \\
2MASS J16014157-2111380 & M4    &  Full 	     &   0.7$\pm$0.5 & 3.51$\pm$0.02 & -1.56$\pm$0.15 & -0.73$(-0.13,+0.14)$ \\
2MASS J16020039-2221237 & M1    &  Debris/Ev. Trans.   &   0.7$\pm$0.5 & 3.57$\pm$0.02 & -0.32$\pm$0.15 & -0.33$(-0.09,+0.08)$ \\
2MASS J16020287-2236139 & M0    &  Debris/Ev. Trans.   &   0.75$\pm$0.33 & 3.59$\pm$0.01 & -1.41$\pm$0.14 & -0.30$(-0.05,+0.05)$ \\
2MASS J16020757-2257467 & M2.5  &  Full 	     &   0.41$\pm$0.33 & 3.54$\pm$0.02 & -0.82$\pm$0.14 & -0.47$(-0.09,+0.08)$ \\
2MASS J16024152-2138245 & M4.75 &  Full 	     &   0.43$\pm$0.37 & 3.50$\pm$0.02 & -1.44$\pm$0.14 & -0.81$(-0.15,+0.10)$ \\
2MASS J16025123-2401574 & K4    &  Debris/Ev. Trans.   &   0.7$\pm$0.5 & 3.65$\pm$0.01 & -0.20$\pm$0.15 & 0.04$(-0.04,+0.05)$ \\
2MASS J16030161-2207523 & M4.75 &  Full 	     &   0.66$\pm$0.44 & 3.50$\pm$0.02 & -1.59$\pm$0.14 & -0.82$(-0.15,+0.09)$ \\
2MASS J16031329-2112569 & M4.75 &  Full 	     &   0.45$\pm$0.42 & 3.50$\pm$0.02 & -1.38$\pm$0.14 & -0.80$(-0.15,+0.11)$ \\
2MASS J16032225-2413111 & M3.5  &  Full 	     &   0.59$\pm$0.32 & 3.52$\pm$0.02 & -0.97$\pm$0.14 & -0.58$(-0.10,+0.11)$ \\
2MASS J16035767-2031055 & K5    &  Full 	     &   0.7$\pm$0.5 & 3.64$\pm$0.01 & -0.17$\pm$0.15 & 0.02$(-0.05,+0.05)$ \\
2MASS J16035793-1942108 & M2    &  Full 	     &   0.7$\pm$0.5 & 3.55$\pm$0.02 & -0.96$\pm$0.15 & -0.44$(-0.10,+0.08)$ \\
2MASS J16041740-1942287 & M3.5  &  Full 	     &   0.36$\pm$0.37 & 3.52$\pm$0.02 & -1.07$\pm$0.14 & -0.60$(-0.10,+0.12)$ \\
2MASS J16042165-2130284 & K2    &  Transitional      &   0.7$\pm$0.5 & 3.69$\pm$0.02 & -0.24$\pm$0.15 & 0.00$(-0.06,+0.05)$ \\
2MASS J16043916-1942459 & M3.25 &  Debris/Ev. Trans.   &   0.37$\pm$0.36 & 3.53$\pm$0.02 & -1.17$\pm$0.14 & -0.59$(-0.11,+0.12)$ \\
2MASS J16050231-1941554 & M4.5  &  Debris/Ev. Trans.   &   -0.07$\pm$0.40 & 3.5$\pm$0.02 & -1.57$\pm$0.14 & -0.79$(-0.15,+0.12)$ \\
2MASS J16052459-1954419 & M3.5  &  Debris/Ev. Trans.   &   0.36$\pm$0.38 & 3.52$\pm$0.02 & -1.08$\pm$0.14 & -0.6$(-0.11,+0.11)$ \\
2MASS J16052556-2035397 & M5    &  Evolved	     &   0.38$\pm$0.42 & 3.49$\pm$0.02 & -1.37$\pm$0.14 & -0.83$(-0.15,+0.09)$ \\
2MASS J16052661-1957050 & M4.5  &  Evolved	     &   0.70$\pm$0.40 & 3.5$\pm$0.02 & -1.13$\pm$0.14 & -0.73$(-0.12,+0.14)$ \\
2MASS J16053215-1933159 & M5    &  Evolved	     &   0.20$\pm$0.43 & 3.49$\pm$0.02 & -1.59$\pm$0.14 & -0.85$(-0.14,+0.08)$ \\
2MASS J16054540-2023088 & M2    &  Full 	     &   1.61$\pm$0.30 & 3.55$\pm$0.02 & -0.90$\pm$0.14 & -0.44$(-0.10,+0.08)$ \\
2MASS J16055863-1949029 & M4    &  Evolved	     &   0.39$\pm$0.35 & 3.51$\pm$0.02 & -1.20$\pm$0.14 & -0.68$(-0.12,+0.13)$ \\
2MASS J16060061-1957114 & M5    &  Evolved	     &   0.22$\pm$0.38 & 3.49$\pm$0.02 & -1.20$\pm$0.14 & -0.80$(-0.14,+0.11)$ \\
2MASS J16061330-2212537 & M4    &  Debris/Ev. Trans.   &   0.7$\pm$0.5 & 3.51$\pm$0.02 & -0.67$\pm$0.15 & -0.59$(-0.09,+0.10)$ \\
2MASS J16062196-1928445 & M0    &  Transitional      &   1.16$\pm$0.26 & 3.59$\pm$0.01 & -0.25$\pm$0.14 & -0.25$(-0.05,+0.04)$ \\
2MASS J16062277-2011243 & M5    &  Transitional      &   -0.20$\pm$0.38 & 3.49$\pm$0.02 & -1.41$\pm$0.14 & -0.83$(-0.15,+0.09)$ \\
2MASS J16063539-2516510 & M4.5  &  Evolved	     &   -0.08$\pm$0.37 & 3.50$\pm$0.02 & -1.60$\pm$0.14 & -0.80$(-0.15,+0.12)$ \\
2MASS J16064102-2455489 & M4.5  &  Evolved	     &   0.7$\pm$0.5 & 3.50$\pm$0.02 & -1.70$\pm$0.15 & -0.80$(-0.15,+0.11)$ \\
2MASS J16064115-2517044 & M3.25 &  Evolved	     &   0.56$\pm$0.31 & 3.53$\pm$0.02 & -1.22$\pm$0.14 & -0.60$(-0.11,+0.12)$ \\
2MASS J16064385-1908056 & K6    &  Evolved	     &   0.75$\pm$0.26 & 3.62$\pm$0.01 & -0.39$\pm$0.14 & -0.05$(-0.04,+0.05)$ \\
2MASS J16070014-2033092 & M2.75 &  Full 	     &   0.04$\pm$0.30 & 3.54$\pm$0.02 & -0.95$\pm$0.14 & -0.51$(-0.09,+0.10)$ \\
2MASS J16070211-2019387 & M5    &  Full 	     &   0.66$\pm$0.44 & 3.49$\pm$0.02 & -1.52$\pm$0.14 & -0.84$(-0.15,+0.08)$ \\
2MASS J16070873-1927341 & M4    &  Debris/Ev. Trans.   &   1.15$\pm$0.37 & 3.51$\pm$0.02 & -1.28$\pm$0.14 & -0.70$(-0.12,+0.13)$ \\
2MASS J16071971-2020555 & M3    &  Debris/Ev. Trans.   &   1.43$\pm$0.36 & 3.53$\pm$0.02 & -1.05$\pm$0.14 & -0.55$(-0.10,+0.11)$ \\
2MASS J16072625-2432079 & M3.5  &  Full 	     &   0.00$\pm$0.37 & 3.52$\pm$0.02 & -0.92$\pm$0.14 & -0.58$(-0.10,+0.11)$ \\
2MASS J16072747-2059442 & M4.75 &  Evolved	     &   0.7$\pm$0.5 & 3.50$\pm$0.02 & -0.99$\pm$0.15 & -0.73$(-0.12,+0.13)$ \\
2MASS J16073939-1917472 & M2    &  Debris/Ev. Trans.   &   0.76$\pm$0.35 & 3.55$\pm$0.02 & -0.76$\pm$0.14 & -0.43$(-0.09,+0.08)$ \\
2MASS J16075796-2040087 & M1    &  Full 	     &   0.7$\pm$0.5 & 3.57$\pm$0.02 & -0.82$\pm$0.15 & -0.35$(-0.10,+0.10)$ \\
2MASS J16080555-2218070 & M3.25 &  Debris/Ev. Trans.   &   0.21$\pm$0.34 & 3.53$\pm$0.02 & -0.82$\pm$0.14 & -0.54$(-0.09,+0.10)$ \\
2MASS J16081566-2222199 & M3.25 &  Full 	     &   0.17$\pm$0.33 & 3.53$\pm$0.02 & -0.85$\pm$0.14 & -0.55$(-0.09,+0.10)$ \\
2MASS J16082324-1930009 & K9    &  Full 	     &   0.7$\pm$0.5 & 3.59$\pm$0.01 & -0.59$\pm$0.15 & -0.18$(-0.04,+0.05)$ \\
2MASS J16082751-1949047 & M5    &  Evolved	     &   0.72$\pm$0.40 & 3.49$\pm$0.02 & -1.16$\pm$0.14 & -0.79$(-0.14,+0.11)$ \\
2MASS J16083455-2211559 & M4.5  &  Evolved	     &   1.07$\pm$0.39 & 3.50$\pm$0.02 & -1.46$\pm$0.14 & -0.78$(-0.14,+0.12)$ \\
2MASS J16084894-2400045 & M3.75 &  Full 	     &   0.57$\pm$0.35 & 3.52$\pm$0.02 & -1.25$\pm$0.14 & -0.66$(-0.12,+0.12)$ \\
2MASS J16090002-1908368 & M5    &  Full 	     &   0.31$\pm$0.40 & 3.49$\pm$0.02 & -1.33$\pm$0.14 & -0.82$(-0.15,+0.09)$ \\
2MASS J16090075-1908526 & K9    &  Full 	     &   0.7$\pm$0.5 & 3.59$\pm$0.01 & -0.45$\pm$0.15 & -0.19$(-0.05,+0.05)$ \\
2MASS J16093558-1828232 & M3    &  Full 	     &   2.00$\pm$0.29 & 3.53$\pm$0.02 & -1.06$\pm$0.14 & -0.55$(-0.09,+0.11)$ \\
2MASS J16094098-2217594 & M0    &  Debris/Ev. Trans.   &   0.7$\pm$0.5 & 3.59$\pm$0.01 & -0.17$\pm$0.15 & -0.25$(-0.04,+0.04)$ \\
2MASS J16095361-1754474 & M3    &  Full 	     &   1.71$\pm$0.37 & 3.53$\pm$0.02 & -1.34$\pm$0.14 & -0.59$(-0.11,+0.12)$ \\
2MASS J16095441-1906551 & M1    &  Debris/Ev. Trans.   &   0.7$\pm$0.5 & 3.57$\pm$0.02 & -0.65$\pm$0.15 & -0.34$(-0.10,+0.09)$ \\
2MASS J16095933-1800090 & M4    &  Full 	     &   0.58$\pm$0.37 & 3.51$\pm$0.02 & -1.00$\pm$0.14 & -0.64$(-0.11,+0.12)$ \\
2MASS J16101473-1919095 & M2    &  Debris/Ev. Trans.   &   0.87$\pm$0.34 & 3.55$\pm$0.02 & -0.84$\pm$0.14 & -0.43$(-0.09,+0.08)$ \\
2MASS J16101888-2502325 & M4.5  &  Transitional      &   0.7$\pm$0.5 & 3.50$\pm$0.02 & -1.35$\pm$0.15 & -0.77$(-0.13,+0.13)$ \\
2MASS J16102174-1904067 & M1    &  Debris/Ev. Trans.   &   0.7$\pm$0.5 & 3.57$\pm$0.02 & -0.67$\pm$0.15 & -0.34$(-0.10,+0.09)$ \\
2MASS J16102819-1910444 & M4    &  Full 	     &   0.7$\pm$0.5 & 3.51$\pm$0.02 & -1.62$\pm$0.15 & -0.74$(-0.13,+0.14)$ \\
2MASS J16102857-1904469 & M3    &  Evolved	     &   0.7$\pm$0.5 & 3.53$\pm$0.02 & -0.35$\pm$0.15 & -0.49$(-0.08,+0.07)$ \\
2MASS J16103956-1916524 & M2    &  Debris/Ev. Trans.   &   0.7$\pm$0.5 & 3.55$\pm$0.02 & -0.94$\pm$0.15 & -0.44$(-0.10,+0.08)$ \\
2MASS J16104202-2101319 & K5    &  Debris/Ev. Trans.   &   0.7$\pm$0.5 & 3.64$\pm$0.01 & -0.14$\pm$0.15 & 0.02$(-0.05,+0.06)$ \\
2MASS J16104636-1840598 & M4.5  &  Full 	     &   0.7$\pm$0.5 & 3.50$\pm$0.02 & -1.57$\pm$0.15 & -0.79$(-0.15,+0.12)$ \\
2MASS J16111330-2019029 & M3    &  Full 	     &   1.68$\pm$0.35 & 3.53$\pm$0.02 & -0.76$\pm$0.14 & -0.52$(-0.08,+0.09)$ \\
2MASS J16111534-1757214 & M1    &  Full 	     &   0.7$\pm$0.5 & 3.57$\pm$0.02 & -0.48$\pm$0.15 & -0.33$(-0.10,+0.08)$ \\
2MASS J16112057-1820549 & K5    &  Debris/Ev. Trans.   &   0.7$\pm$0.5 & 3.64$\pm$0.01 & -0.13$\pm$0.15 & 0.03$(-0.05,+0.06)$ \\
2MASS J16113134-1838259 & K5    &  Full 	     &   0.7$\pm$0.5 & 3.64$\pm$0.01 & 0.45$\pm$0.15 & 0.05$(-0.09,+0.10)$ \\
2MASS J16115091-2012098 & M3.5  &  Full 	     &   0.65$\pm$0.39 & 3.52$\pm$0.02 & -1.04$\pm$0.14 & -0.60$(-0.11,+0.11)$ \\
2MASS J16122737-2009596 & M4.5  &  Full 	     &   1.24$\pm$0.45 & 3.50$\pm$0.02 & -1.44$\pm$0.14 & -0.78$(-0.14,+0.13)$ \\
2MASS J16123916-1859284 & M0.5  &  Full 	     &   0.7$\pm$0.5 & 3.58$\pm$0.01 & -0.50$\pm$0.15 & -0.29$(-0.07,+0.07)$ \\
2MASS J16124893-1800525 & M3    &  Debris/Ev. Trans.   &   0.81$\pm$0.38 & 3.53$\pm$0.02 & -0.96$\pm$0.14 & -0.54$(-0.09,+0.10)$ \\
2MASS J16125533-2319456 & G2    &  Debris/Ev. Trans.   &   0.7$\pm$0.5 & 3.77$\pm$0.00 & 0.78$\pm$0.15 & 0.21$(-0.07,+0.09)$ \\
2MASS J16130996-1904269 & M4    &  Full 	     &   1.13$\pm$0.38 & 3.51$\pm$0.02 & -1.11$\pm$0.14 & -0.67$(-0.12,+0.12)$ \\
2MASS J16133650-2503473 & M3.5  &  Full 	     &   0.7$\pm$0.5 & 3.52$\pm$0.02 & -1.00$\pm$0.15 & -0.59$(-0.10,+0.11)$ \\
2MASS J16135434-2320342 & M4.5  &  Full 	     &   -0.55$\pm$0.37 & 3.50$\pm$0.02 & -1.07$\pm$0.14 & -0.72$(-0.12,+0.13)$ \\
2MASS J16141107-2305362 & K2    &  Full 	     &   0.7$\pm$0.5 & 3.69$\pm$0.02 & 0.43$\pm$0.15 & 0.23$(-0.05,+0.05)$ \\
2MASS J16142029-1906481 & M0    &  Full 	     &   2.0$\pm$0.5 & 3.59$\pm$0.01 & -0.33$\pm$0.15 & -0.25$(-0.05,+0.04)$ \\
2MASS J16142893-1857224 & M2.5  &  Debris/Ev. Trans.   &   0.7$\pm$0.5 & 3.54$\pm$0.02 & -0.61$\pm$0.15 & -0.46$(-0.08,+0.08)$ \\
2MASS J16143367-1900133 & M3    &  Full 	     &   0.7$\pm$0.5 & 3.53$\pm$0.02 & -0.47$\pm$0.15 & -0.50$(-0.08,+0.07)$ \\
2MASS J16145918-2750230 & G8    &  Debris/Ev. Trans.   &   0.7$\pm$0.5 & 3.74$\pm$0.01 & 0.07$\pm$0.15 & 0.03$(-0.04,+0.02)$ \\
2MASS J16145928-2459308 & M4.25 &  Full 	     &   4.29$\pm$0.24 & 3.51$\pm$0.02 & -0.92$\pm$0.14 & -0.66$(-0.11,+0.12)$ \\
2MASS J16151239-2420091 & M4    &  Transitional      &   1.39$\pm$0.36 & 3.51$\pm$0.02 & -1.62$\pm$0.14 & -0.74$(-0.13,+0.13)$ \\
2MASS J16153456-2242421 & M0    &  Full 	     &   0.7$\pm$0.5 & 3.59$\pm$0.01 & -0.13$\pm$0.15 & -0.25$(-0.04,+0.04)$ \\
2MASS J16154416-1921171 & K5    &  Full 	     &   0.7$\pm$0.5 & 3.64$\pm$0.01 & -0.31$\pm$0.15 & -0.01$(-0.04,+0.05)$ \\
2MASS J16163345-2521505 & M0.5  &  Full 	     &   1.13$\pm$0.29 & 3.58$\pm$0.01 & -0.83$\pm$0.14 & -0.29$(-0.08,+0.08)$ \\
2MASS J16181618-2619080 & M4.5  &  Evolved	     &   1.64$\pm$0.36 & 3.5$\pm$0.02 & -1.26$\pm$0.14 & -0.75$(-0.13,+0.13)$ \\
2MASS J16181904-2028479 & M4.75 &  Evolved	     &   1.86$\pm$0.39 & 3.5$\pm$0.02 & -1.32$\pm$0.14 & -0.79$(-0.14,+0.12)$ \\
2MASS J16215466-2043091 & K7    &  Debris/Ev. Trans.   &   0.7$\pm$0.5 & 3.61$\pm$0.01 & -0.35$\pm$0.15 & -0.10$(-0.04,+0.04)$ \\
2MASS J16220961-1953005 & M3.75 &  Debris/Ev. Trans.   &   0.7$\pm$0.5 & 3.52$\pm$0.02 & -0.50$\pm$0.15 & -0.56$(-0.08,+0.08)$ \\
2MASS J16230783-2300596 & K3.5  &  Debris/Ev. Trans.   &   0.7$\pm$0.5 & 3.66$\pm$0.01 & 0.09$\pm$0.15 & 0.12$(-0.04,+0.05)$ \\
2MASS J16235385-2946401 & G2.5  &  Debris/Ev. Trans.   &   0.7$\pm$0.5 & 3.77$\pm$0.00 & 0.66$\pm$0.15 & 0.16$(-0.11,+0.10)$ \\
2MASS J16270942-2148457 & M4.5  &  Full 	     &   1.8$\pm$0.38 & 3.50$\pm$0.02 & -1.55$\pm$0.14 & -0.79$(-0.15,+0.12)$ \\
2MASS J16303390-2428062 & M4    &  Full 	     &   0.7$\pm$0.5 & 3.51$\pm$0.02 & -1.11$\pm$0.15 & -0.66$(-0.12,+0.12)$ \\

\enddata
\label{tab:starProp}
\end{deluxetable}

\begin{deluxetable}{lcccccc}
\tabletypesize{\footnotesize}
\tablecolumns{7} 
\tablewidth{0pt} 
\tablecaption{Observations}
\tablehead{
  \colhead{UT Date}   &
  \colhead{Number}  &
  \colhead{Baseline Range}  &
  \colhead{pwv}  &
  \multicolumn{3}{c}{Calibrators}  \\
  \cline{5-7}
  \colhead{} & 
  \colhead{Antennas} & 
  \colhead{(m)} & 
  \colhead{(mm)} & 
  \colhead{Flux} & 
  \colhead{Passband} & 
  \colhead{Gain} 
}
\startdata
2012 Aug 24 & 25      &     17-375     & 0.77 & Neptune           & J1924-0939 & J1625-2527\\
2012 Aug 28 & 28      &     12-386     & 0.68 & Titan             & J1924-0939 & J1625-2527\\
2012 Dec 16 & 17      &     16-402     & 1.16 & Titan             & J1924-0939 & J1625-2527\\
2014 Jun 15 & 34      &     16-650     & 0.78 & Titan, J1733-130  & J1517-2422 & J1517-2422\\
2014 Jun 16 & 36      &     16-650     & 0.56 & Titan             & J1517-2422 & J1517-2422\\
2014 Jun 30 & 36      &     16-650     & 0.52 & Titan             & J1517-2422 & J1517-2422\\
2014 Jul 07 & 36      &     19-650     & 0.60 & Titan             & J1517-2422 & J1517-2422\\
\enddata
\label{tab:obs}
\end{deluxetable}

\begin{deluxetable}{lccccc}
\tabletypesize{\footnotesize}
\tablecolumns{6} 
\tablewidth{0pt} 
\tablecaption{Secondary Source Properties}
\tablehead{
  \colhead{Field}   &
  \multicolumn{2}{c}{Secondary Source Position (J2000)} &
  \colhead{$S_{tot}$} & 
  \colhead{$\Delta\alpha$} & 
  \colhead{$\Delta\delta$} \\
  \cline{2-3}
  \colhead{} & 
  \colhead{Right Ascension} & 
  \colhead{Declination} & 
  \colhead{(mJy)} & 
  \colhead{(arcsec)} & 
  \colhead{(arcsec)} 
}
\startdata
2MASS J15584772-1757595  & 15$^\mathrm{h}$58$^\mathrm{m}$47$\overset{^\mathrm{s}}{.}$49 & -17$^{\circ}$57$'$59$\overset{''}{.}$11 & 1.33  $\pm$  0.15    & -3.19  $\pm$  0.14    & 0.81 $\pm$ 0.17  \\
2MASS J16020287-2236139  & 16$^\mathrm{h}$02$^\mathrm{m}$03$\overset{^\mathrm{s}}{.}$15 & -22$^{\circ}$36$'$11$\overset{''}{.}$75 & 2.19  $\pm$  0.15    & 4.02  $\pm$  0.13    & 2.67 $\pm$ 0.14  \\
2MASS J16025123-2401574  & 16$^\mathrm{h}$02$^\mathrm{m}$51$\overset{^\mathrm{s}}{.}$50 & -24$^{\circ}$01$'$54$\overset{''}{.}$04 & 1.35  $\pm$  0.15    & 3.87  $\pm$  0.13    & 3.78 $\pm$ 0.17  \\
2MASS J16032225-2413111  & 16$^\mathrm{h}$03$^\mathrm{m}$21$\overset{^\mathrm{s}}{.}$75 & -24$^{\circ}$13$'$11$\overset{''}{.}$71 & 1.54  $\pm$  0.15    & -6.62  $\pm$  0.13    & -0.15 $\pm$ 0.14  \\
2MASS J16032225-2413111  & 16$^\mathrm{h}$03$^\mathrm{m}$22$\overset{^\mathrm{s}}{.}$30 & -24$^{\circ}$13$'$11$\overset{''}{.}$46 & 0.86  $\pm$  0.15    & 0.84  $\pm$  0.13    & 0.10 $\pm$ 0.14  \\
2MASS J16071971-2020555  & 16$^\mathrm{h}$07$^\mathrm{m}$19$\overset{^\mathrm{s}}{.}$42 & -20$^{\circ}$20$'$57$\overset{''}{.}$99 & 0.84  $\pm$  0.16    & -4.12  $\pm$  0.13    & -2.13 $\pm$ 0.14 \\
2MASS J16113134-1838259\tablenotemark{a}  & 16$^\mathrm{h}$11$^\mathrm{m}$31$\overset{^\mathrm{s}}{.}$30 & -18$^{\circ}$38$'$27$\overset{''}{.}$26 & 76.95 $\pm$  0.31    & -0.42  $\pm$  0.12    & -0.88 $\pm$ 0.13  \\
2MASS J16123916-1859284  & 16$^\mathrm{h}$12$^\mathrm{m}$39$\overset{^\mathrm{s}}{.}$21 & -18$^{\circ}$59$'$28$\overset{''}{.}$98 & 1.09  $\pm$  0.16    & 0.63  $\pm$  0.14    & -0.21 $\pm$ 0.15  \\
2MASS J16125533-2319456  & 16$^\mathrm{h}$12$^\mathrm{m}$54$\overset{^\mathrm{s}}{.}$97 & -23$^{\circ}$19$'$36$\overset{''}{.}$97 & 0.94  $\pm$  0.13    & -4.87  $\pm$  0.12    & 9.02 $\pm$ 0.12  \\
2MASS J16135434-2320342\tablenotemark{a}  & 16$^\mathrm{h}$13$^\mathrm{m}$54$\overset{^\mathrm{s}}{.}$36 & -23$^{\circ}$20$'$34$\overset{''}{.}$76 & 5.82  $\pm$  0.13    & 0.41  $\pm$  0.13    & -0.13 $\pm$ 0.14  \\
\tablenotetext{a}{Secondary source also detected in CO at the same velocity as the primary source.}
\enddata
\label{tab:sec_fluxes}
\end{deluxetable}

\begin{deluxetable}{@{\extracolsep{4pt}}lccccccc@{}}
\tabletypesize{\tiny}
\tablecolumns{8} 
\tablewidth{0pt} 
\tablecaption{Continuum and CO $J$ = 3$-$2 Flux Measurements}
\tablehead{
  \colhead{Source}   &
  \multicolumn{4}{c}{0.88 mm Continuum} & 
  \multicolumn{3}{c}{CO $J$ = 3$-$2} \\

  \cline{2-5}
  \cline{6-8}
  \colhead{} &
  \colhead{$S_{\nu}$}  &
  \colhead{$\Delta\alpha$\tablenotemark{a}}  &
  \colhead{$\Delta\delta$\tablenotemark{a}}  &
  \colhead{FWHM\tablenotemark{b}} & 
  \colhead{Flux} & 
  \colhead{Velocity Range} & 
  \colhead{Aperture Radius} \\

  \colhead{} & 
  \colhead{(mJy)} & 
  \colhead{(arcsec)} & 
  \colhead{(arcsec)} & 
  \colhead{(arcsec)} & 
  \colhead{(mJy km s$^{-1}$)} & 
  \colhead{(km s$^{-1}$)} & 
  \colhead{(arcsec)}
}
\startdata
2MASS J15354856-2958551 & 1.92  $\pm$  0.15    & -0.40  $\pm$  0.14    & -0.04 $\pm$ 0.15   &  ...                  &  55 $\pm$ 34  &  -1.5 $-$ 10.5  &  0.3\\
2MASS J15514032-2146103 & 0.76  $\pm$  0.16    & 0.01  $\pm$  0.14     & 0.06 $\pm$ 0.16    &  ...                  &  87 $\pm$ 38  &  -1.5 $-$ 10.5  &  0.3\\
2MASS J15521088-2125372 & -0.10  $\pm$  0.15   &   ...  	       &  ...		    &  ...                  &  285 $\pm$ 45  &  -2.5 $-$ 7.5  &  0.3\\
2MASS J15530132-2114135 & 5.78  $\pm$  0.14    & -0.15  $\pm$  0.13    & 0.02 $\pm$ 0.14    &  ...                  &  160 $\pm$ 28  &  -1.5 $-$ 10.5  &  0.3\\
2MASS J15534211-2049282 & 2.93  $\pm$  0.29    & -0.52  $\pm$  0.14    & -0.03 $\pm$ 0.15   &  0.478 $\pm$ 0.068    &  511 $\pm$ 59  &  0.0 $-$ 17.0  &  0.4\\
2MASS J15551704-2322165 & 0.11  $\pm$  0.15    &   ...  	       &  ...		    &  ... 		    &  5 $\pm$ 37  &  -1.5 $-$ 10.5  &  0.3\\
2MASS J15554883-2512240 & -0.14  $\pm$  0.15   &   ...  	       &  ...		    &  ... 		    &  -14 $\pm$ 44  &  -1.5 $-$ 10.5  &  0.3\\
2MASS J15562477-2225552 & 0.28  $\pm$  0.18    &   ...  	       &  ...		    &  ... 		    &  133 $\pm$ 19  &  -1.5 $-$ 10.5  &  0.3\\
2MASS J15570641-2206060 & 0.32  $\pm$  0.20    &   ...  	       &  ...		    &  ... 		    &  -9 $\pm$ 23  &  -1.5 $-$ 10.5  &  0.3\\
2MASS J15572986-2258438 & -0.04  $\pm$  0.20   &   ...  	       &  ...		    &  ... 		    &  56 $\pm$ 36  &  -1.5 $-$ 10.5  &  0.3\\
2MASS J15581270-2328364 & 0.00  $\pm$  0.15    &   ...  	       &  ...		    &  ... 		    &  30 $\pm$ 37  &  -1.5 $-$ 10.5  &  0.3\\
2MASS J15582981-2310077 & 5.86  $\pm$  0.18    & 0.10  $\pm$  0.11     & -0.01 $\pm$ 0.11   &  ... 		    &  56 $\pm$ 23  &  -1.5 $-$ 10.5  &  0.3\\
2MASS J15583692-2257153\tablenotemark{c} & 174.92  $\pm$  0.27  & -0.12  $\pm$  0.11    & 0.06 $\pm$ 0.12  &  ...   &  4607 $\pm$ 75  &  -1.0 $-$ 14.0  &  1.0\\
2MASS J15584772-1757595 & -0.20  $\pm$  0.15   &   ...  	       &  ...		    &  ... 		    &  -75 $\pm$ 30  &  -1.5 $-$ 10.5  &  0.3\\
2MASS J16001330-2418106 & 0.05  $\pm$  0.15    &   ...  	       &  ...		    &  ... 		    &  -32 $\pm$ 40  &  -1.5 $-$ 10.5  &  0.3\\
2MASS J16001730-2236504 & 0.10  $\pm$  0.15    &   ...  	       &  ...		    &  ... 		    &  -35 $\pm$ 33  &  -1.5 $-$ 10.5  &  0.3\\
2MASS J16001844-2230114 & 3.89  $\pm$  0.15    & -0.14  $\pm$  0.13    & 0.08 $\pm$ 0.13    &  ... 		    &  1835 $\pm$ 69  &  3.5 $-$ 24.0  &  0.6\\
2MASS J16014086-2258103 & 3.45  $\pm$  0.14    & -0.03  $\pm$  0.14    & -0.24 $\pm$ 0.15   &  ... 		    &  507 $\pm$ 39  &  -5.0 $-$ 8.5  &  0.4\\
2MASS J16014157-2111380 & 0.66  $\pm$  0.14    & -0.01  $\pm$  0.14    & 0.01 $\pm$ 0.14    &  ... 		    &  9 $\pm$ 35  &  -1.5 $-$ 10.5  &  0.3\\
2MASS J16020039-2221237 & -0.08  $\pm$  0.14   &   ...  	       &  ...		    &  ... 		    &  60 $\pm$ 27  &  -1.5 $-$ 10.5  &  0.3\\
2MASS J16020287-2236139 & 0.04  $\pm$  0.15    &   ...  	       &  ...		    &  ... 		    &  -30 $\pm$ 32  &  -1.5 $-$ 10.5  &  0.3\\
2MASS J16020757-2257467 & 5.26  $\pm$  0.27    & 0.12  $\pm$  0.14     & -0.06 $\pm$ 0.15   &  0.257 $\pm$ 0.029    &  632 $\pm$ 63  &  -2.0 $-$ 10.0  &  0.6\\
2MASS J16024152-2138245 & 10.25  $\pm$  0.19   & -0.03  $\pm$  0.13    & -0.06 $\pm$ 0.14   &  0.142 $\pm$ 0.011    &  40 $\pm$ 26  &  -1.5 $-$ 10.5  &  0.3\\
2MASS J16025123-2401574 & 0.07  $\pm$  0.15    & ...		       & ...		    &  ... 		    &  -24 $\pm$ 30  &  -1.5 $-$ 10.5  &  0.3\\
2MASS J16030161-2207523 & 2.81  $\pm$  0.12    & -0.03  $\pm$  0.14    & -0.08 $\pm$ 0.15   &  ... 		    &  55 $\pm$ 25  &  -1.5 $-$ 10.5  &  0.3\\
2MASS J16031329-2112569 & 0.06  $\pm$  0.12    &   ...  	       &  ...		    &  ... 		    &  -12 $\pm$ 25  &  -1.5 $-$ 10.5  &  0.3\\
2MASS J16032225-2413111 & 2.42  $\pm$  0.15    & 0.03  $\pm$  0.13     & 0.04 $\pm$ 0.14    &  ... 		    &  40 $\pm$ 17  &  -1.5 $-$ 10.5  &  0.3\\
2MASS J16035767-2031055 & 4.30  $\pm$  0.39    & 0.01  $\pm$  0.08     & 0.06 $\pm$ 0.08    &  ... 		    &  180 $\pm$ 26  &  -1.5 $-$ 10.5  &  0.3\\
2MASS J16035793-1942108 & 1.17  $\pm$  0.14    & 0.02  $\pm$  0.13     & -0.05 $\pm$ 0.14   &  ... 		    &  1490 $\pm$ 158  &  -1.0 $-$ 15.5  &  0.9\\
2MASS J16041740-1942287 & 0.89  $\pm$  0.14    & 0.09  $\pm$  0.14     & 0.03 $\pm$ 0.15    &  ... 		    &  67 $\pm$ 44  &  -1.5 $-$ 10.5  &  0.3\\
2MASS J16042165-2130284\tablenotemark{c} & 218.76  $\pm$  0.81  & 0.01  $\pm$  0.11     & -0.03 $\pm$ 0.11  &  ...  &  20268 $\pm$ 67  &  2.5 $-$ 6.0  &  2.1\\
2MASS J16043916-1942459 & 0.49  $\pm$  0.15    & -0.03  $\pm$  0.15    & 0.08 $\pm$ 0.15    &  ... 		    &  -31 $\pm$ 37  &  -1.5 $-$ 10.5  &  0.3\\
2MASS J16050231-1941554 & -0.16  $\pm$  0.15   &   ...  	       &  ...		    &  ... 		    &  -14 $\pm$ 41  &  -1.5 $-$ 10.5  &  0.3\\
2MASS J16052459-1954419 & 0.22  $\pm$  0.15    &   ...  	       &  ...		    &  ... 		    &  -43 $\pm$ 34  &  -1.5 $-$ 10.5  &  0.3\\
2MASS J16052556-2035397 & 1.53  $\pm$  0.20    & -0.09  $\pm$  0.19    & 0.52 $\pm$ 0.19    &  ... 		    &  8 $\pm$ 31  &  -1.5 $-$ 10.5  &  0.3\\
2MASS J16052661-1957050 & 0.07  $\pm$  0.15    &   ...  	       &  ...		    &  ... 		    &  111 $\pm$ 37  &  -1.5 $-$ 10.5  &  0.3\\
2MASS J16053215-1933159 & 0.25  $\pm$  0.20    &   ...  	       &  ...		    &  ... 		    &  2 $\pm$ 25  &  -1.5 $-$ 10.5  &  0.3\\
2MASS J16054540-2023088 & 7.64  $\pm$  0.15    & 0.09  $\pm$  0.13     & -0.02 $\pm$ 0.13   &  ... 		    &  101 $\pm$ 39  &  -1.5 $-$ 10.5  &  0.3\\
2MASS J16055863-1949029 & -0.08  $\pm$  0.15   &   ...  	       &  ...		    &  ... 		    &  -59 $\pm$ 37  &  -1.5 $-$ 10.5  &  0.3\\
2MASS J16060061-1957114 & 0.00  $\pm$  0.13    &   ...  	       &  ...		    &  ... 		    &  3 $\pm$ 31  &  -1.5 $-$ 10.5  &  0.3\\
2MASS J16061330-2212537 & -0.20  $\pm$  0.12   &   ...  	       &  ...		    &  ... 		    &  -13 $\pm$ 31  &  -1.5 $-$ 10.5  &  0.3\\
2MASS J16062196-1928445 & 4.08  $\pm$  0.52    & 0.02  $\pm$  0.22     & 0.50 $\pm$ 0.22    &  ... 		    &  23 $\pm$ 50  &  -1.5 $-$ 10.5  &  0.3\\
2MASS J16062277-2011243 & 0.59  $\pm$  0.14    & 0.09  $\pm$  0.19     & 0.05 $\pm$ 0.19    &  ... 		    &  151 $\pm$ 27  &  2.0 $-$ 11.5  &  0.4\\
2MASS J16063539-2516510 & 1.69  $\pm$  0.15    & 0.04  $\pm$  0.13     & 0.00 $\pm$ 0.14    &  ... 		    &  48 $\pm$ 31  &  -1.5 $-$ 10.5  &  0.3\\
2MASS J16064102-2455489 & 3.05  $\pm$  0.14    & -0.15  $\pm$  0.13    & -0.06 $\pm$ 0.14   &  ... 		    &  14 $\pm$ 31  &  -1.5 $-$ 10.5  &  0.3\\
2MASS J16064115-2517044 & 0.20  $\pm$  0.15    &   ...  	       &  ...		    &  ... 		    &  -46 $\pm$ 23  &  -1.5 $-$ 10.5  &  0.3\\
2MASS J16064385-1908056 & 0.84  $\pm$  0.15    & -0.04  $\pm$  0.15    & -0.15 $\pm$ 0.15   &  ... 		    &  60 $\pm$ 29  &  -1.5 $-$ 10.5  &  0.3\\
2MASS J16070014-2033092 & 0.22  $\pm$  0.15    &   ...  	       &  ...		    &  ... 		    &  16 $\pm$ 44  &  -1.5 $-$ 10.5  &  0.3\\
2MASS J16070211-2019387 & -0.09  $\pm$  0.20   &   ...  	       &  ...		    &  ... 		    &  45 $\pm$ 24  &  -1.5 $-$ 10.5  &  0.3\\
2MASS J16070873-1927341 & -0.09  $\pm$  0.15   &   ...  	       &  ...		    &  ... 		    &  53 $\pm$ 45  &  -1.5 $-$ 10.5  &  0.3\\
2MASS J16071971-2020555 & 0.16  $\pm$  0.16    &   ...  	       &  ...		    &  ... 		    &  18 $\pm$ 36  &  -1.5 $-$ 10.5  &  0.3\\
2MASS J16072625-2432079 & 13.12  $\pm$  0.24   & -0.03  $\pm$  0.14    & 0.12 $\pm$ 0.15    &  0.140 $\pm$ 0.013    &  171 $\pm$ 49  &  -1.5 $-$ 10.5  &  0.3\\
2MASS J16072747-2059442 & 2.13  $\pm$  0.12    & -0.21  $\pm$  0.13    & 0.13 $\pm$ 0.13    &  ... 		    &  34 $\pm$ 48  &  -1.5 $-$ 10.5  &  0.3\\
2MASS J16073939-1917472 & 0.58  $\pm$  0.16    & -0.32  $\pm$  0.15    & -0.35 $\pm$ 0.15   &  ... 		    &  -18 $\pm$ 42  &  -1.5 $-$ 10.5  &  0.3\\
2MASS J16075796-2040087 & 23.49  $\pm$  0.12   & -0.07  $\pm$  0.13    & 0.16 $\pm$ 0.14    &  ... 		    &  3258 $\pm$ 73  &  -17.0 $-$ 17.0  &  0.6\\
2MASS J16080555-2218070 & 0.02  $\pm$  0.12    &   ...  	       &  ...		    &  ... 		    &  17 $\pm$ 33  &  -1.5 $-$ 10.5  &  0.3\\
2MASS J16081566-2222199 & 0.97  $\pm$  0.12    & 0.09  $\pm$  0.14     & -0.01 $\pm$ 0.15   &  ... 		    &  191 $\pm$ 31  &  -1.5 $-$ 10.5  &  0.3\\
2MASS J16082324-1930009 & 43.19  $\pm$  0.81   & 0.21  $\pm$  0.20     & 0.29 $\pm$ 0.21    &  0.400 $\pm$ 0.015    &  246 $\pm$ 42  &  -1.5 $-$ 10.5  &  0.3\\
2MASS J16082751-1949047 & 0.76  $\pm$  0.13    & 0.01  $\pm$  0.15     & -0.03 $\pm$ 0.15   &  ... 		    &  21 $\pm$ 35  &  -1.5 $-$ 10.5  &  0.3\\
2MASS J16083455-2211559 & 0.01  $\pm$  0.12    &   ...  	       &  ...		    &  ... 		    &  23 $\pm$ 28  &  -1.5 $-$ 10.5  &  0.3\\
2MASS J16084894-2400045 & -0.06  $\pm$  0.15   &   ...  	       &  ...		    &  ... 		    &  -8 $\pm$ 23  &  -1.5 $-$ 10.5  &  0.3\\
2MASS J16090002-1908368 & 1.73  $\pm$  0.13    & 0.04  $\pm$  0.12     & 0.09 $\pm$ 0.12    &  ... 		    &  35 $\pm$ 16  &  -1.5 $-$ 10.5  &  0.3\\
2MASS J16090075-1908526 & 47.28  $\pm$  0.91   & 0.42  $\pm$  0.20     & -0.27 $\pm$ 0.21   &  0.315 $\pm$ 0.018    &  815 $\pm$ 64  &  -0.5 $-$ 15.5  &  0.5\\
2MASS J16093558-1828232 & 0.69  $\pm$  0.15    & 0.08  $\pm$  0.14     & 0.14 $\pm$ 0.14    &  ... 		    &  55 $\pm$ 38  &  -1.5 $-$ 10.5  &  0.3\\
2MASS J16094098-2217594 & 0.44  $\pm$  0.12    & 0.16  $\pm$  0.14     & -0.10 $\pm$ 0.15   &  ... 		    &  -15 $\pm$ 37  &  -1.5 $-$ 10.5  &  0.3\\
2MASS J16095361-1754474 & 0.87  $\pm$  0.16    & -0.12  $\pm$  0.13    & -0.02 $\pm$ 0.17   &  ... 		    &  60 $\pm$ 44  &  -1.5 $-$ 10.5  &  0.3\\
2MASS J16095441-1906551 & 0.50  $\pm$  0.16    & -0.48  $\pm$  0.16    & 0.43 $\pm$ 0.16    &  ... 		    &  56 $\pm$ 34  &  -1.5 $-$ 10.5  &  0.3\\
2MASS J16095933-1800090 & 0.67  $\pm$  0.18    & -0.19  $\pm$  0.26    & -0.13 $\pm$ 0.26   &  ... 		    &  460 $\pm$ 91  &  -0.5 $-$ 10.5  &  0.9\\
2MASS J16101473-1919095 & 0.01  $\pm$  0.16    &   ...  	       &  ...		    &  ... 		    &  -4 $\pm$ 18  &  -1.5 $-$ 10.5  &  0.3\\
2MASS J16101888-2502325 & 0.30  $\pm$  0.14    &   ...  	       &  ...		    &  ... 		    &  63 $\pm$ 30  &  -1.5 $-$ 10.5  &  0.3\\
2MASS J16102174-1904067 & -0.05  $\pm$  0.16   &   ...  	       &  ...		    &  ... 		    &  -7 $\pm$ 32  &  -1.5 $-$ 10.5  &  0.3\\
2MASS J16102819-1910444 & 0.05  $\pm$  0.16    &   ...  	       &  ...		    &  ... 		    &  -18 $\pm$ 30  &  -1.5 $-$ 10.5  &  0.3\\
2MASS J16102857-1904469 & 0.66  $\pm$  0.16    & -0.22  $\pm$  0.15    & -0.30 $\pm$ 0.15   &  ... 		    &  -86 $\pm$ 30  &  -1.5 $-$ 10.5  &  0.3\\
2MASS J16103956-1916524 & 0.07  $\pm$  0.16    &   ...  	       &  ...		    &  ... 		    &  63 $\pm$ 26  &  -1.5 $-$ 10.5  &  0.3\\
2MASS J16104202-2101319 & 0.17  $\pm$  0.12    &   ...  	       &  ...		    &  ... 		    &  20 $\pm$ 19  &  -1.5 $-$ 10.5  &  0.3\\
2MASS J16104636-1840598 & 1.78  $\pm$  0.16    & 0.10  $\pm$  0.14     & 0.03 $\pm$ 0.14    &  ... 		    &  216 $\pm$ 40  &  -1.5 $-$ 10.5  &  0.3\\
2MASS J16111330-2019029 & 4.88  $\pm$  0.16    & 0.03  $\pm$  0.14     & -0.08 $\pm$ 0.14   &  ... 		    &  59 $\pm$ 29  &  -1.5 $-$ 10.5  &  0.3\\
2MASS J16111534-1757214 & 0.18  $\pm$  0.16    &   ...  	       &  ...		    &  ... 		    &  97 $\pm$ 39  &  -1.5 $-$ 10.5  &  0.3\\
2MASS J16112057-1820549 & -0.06  $\pm$  0.16   &   ...  	       &  ...		    &  ... 		    &  -2 $\pm$ 33  &  -1.5 $-$ 10.5  &  0.3\\
2MASS J16113134-1838259 & 903.56  $\pm$  0.85  & 0.38  $\pm$  0.12     & 0.17 $\pm$ 0.13    &  0.401 $\pm$ 0.001    &  22748 $\pm$ 91  &  -1.0 $-$ 11.5  &  0.8\\
2MASS J16115091-2012098 & 0.66  $\pm$  0.16    & 0.15  $\pm$  0.14     & -0.01 $\pm$ 0.14   &  ... 		    &  235 $\pm$ 45  &  -1.5 $-$ 10.5  &  0.3\\
2MASS J16122737-2009596 & 0.53  $\pm$  0.16    & -0.09  $\pm$  0.16    & -0.15 $\pm$ 0.17   &  ... 		    &  55 $\pm$ 38  &  -1.5 $-$ 10.5  &  0.3\\
2MASS J16123916-1859284 & 6.01  $\pm$  0.29    & -0.12  $\pm$  0.14    & -0.06 $\pm$ 0.14   &  ... 		    &  1554 $\pm$ 125  &  -1.5 $-$ 8.5  &  1.3\\
2MASS J16124893-1800525 & 0.11  $\pm$  0.16    &   ...  	       &  ...		    &  ... 		    &  24 $\pm$ 31  &  -1.5 $-$ 10.5  &  0.3\\
2MASS J16125533-2319456 & 0.08  $\pm$  0.13    &   ...  	       &  ...		    &  ... 		    &  31 $\pm$ 25  &  -1.5 $-$ 10.5  &  0.3\\
2MASS J16130996-1904269 & -0.05  $\pm$  0.16   &   ...  	       &  ...		    &  ... 		    &  60 $\pm$ 31  &  -1.5 $-$ 10.5  &  0.3\\
2MASS J16133650-2503473 & 0.88  $\pm$  0.19    & 0.17  $\pm$  0.14     & 0.02 $\pm$ 0.14    &  ... 		    &  21 $\pm$ 41  &  -1.5 $-$ 10.5  &  0.3\\
2MASS J16135434-2320342 & 7.53  $\pm$  0.13    & -0.17  $\pm$  0.13    & 0.06 $\pm$ 0.14    &  ... 		    &  110 $\pm$ 29  &  -1.5 $-$ 10.5  &  0.3\\
2MASS J16141107-2305362 & 4.77  $\pm$  0.14    & 0.09  $\pm$  0.04     & -0.07 $\pm$ 0.04   &  ... 		    &  -14 $\pm$ 18  &  -1.5 $-$ 10.5  &  0.3\\
2MASS J16142029-1906481 & 40.69  $\pm$  0.22   & -0.12  $\pm$  0.20    & 0.11 $\pm$ 0.20    &  0.169 $\pm$ 0.005    &  4681 $\pm$ 118  &  -17.0 $-$ 15.0  &  1.0\\
2MASS J16142893-1857224 & 0.10  $\pm$  0.16    &   ...  	       &  ...		    &  ... 		    &  14 $\pm$ 29  &  -1.5 $-$ 10.5  &  0.3\\
2MASS J16143367-1900133 & 1.24  $\pm$  0.16    & -0.16  $\pm$  0.14    & -0.22 $\pm$ 0.14   &  ... 		    &  339 $\pm$ 49  &  -3.0 $-$ 8.5  &  0.3\\
2MASS J16145918-2750230 & 0.03  $\pm$  0.19    &   ...  	       &  ...		    &  ... 		    &  -53 $\pm$ 33  &  -1.5 $-$ 10.5  &  0.3\\
2MASS J16145928-2459308 & -0.03  $\pm$  0.12   &   ...  	       &  ...		    &  ... 		    &  110 $\pm$ 29  &  -1.5 $-$ 10.5  &  0.3\\
2MASS J16151239-2420091 & 0.22  $\pm$  0.12    &   ...  	       &  ...		    &  ... 		    &  -8 $\pm$ 25  &  -1.5 $-$ 10.5  &  0.3\\
2MASS J16153456-2242421 & 11.75  $\pm$  0.12   & 0.26  $\pm$  0.14     & -0.55 $\pm$ 0.15   &  ... 		    &  139 $\pm$ 36  &  -1.5 $-$ 10.5  &  0.3\\
2MASS J16154416-1921171 & 23.57  $\pm$  0.16   & 0.14  $\pm$  0.14     & -0.17 $\pm$ 0.14   &  ... 		    &  14147 $\pm$ 138  &  -3.0 $-$ 11.5  &  1.5\\
2MASS J16163345-2521505 & 2.88  $\pm$  0.30    & 0.00  $\pm$  0.13     & 0.01 $\pm$ 0.14    &  0.492 $\pm$ 0.067    &  164 $\pm$ 30  &  -1.5 $-$ 10.5  &  0.3\\
2MASS J16181618-2619080 & -0.07  $\pm$  0.12   &   ...  	       &  ...		    &  ... 		    &  82 $\pm$ 29  &  -1.5 $-$ 10.5  &  0.3\\
2MASS J16181904-2028479 & 4.62  $\pm$  0.12    & 0.11  $\pm$  0.13     & 0.19 $\pm$ 0.13    &  ... 		    &  177 $\pm$ 31  &  -1.5 $-$ 10.5  &  0.3\\
2MASS J16215466-2043091 & 0.49  $\pm$  0.12    & 0.10  $\pm$  0.14     & 0.25 $\pm$ 0.22    &  ... 		    &  -56 $\pm$ 31  &  -1.5 $-$ 10.5  &  0.3\\
2MASS J16220961-1953005 & 0.07  $\pm$  0.16    &   ...  	       &  ...		    &  ... 		    &  15 $\pm$ 45  &  -1.5 $-$ 10.5  &  0.3\\
2MASS J16230783-2300596 & -0.35  $\pm$  0.12   &   ...  	       &  ...		    &  ... 		    &  75 $\pm$ 32  &  -1.5 $-$ 10.5  &  0.3\\
2MASS J16235385-2946401 & 0.11  $\pm$  0.12    &   ...  	       &  ...		    &  ... 		    &  -24 $\pm$ 28  &  -1.5 $-$ 10.5  &  0.3\\
2MASS J16270942-2148457 & 2.87  $\pm$  0.12    & -0.02  $\pm$  0.14    & 0.08 $\pm$ 0.16    &  ... 		    &  109 $\pm$ 32  &  -1.5 $-$ 10.5  &  0.3\\
2MASS J16303390-2428062 & 0.60  $\pm$  0.12    & 0.07  $\pm$  0.13     & -0.02 $\pm$ 0.14   &  ... 		    &  6 $\pm$ 31  &  -1.5 $-$ 10.5  &  0.3\\

\enddata
\tablenotetext{a}{Offsets of the continuum source from the expected stellar position.  Ellipses indicate a non-detection, for which 
the fit position is held fixed at the expected stellar position.}
\tablenotetext{b}{Full width at half maximum for sources fitted with an elliptical Gaussian. Ellipses indicate point sources and sources measured with 
aperture photometry.}
\tablenotetext{c}{Continuum flux density measured using aperture photometry.}
\label{tab:fluxes}
\end{deluxetable}

\begin{deluxetable}{lc}
\tabletypesize{\footnotesize}
\tablecolumns{2} 
\tablewidth{0pt} 
\tablecaption{Derived Dust Masses}
\tablehead{
  \colhead{Source}   &
  \colhead{$M_{\mathrm{dust}}/M_{\oplus}$}
}
\startdata
2MASS J15354856-2958551 & $0.62\pm0.16$ \\
2MASS J15514032-2146103 & $0.49\pm0.15$ \\
2MASS J15521088-2125372 & $<0.52$ \\
2MASS J15530132-2114135 & $3.34\pm0.83$ \\
2MASS J15534211-2049282 & $1.18\pm0.31$ \\
2MASS J15551704-2322165 & $<0.17$ \\
2MASS J15554883-2512240 & $<0.07$ \\
2MASS J15562477-2225552 & $<0.46$ \\
2MASS J15570641-2206060 & $<0.69$ \\
2MASS J15572986-2258438 & $<0.40$ \\
2MASS J15581270-2328364 & $<0.07$ \\
2MASS J15582981-2310077 & $3.77\pm0.94$ \\
2MASS J15583692-2257153 & $24.30\pm5.99$ \\
2MASS J15584772-1757595 & $<0.09$ \\
2MASS J16001330-2418106 & $<0.16$ \\
2MASS J16001730-2236504 & $<0.22$ \\
2MASS J16001844-2230114 & $2.08\pm0.52$ \\
2MASS J16014086-2258103 & $1.48\pm0.37$ \\
2MASS J16014157-2111380 & $0.56\pm0.17$ \\
2MASS J16020039-2221237 & $<0.11$ \\
2MASS J16020287-2236139 & $<0.35$ \\
2MASS J16020757-2257467 & $2.08\pm0.52$ \\
2MASS J16024152-2138245 & $7.63\pm1.89$ \\
2MASS J16025123-2401574 & $<0.12$ \\
2MASS J16030161-2207523 & $2.48\pm0.62$ \\
2MASS J16031329-2112569 & $<0.29$ \\
2MASS J16032225-2413111 & $1.10\pm0.28$ \\
2MASS J16035767-2031055 & $0.98\pm0.25$ \\
2MASS J16035793-1942108 & $0.53\pm0.14$ \\
2MASS J16041740-1942287 & $0.45\pm0.13$ \\
2MASS J16042165-2130284 & $52.29\pm12.90$ \\
2MASS J16043916-1942459 & $0.27\pm0.10$ \\
2MASS J16050231-1941554 & $<0.39$ \\
2MASS J16052459-1954419 & $<0.34$ \\
2MASS J16052556-2035397 & $1.05\pm0.28$ \\
2MASS J16052661-1957050 & $<0.28$ \\
2MASS J16053215-1933159 & $<0.75$ \\
2MASS J16054540-2023088 & $3.27\pm0.81$ \\
2MASS J16055863-1949029 & $<0.26$ \\
2MASS J16060061-1957114 & $<0.23$ \\
2MASS J16061330-2212537 & $<0.12$ \\
2MASS J16062196-1928445 & $0.99\pm0.27$ \\
2MASS J16062277-2011243 & $0.43\pm0.14$ \\
2MASS J16063539-2516510 & $1.51\pm0.39$ \\
2MASS J16064102-2455489 & $3.06\pm0.76$ \\
2MASS J16064115-2517044 & $<0.38$ \\
2MASS J16064385-1908056 & $0.23\pm0.07$ \\
2MASS J16070014-2033092 & $<0.30$ \\
2MASS J16070211-2019387 & $<0.49$ \\
2MASS J16070873-1927341 & $<0.28$ \\
2MASS J16071971-2020555 & $<0.32$ \\
2MASS J16072625-2432079 & $5.71\pm1.41$ \\
2MASS J16072747-2059442 & $0.99\pm0.25$ \\
2MASS J16073939-1917472 & $0.22\pm0.07$ \\
2MASS J16075796-2040087 & $9.31\pm2.30$ \\
2MASS J16080555-2218070 & $<0.15$ \\
2MASS J16081566-2222199 & $0.39\pm0.11$ \\
2MASS J16082324-1930009 & $13.94\pm3.45$ \\
2MASS J16082751-1949047 & $0.42\pm0.12$ \\
2MASS J16083455-2211559 & $<0.28$ \\
2MASS J16084894-2400045 & $<0.27$ \\
2MASS J16090002-1908368 & $1.15\pm0.29$ \\
2MASS J16090075-1908526 & $13.50\pm3.34$ \\
2MASS J16093558-1828232 & $0.34\pm0.11$ \\
2MASS J16094098-2217594 & $0.10\pm0.03$ \\
2MASS J16095361-1754474 & $0.58\pm0.17$ \\
2MASS J16095441-1906551 & $0.17\pm0.06$ \\
2MASS J16095933-1800090 & $0.32\pm0.11$ \\
2MASS J16101473-1919095 & $<0.20$ \\
2MASS J16101888-2502325 & $<0.49$ \\
2MASS J16102174-1904067 & $<0.17$ \\
2MASS J16102819-1910444 & $<0.48$ \\
2MASS J16102857-1904469 & $0.17\pm0.06$ \\
2MASS J16103956-1916524 & $<0.24$ \\
2MASS J16104202-2101319 & $<0.12$ \\
2MASS J16104636-1840598 & $1.53\pm0.39$ \\
2MASS J16111330-2019029 & $1.83\pm0.45$ \\
2MASS J16111534-1757214 & $<0.19$ \\
2MASS J16112057-1820549 & $<0.10$ \\
2MASS J16113134-1838259 & $127.28\pm31.39$ \\
2MASS J16115091-2012098 & $0.32\pm0.10$ \\
2MASS J16122737-2009596 & $0.39\pm0.14$ \\
2MASS J16123916-1859284 & $1.79\pm0.45$ \\
2MASS J16124893-1800525 & $<0.27$ \\
2MASS J16125533-2319456 & $<0.05$ \\
2MASS J16130996-1904269 & $<0.25$ \\
2MASS J16133650-2503473 & $0.41\pm0.13$ \\
2MASS J16135434-2320342 & $3.81\pm0.94$ \\
2MASS J16141107-2305362 & $0.68\pm0.17$ \\
2MASS J16142029-1906481 & $10.52\pm2.59$ \\
2MASS J16142893-1857224 & $<0.19$ \\
2MASS J16143367-1900133 & $0.36\pm0.10$ \\
2MASS J16145918-2750230 & $<0.11$ \\
2MASS J16145928-2459308 & $<0.16$ \\
2MASS J16151239-2420091 & $<0.53$ \\
2MASS J16153456-2242421 & $2.57\pm0.63$ \\
2MASS J16154416-1921171 & $5.99\pm1.48$ \\
2MASS J16163345-2521505 & $1.15\pm0.30$ \\
2MASS J16181618-2619080 & $<0.22$ \\
2MASS J16181904-2028479 & $3.02\pm0.75$ \\
2MASS J16215466-2043091 & $0.13\pm0.04$ \\
2MASS J16220961-1953005 & $<0.16$ \\
2MASS J16230783-2300596 & $<0.07$ \\
2MASS J16235385-2946401 & $<0.06$ \\
2MASS J16270942-2148457 & $2.41\pm0.60$ \\
2MASS J16303390-2428062 & $0.32\pm0.09$ \\
\enddata
\label{tab:dust_masses}
\end{deluxetable}


\clearpage
\begin{thebibliography}{}

\bibitem[Andre 
\& Montmerle(1994)]{Andre1994} Andre, P., \& Montmerle, T.\ 1994, \apj, 420, 837 

\bibitem[Andrews 
\& Williams(2005)]{Andrews2005} Andrews, S.~M., \& Williams, J.~P.\ 2005, \apj, 631, 1134 

\bibitem[Andrews 
\& Williams(2007)]{Andrews2007} Andrews, S.~M., \& Williams, J.~P.\ 2007, \apj, 659, 705
 

\bibitem[Andrews et al.(2009)]{Andrews2009} Andrews, S.~M., Wilner, 
D.~J., Hughes, A.~M., Qi, C., \& Dullemond, C.~P.\ 2009, \apj, 700, 1502 

\bibitem[Andrews et al.(2010)]{Andrews2010} Andrews, S.~M., Wilner, D.~J., Hughes, A.~M., 
Qi, C., \& Dullemond, C.~P.\ 2010, \apj, 723, 1241 

\bibitem[Andrews et al.(2013)]{Andrews2013} Andrews, S.~M., 
Rosenfeld, K.~A., Kraus, A.~L., \& Wilner, D.~J.\ 2013, \apj, 771, 129 

\bibitem[Ansdell et al.(2015)]{Ansdell2015} Ansdell, M., Williams, 
J.~P., \& Cieza, L.~A.\ 2015, \apj, 806, 221 

\bibitem[Balog et al.(2007)]{Balog2007} Balog, Z., Muzerolle, J., 
Rieke, G.~H., et al.\ 2007, \apj, 660, 1532 

\bibitem[Baraffe et 
al.(2015)]{Baraffe2015} Baraffe, I., Homeier, D., Allard, F., \& Chabrier, G.\ 2015, \aap, 577, A42 

\bibitem[Barrado y Navascu{\'e}s et al.(2007)]{Barrado2007} Barrado 
y Navascu{\'e}s, D., Stauffer, J.~R., Morales-Calder{\'o}n, M., et al.\ 
2007, \apj, 664, 481 

\bibitem[Beckwith et al.(1990)]{Beckwith1990} Beckwith, S.~V.~W., 
Sargent, A.~I., Chini, R.~S., \& Guesten, R.\ 1990, \aj, 99, 924 

\bibitem[Beckwith \& Sargent(1993)]{Beckwith1993} Beckwith, S.~V.~W., \& Sargent, A.~I.\ 1993, \apj, 402, 280 

\bibitem[Bertout et 
al.(2007)]{Bertout2007} Bertout, C., Siess, L., \& Cabrit, S.\ 2007, \aap, 473, L21 

\bibitem[Bisschop et 
al.(2006)]{Bisschop2006} Bisschop, S.~E., Fraser, H.~J., {\"O}berg, K.~I., van Dishoeck, E.~F., \& Schlemmer, S.\ 2006, \aap, 449, 1297 

\bibitem[Cardelli et al.(1989)]{Cardelli1989} Cardelli, J.~A., 
Clayton, G.~C., \& Mathis, J.~S.\ 1989, \apj, 345, 245 

\bibitem[Carpenter(2002)]{Carpenter2002} Carpenter, J.~M.\ 2002, \aj, 
124, 1593 

\bibitem[Carpenter et al.(2006)]{Carpenter2006} Carpenter, J.~M., 
Mamajek, E.~E., Hillenbrand, L.~A., \& Meyer, M.~R.\ 2006, ApJL, 651, L49 

\bibitem[Carpenter et al.(2014)]{Carpenter2014} Carpenter, J.~M., 
Ricci, L., \& Isella, A.\ 2014, \apj, 787, 42 

\bibitem[Chen et al.(2011)]{Chen2011} Chen, C.~H., Mamajek, 
E.~E., Bitner, M.~A., et al.\ 2011, \apj, 738, 122 

\bibitem[Chapillon et al.(2008)]{Chapillon2008} Chapillon, E., Guilloteau, S., Dutrey, A., \& Pi{\'e}tu, V.\ 2008, \aap, 488, 565 

\bibitem[Chapillon et al.(2010)]{Chapillon2010} Chapillon, E., Parise, B., Guilloteau, S., Dutrey, A., \& Wakelam, V.\ 2010, \aap, 520, A61 

\bibitem[Cieza 
\& Baliber(2007)]{Cieza2007} Cieza, L., \& Baliber, N.\ 2007, \apj, 671, 605 

\bibitem[Collings et al.(2003)]{Collings2003} Collings, M.~P., 
Dever, J.~W., Fraser, H.~J., McCoustra, M.~R.~S., 
\& Williams, D.~A.\ 2003, \apj, 583, 1058 

\bibitem[Cutri et al.(2003)]{Cutri2003} Cutri, R.~M., Skrutskie, M.~F., van Dyk, S., et al.\ 2003, VizieR Online Data Catalog, 2246

\bibitem[Dahm 
\& Hillenbrand(2007)]{Dahm2007} Dahm, S.~E., \& Hillenbrand, L.~A.\ 2007, \aj, 133, 2072 

\bibitem[Dahm 
\& Carpenter(2009)]{Dahm2009} Dahm, S.~E., \& Carpenter, J.~M.\ 2009, \aj, 137, 4024 

\bibitem[Dahm(2010)]{Dahm2010} Dahm, S.~E.\ 2010, \aj, 140, 1444 

\bibitem[Dahm et al.(2012)]{Dahm2012} Dahm, S.~E., Slesnick, 
C.~L., \& White, R.~J.\ 2012, \apj, 745, 56 

\bibitem[David et al.(2015)]{David2015} David, T.~J., 
Hillenbrand, L.~A., Cody, A.~M., Carpenter, J.~M., 
\& Howard, A.~W.\ 2015, arXiv:1510.08087 

\bibitem[de Geus et 
al.(1989)]{deGeus1989} de Geus, E.~J., de Zeeuw, P.~T., \& Lub, J.\ 1989, \aap, 216, 44 

\bibitem[de Zeeuw et al.(1999)]{deZeeuw1999} de Zeeuw, P.~T., 
Hoogerwerf, R., de Bruijne, J.~H.~J., Brown, A.~G.~A., 
\& Blaauw, A.\ 1999, \aj, 117, 354 

\bibitem[The DENIS Consortium(2005)]{DENIS2005} The DENIS Consortium. 2005, yCat, 2263, 0

\bibitem[Draine(2006)]{Draine2006} Draine, B.~T.\ 2006, \apj, 636, 
1114 

\bibitem[Dutrey et al.(1996)]{Dutrey1996} Dutrey, A., Guilloteau, S., Duvert, G., et al.\ 1996, \aap, 309, 493 

\bibitem[Dutrey et al.(2003)]{Dutrey2003} Dutrey, A., Guilloteau, S., \& Simon, M.\ 2003, \aap, 402, 1003 

\bibitem[Eisner et al.(2005)]{Eisner2005} Eisner, J.~A., 
Hillenbrand, L.~A., White, R.~J., Akeson, R.~L., 
\& Sargent, A.~I.\ 2005, \apj, 623, 952 

\bibitem[Fedele et 
al.(2010)]{Fedele2010} Fedele, D., van den Ancker, M.~E., Henning, T., Jayawardhana, R., \& Oliveira, J.~M.\ 2010, \aap, 510, A72 

\bibitem[Feiden et al.(2015)]{Feiden2015} Feiden, G.~A., Jones, 
J., 
\& Chaboyer, B.\ 2015, 18th Cambridge Workshop on Cool Stars, Stellar Systems, and the Sun, 18, 171 

\bibitem[Flaherty 
\& Muzerolle(2008)]{Flaherty2008} Flaherty, K.~M., \& Muzerolle, J.\ 2008, \aj, 135, 966 

\bibitem[Guilloteau et 
al.(2011)]{Guilloteau2011} Guilloteau, S., Dutrey, A., Pi{\'e}tu, V., \& Boehler, Y.\ 2011, \aap, 529, A105 

\bibitem[Gutermuth et al.(2004)]{Gutermuth2004} Gutermuth, R.~A., 
Megeath, S.~T., Muzerolle, J., et al.\ 2004, \apjs, 154, 374 

\bibitem[Gutermuth et al.(2008)]{Gutermuth2008} Gutermuth, R.~A., 
Myers, P.~C., Megeath, S.~T., et al.\ 2008, \apj, 674, 336 

\bibitem[Hardy et al.(2015)]{Hardy2015} Hardy, A., Caceres, C., Schreiber, M.~R., et al.\ 2015, \aap, 583, A66 

\bibitem[Hartmann(2001)]{Hartmann2001} Hartmann, L.\ 2001, \aj, 121, 
1030 

\bibitem[Hartmann et al.(2005)]{Hartmann2005} Hartmann, L., Megeath, 
S.~T., Allen, L., et al.\ 2005, \apj, 629, 881 

\bibitem[Herczeg 
\& Hillenbrand(2015)]{Herczeg2015} Herczeg, G.~J., \& Hillenbrand, L.~A.\ 2015, \apj, 808, 23 

\bibitem[Hern{\'a}ndez et al.(2007a)]{Hernandez2007a} Hern{\'a}ndez, 
J., Hartmann, L., Megeath, T., et al.\ 2007, \apj, 662, 1067

\bibitem[Hern{\'a}ndez et al.(2007b)]{Hernandez2007b} Hern{\'a}ndez, 
J., Calvet, N., Brice{\~n}o, C., et al.\ 2007, \apj, 671, 1784 

\bibitem[Hern{\'a}ndez et al.(2008)]{Hernandez2008} Hern{\'a}ndez, 
J., Hartmann, L., Calvet, N., et al.\ 2008, \apj, 686, 1195

\bibitem[Howell et al.(2014)]{Howell2014} Howell, S.~B., Sobeck, 
C., Haas, M., et al.\ 2014, \pasp, 126, 398 

\bibitem[Huber \& Herzberg(1979)]{Huber1979} K. P. Huber and G. Herzberg, Molecular Spectra and Molecular Structure IV. Constants of Diatomic Molecules (Van Nostrand Reinhold, New York, 1979)

\bibitem[Isella et al.(2009)]{Isella2009} Isella, A., Carpenter, 
J.~M., \& Sargent, A.~I.\ 2009, \apj, 701, 260 

\bibitem[Isella et al.(2010)]{Isella2010} Isella, A., Carpenter, 
J.~M., \& Sargent, A.~I.\ 2010, \apj, 714, 1746 

\bibitem[Kelly(2007)]{Kelly2007} Kelly, B.~C.\ 2007, \apj, 665, 
1489 

\bibitem[Kenyon 
\& Hartmann(1995)]{Kenyon1995} Kenyon, S.~J., \& Hartmann, L.\ 1995, \apjs, 101, 117 

\bibitem[Kitamura et al.(2002)]{Kitamura2002} Kitamura, Y., Momose, 
M., Yokogawa, S., et al.\ 2002, \apj, 581, 357 

\bibitem[Kraus et al.(2015)]{Kraus2015} Kraus, A.~L., Cody, 
A.~M., Covey, K.~R., et al.\ 2015, \apj, 807, 3 

\bibitem[Lada et al.(2006)]{Lada2006} Lada, C.~J., Muench, 
A.~A., Luhman, K.~L., et al.\ 2006, \aj, 131, 1574 

\bibitem[Lawrence et al.(2007)]{Lawrence2007} Lawrence, A., Warren, 
S.~J., Almaini, O., et al.\ 2007, \mnras, 379, 1599 

\bibitem[Lee et al.(2011)]{Lee2011} Lee, N., Williams, J.~P., 
\& Cieza, L.~A.\ 2011, \apj, 736, 135 

\bibitem[Lodieu et al.(2015)]{Lodieu2015} Lodieu, N., Alonso, R., 
Gonzalez Hernandez, J.~I., et al.\ 2015, arXiv:1511.03083 

\bibitem[Luhman(1999)]{Luhman1999} Luhman, K.~L.\ 1999, \apj, 525, 
466 

\bibitem[Luhman et al.(2010)]{Luhman2010} Luhman, K.~L., Allen, 
P.~R., Espaillat, C., Hartmann, L., \& Calvet, N.\ 2010, \apjs, 186, 111 

\bibitem[Luhman 
\& Mamajek(2012)]{Luhman2012} Luhman, K.~L., \& Mamajek, E.~E.\ 2012, \apj, 758, 31 

\bibitem[Mangum 
\& Shirley(2015)]{Mangum2015} Mangum, J.~G., \& Shirley, Y.~L.\ 2015, \pasp, 127, 266 

\bibitem[Mart{\'{\i}}-Vidal et 
al.(2014)]{Marti-Vidal2014} Mart{\'{\i}}-Vidal, I., Vlemmings, W.~H.~T., Muller, S., \& Casey, S.\ 2014, \aap, 563, A136 

\bibitem[Mathews et al.(2012)]{Mathews2012} Mathews, G.~S., 
Williams, J.~P., M{\'e}nard, F., et al.\ 2012, \apj, 745, 23 

\bibitem[Mathews et al.(2013)]{Mathews2013} Mathews, G.~S., Pinte, C., Duch{\^e}ne, G., Williams, J.~P., \& M{\'e}nard, F.\ 2013, \aap, 558, A66 

\bibitem[McMullin et al.(2007)]{McMullin2007} McMullin, J.~P., 
Waters, B., Schiebel, D., Young, W., 
\& Golap, K.\ 2007, adass XVI, 376, 127 

\bibitem[Megeath et al.(2005)]{Megeath2005} Megeath, S.~T., 
Hartmann, L., Luhman, K.~L., \& Fazio, G.~G.\ 2005, ApJL, 634, L113 

\bibitem[Mordasini et al.(2010)]{Mordasini2010} Mordasini, C., Klahr, 
H., Alibert, Y., Benz, W., \& Dittkrist, K.-M.\ 2010, arXiv:1012.5281 

\bibitem[Motte et 
al.(1998)]{Motte1998} Motte, F., Andre, P., \& Neri, R.\ 1998, \aap, 336, 150 

\bibitem[M{\"u}ller et 
al.(2001)]{Muller2001} M{\"u}ller, H.~S.~P., Thorwirth, S., Roth, D.~A., \& Winnewisser, G.\ 2001, \aap, 370, L49 

\bibitem[M{\"u}ller et al.(2005)]{Muller2005} M{\"u}ller, 
H.~S.~P., Schl{\"o}der, F., Stutzki, J., 
\& Winnewisser, G.\ 2005, Journal of Molecular Structure, 742, 215 

\bibitem[Natta 
\& Testi(2004)]{Natta2004} Natta, A., \& Testi, L.\ 2004, Star Formation in the Interstellar Medium: In Honor of David Hollenbach, 323, 279 

\bibitem[Nuernberger et 
al.(1997)]{Nuernberger1997} Nuernberger, D., Chini, R., \& Zinnecker, H.\ 1997, \aap, 324, 1036 

\bibitem[Pecaut et al.(2012)]{Pecaut2012} Pecaut, M.~J., Mamajek, 
E.~E., \& Bubar, E.~J.\ 2012, \apj, 746, 154 

\bibitem[Pecaut 
\& Mamajek(2013)]{Pecaut2013} Pecaut, M.~J., \& Mamajek, E.~E.\ 2013, \apjs, 208, 9 

\bibitem[Pi{\'e}tu et al.(2014)]{Pietu2014} Pi{\'e}tu, V., Guilloteau, S., Di Folco, E., Dutrey, A., \& Boehler, Y.\ 2014, \aap, 564, A95 

\bibitem[Prato et al.(2003)]{Prato2003} Prato, L., Greene, T.~P., 
\& Simon, M.\ 2003, \apj, 584, 853 

\bibitem[Preibisch et al.(2002)]{Preibisch2002} Preibisch, T., Brown, 
A.~G.~A., Bridges, T., Guenther, E., \& Zinnecker, H.\ 2002, \aj, 124, 404 

\bibitem[Preibisch 
\& Mamajek(2008)]{Preibisch2008} Preibisch, T., \& Mamajek, E.\ 2008, Handbook of Star Forming Regions, Volume II, 5, 235 

\bibitem[R Development Core Team(2008)]{R} R Development Core Team(2008). R: A language and environment for
statistical computing. R Foundation for Statistical Computing,
Vienna, Austria. ISBN 3-900051-07-0, URL http://www.R-project.org.

\bibitem[Reboussin et 
al.(2015)]{Reboussin2015} Reboussin, L., Guilloteau, S., Simon, M., et al.\ 2015, \aap, 578, A31 

\bibitem[Rebull et al.(2010)]{Rebull2010} Rebull, L.~M., Padgett, 
D.~L., McCabe, C.-E., et al.\ 2010, \apjs, 186, 259 

\bibitem[Ricci et 
al.(2010)]{Ricci2010} Ricci, L., Testi, L., Natta, A., et al.\ 2010, \aap, 512, A15 

\bibitem[Roeser et al.(2010)]{Roeser2010} Roeser, S., Demleitner, 
M., \& Schilbach, E.\ 2010, \aj, 139, 2440 

\bibitem[Schmidt-Kaler(1982)]{SchmidtKaler1982} Schmidt-Kaler, T.\ 1982, 
Bulletin d'Information du Centre de Donnees Stellaires, 23, 2 

\bibitem[Sicilia-Aguilar et al.(2006)]{Sicilia-Aguilar2006} 
Sicilia-Aguilar, A., Hartmann, L., Calvet, N., et al.\ 2006, \apj, 638, 897 

\bibitem[Siess et 
al.(2000)]{Siess2000} Siess, L., Dufour, E., \& Forestini, M.\ 2000, \aap, 358, 593 

\bibitem[Skrutskie et al.(2006)]{Skrutskie2006} Skrutskie, M.~F., 
Cutri, R.~M., Stiening, R., et al.\ 2006, \aj, 131, 1163 

\bibitem[Slesnick et al.(2006)]{Slesnick2006} Slesnick, C.~L., 
Carpenter, J.~M., \& Hillenbrand, L.~A.\ 2006, \aj, 131, 3016 

\bibitem[Slesnick et al.(2008)]{Slesnick2008} Slesnick, C.~L., 
Hillenbrand, L.~A., \& Carpenter, J.~M.\ 2008, \apj, 688, 377 

\bibitem[Strai{\v z}ys(1992)]{Straizys1992} Strai{\v z}ys, V.\ 1992, Tucson : Pachart Pub.~House, c1992.,  

\bibitem[Testi et al.(2014)]{Testi2014} Testi, L., Birnstiel, T., Ricci, L., et al.\ 2014, Protostars and Planets VI, 339 

\bibitem[Tognelli et 
al.(2011)]{Tognelli2011} Tognelli, E., Prada Moroni, P.~G., \& Degl'Innocenti, S.\ 2011, \aap, 533, A109 

\bibitem[Ubach et al.(2012)]{Ubach2012} Ubach, C., Maddison, 
S.~T., Wright, C.~M., et al.\ 2012, \mnras, 425, 3137 

\bibitem[Werner et al.(2004)]{Werner2004} Werner, M.~W., Roellig, 
T.~L., Low, F.~J., et al.\ 2004, \apjs, 154, 1 

\bibitem[Williams et al.(2013)]{Williams2013} Williams, J.~P., 
Cieza, L.~A., Andrews, S.~M., et al.\ 2013, \mnras, 435, 1671 

\bibitem[Williams \& Best(2014)]{Williams2014} Williams, J.~P., \& Best, W.~M.~J.\ 2014, \apj, 788, 59 

\bibitem[Woitke et al.(2011)]{Woitke2011} Woitke, P., Riaz, B., Duch{\^e}ne, G., et al.\ 2011, \aap, 534, A44 

\bibitem[Wright et al.(2010)]{Wright2010} Wright, E.~L., 
Eisenhardt, P.~R.~M., Mainzer, A.~K., et al.\ 2010, \aj, 140, 1868-1881

\bibitem[van der Plas et al.(2016)]{VanDerPlas2016} van der Plas, G., M{\'e}nard, F., Ward-Duong, K., et al.\ 2016, \apj, 819, 102  

\bibitem[Zhang et al.(2014)]{Zhang2014} Zhang, K., Isella, A., 
Carpenter, J.~M., \& Blake, G.~A.\ 2014, \apj, 791, 42 

\end{thebibliography}
\end{document}